\documentclass[aps,manuscript,showpacs,showkeys,superscriptaddress,nofootinbib]{revtex4-1}
\usepackage{graphicx}
\usepackage{epsfig}  
\usepackage{epsf}    
\usepackage{dcolumn}
\usepackage{bm}
\usepackage{dcolumn}
\usepackage{textcomp}
\usepackage[tbtags]{amsmath}
\usepackage{amsfonts}
\usepackage{float}
\usepackage{subfig}
\usepackage[]{hyperref}
  \hypersetup{
  unicode=false,          
  pdftoolbar=true,        
  pdfmenubar=true,        
  pdffitwindow=true,     
  pdfstartview={FitH},    
  pdfsubject={Getting the best out of T2K and NOvA},   
  pdfnewwindow=true,      
  pdfcreator={RevTeX},
  colorlinks=true,       
  linkcolor=red,          
  citecolor=blue,        
  urlcolor=blue,           
  }
\usepackage{hypcap}

\def\anu{{\bar\nu}}

\newcommand{\beq}{\begin{equation}}
\newcommand{\eeq}{\end{equation}}
\newcommand{\beqa}{\begin{eqnarray}}
\newcommand{\eeqa}{\end{eqnarray}}

\newcommand{\tx}{{\theta_{12}}}
\newcommand{\ty}{{\theta_{13}}}
\newcommand{\tz}{{\theta_{23}}}

\newcommand{\dl}{{\Delta_{31}}}
\newcommand{\ds}{{\Delta_{21}}}

\newcommand{\ahat}{\hat{A}}
\newcommand{\dhat}{\hat{\Delta}}

\newcommand{\dcp}{\delta_{\mathrm{CP}}}
\newcommand{\nova}{NO$\nu$A~}

\newcommand{\pme}{P_{\mu e}}
\newcommand{\pmuebar}{P(\bar{\nu}_{\mu} \rightarrow \bar{\nu}_e)}

\newcommand{\pmebar}{P_{\bar{\mu} \bar{e}}}

\newcommand{\dchsq}{\Delta\chi^2}

\newcommand{\dmm}{\Delta m^2_{\mu\mu}}

\begin{document}


\title{A review of the tension between the T2K and NO$\nu$A appearance data and hints to new physics}

\author{Ushak Rahaman}
\email[Email Address: ]{ushakr@uj.ac.za}
\affiliation{Centre for Astro-Particle Physics (CAPP) and Department of Physics, University of Johannesburg, PO Box 524, Auckland Park 2006, South Africa}
\author{Soebur Razzaque}
\email[Email Address: ]{srazzaque@uj.ac.za}
\affiliation{Centre for Astro-Particle Physics (CAPP) and Department of Physics, University of Johannesburg, PO Box 524, Auckland Park 2006, South Africa}

\author{S. Uma Sankar}
\email[Email Address: ]{uma@phy.iitb.ac.in}
\affiliation{Dept. of Physics, IIT Bombay, Powai, Mumbai-400076, India}
\date{\today}
\begin{abstract}
In this article, we review the status of the tension between the long-baseline accelerator neutrino experiments T2K and NO$\nu$A. The tension arises mostly due to the mismatch in the appearance data of the two experiments. We explain how this tension arises based on $\nu_\mu \to \nu_e$ and $\bar{\nu}_\mu \to \bar{\nu}_e$ oscillation probabilities. We define the reference point of vacuum oscillation, maximal $\tz$ and $\dcp=0$ and compute the $\nu_e/\bar{\nu}_e$ appearance events for each experiment. We then study the effects of deviating the unknown parameters from the reference point and the compatibility of any given set of values of
unknown parameters with the data from T2K and NO$\nu$A. 
T2K observes a large excess in the $\nu_e$ appearance event sample compared to the expected $\nu_e$ events  at the reference point, whereas \nova observes a moderate excess. The large excess in T2K dictates that $\dcp$ be anchored at $-90^\circ$ and that $\tz>\pi/4$ with a preference for normal hierarchy. The moderate excess at \nova leads to two degenerate solutions: A) NH, $0<\dcp<180^\circ$, and $\tz>\pi/4$; B)  IH, $-180^\circ<\dcp<0$, and $\tz>\pi/4$. This is 
the main cause of the tension between the two experiments. We have reviewed the status of three beyond standard model (BSM) physics scenarios, (a) non-unitary mixing, (b) Lorentz invariance  violation and (c) non-standard neutrino interactions, to resolve the tension.
\end{abstract}

\maketitle

\section{Introduction}
Neutrino oscillations have provided the first signal for physics beyond the standard model (SM). They were first proposed to explain the deficit in the solar neutrino flux observed by the pioneering Homestake experiment \cite{Cleveland:1998nv}. Oscillations between two neutrino flavours require them to mix and form two mass eigenstates. The survival probability of a neutrino with energy $E$ and given  flavour $\alpha$, after propagation over a distance $L$ in vacuum, is given by 
\begin{equation}
    P(\nu_\alpha \to \nu_\alpha) = 1 - \sin^2 2 \theta
    \sin^2 \left( 1.27 \frac{\Delta m^2 L}{E} \right),
    \label{2flsurprob}
\end{equation}
where $\Delta m^2$ is the difference between the squares of the neutrino masses and $\theta$ is the mixing angle. In eq.~(\ref{2flsurprob}), the units are chosen such that $\Delta m^2$ should be specified in eV$^2$, $L$ in meters and $E$ in MeV. The solar neutrino deficit was confirmed by the water Cerenkov detector Kamiokande \cite{Kamiokande-II:1991pyu}, which detected the solar neutrinos in real time. Radio-chemical Gallium experiments, GALLEX \cite{Hampel:1998xg}, SAGE \cite{Abdurashitov:2009tn} and GNO \cite{GNO:2005bds}, which were mostly sensitive to the low energy $pp$ solar neutrinos, also observed a deficit. The high statistics water Cerenkov detector Super-Kamiokande \cite{Super-Kamiokande:2016yck} and the heavy water Cerenkov detector SNO \cite{SNO:2011ajh} and Borexino  \cite{Borexino:2017rsf} also have made detailed spectral measurements of the solar neutrino fluxes. Analysing the solar neutrino data in a two flavour oscillation framework gives the oscillation parameters
\begin{equation}
    \Delta m^2_{\rm sol} \sim 10^{-4}~{\rm eV}^2~~~~{\rm and}~~~~
    \sin^2 \theta_{\rm sol} \sim 0.33.
\end{equation}

Observation of proton decay is one of the main physics motivations for the construction of the water Cerenkov detectors, IMB \cite{Casper:1990ac, BeckerSzendy:1992hq} and Kamiokande \cite{Hirata:1992ku, Fukuda:1994mc}. The interactions of atmospheric neutrinos in the detector, especially those of $\nu_e$ and $\bar{\nu}_e$, could mimic the proton decay signal. Hence, these experiments made a detailed study of the atmospheric neutrino interactions. They did not find any signal for proton decay but instead observed a deficit of up-going atmospheric $(\nu_\mu + \bar{\nu}_\mu)$ flux, relative to the down-going flux. It was proposed that the up-going neutrinos, which travel thousands of km inside the Earth, oscillate into another flavour whereas the down-going neutrinos, which travel tens of km, do not. Super-Kamiokande experiment \cite{Fukuda:1998mi} observed a zenith angle dependence of the deficit, which is expected from neutrino oscillations. An analysis of the atmospheric neutrino data in a two-flavour oscillation framework gives the oscillation parameters
\begin{equation}
    |\Delta m^2_{\rm atm}| \sim 3 \times 10^{-3}~{\rm eV}^2~~~~{\rm and}~~~~
    \sin^2 \theta_{\rm atm} \sim 0.5.
\end{equation}
We note that $\Delta m^2_{\rm sol} \ll \Delta m^2_{\rm atm}$.

It is known that there are three flavours of neutrinos, $\nu_e$, $\nu_\mu$ and $\nu_\tau$ \cite{Steinberger:1990hr}. They mix to form three mass eigenstates, $\nu_1$, $\nu_2$ and $\nu_3$, with mass eigenvalues $m_1$, $m_2$ and $m_3$. The $3 \times 3$ unitary matrix $U$, connecting the flavour basis to the mass basis,
\begin{equation}
    \left[ \begin{array}{c} \nu_e \\ \nu_\mu \\ \nu_\tau 
    \end{array} \right]
    =\left[ \begin{array}{ccc} U_{e1} & U_{e2} & U_{e3} \\ 
     U_{\mu 1} & U_{\mu 2} & U_{\mu 3} \\ 
     U_{\tau 1} & U_{\tau 2} & U_{\tau 3} 
    \end{array} \right]
    \left[ \begin{array}{c} \nu_1 \\ \nu_2 \\ \nu_3
    \end{array} \right],
\end{equation}
is called the Pontecorvo-Maki-Nakagawa-Sakata (PMNS) matrix \cite{Maki:1962mu, Bilenky:1978nj}. Naively, it seems desirable to label the lightest mass $m_1$, the middle mass $m_2$ and the heaviest mass $m_3$. However, no method exists at present to directly measure these masses. What can be measured in oscillation experiments are the mixing matrix elements $U_{\alpha j}$, where $\alpha$ is a flavour index and $j$ is a mass index. In particular, the three elements of the first row $U_{e1}$, $U_{e2}$ and $U_{e3}$ are well measured. The labels $1$, $2$ and $3$ are chosen such that $|U_{e3}| < |U_{e2}| < |U_{e1}|$. 

Given the three masses $m_1$, $m_2$ and $m_3$, it is possible to define two independent mass-squared differences, $\Delta_{31} = m_3^2 - m_1^2$ and $\Delta_{21} = m_2^2 - m_1^2$. The third mass-squared difference is then $\Delta_{32} = m_3^2 - m_2^2 = \Delta_{31}- \Delta_{21}$. Without loss of generality, we can choose $\Delta_{21} = \Delta m^2_{\rm sol}$ and $\Delta_{31} = \Delta m^2_{\rm atm}$. Since $\Delta m^2_{\rm sol} \ll \Delta m^2_{\rm atm}$, we find that $\Delta_{32} \approx \Delta m^2_{\rm atm}$. Because of the way the labels $1$, $2$ and $3$ are chosen, these mass-squared differences, in principle, can be either positive or negative. Their signs have to be determined by experiments. The solar neutrinos, produced at the core of the sun, undergo forward elastic scattering as they travel through the solar matter. This scattering leads to matter effect \cite{msw1, Wolfenstein:1979ni}, which modifies the solar electron neutrino survival probability $P_{\rm ee}$. Super-Kamiokande  \cite{Super-Kamiokande:2017yvm} and SNO \cite{SNO:2011ajh} have measured $P_{\rm ee}$ as a function of neutrino energy for $E > 5$ MeV and found it to be of a constant value $\simeq 0.3$. SNO has also measured \cite{SNO:2011ajh} the neutral current interaction rate of solar neutrinos to be consistent with predictions of the standard solar model \cite{Bahcall:2004fg}. The measurements of the Gallium experiments imply that
$P_{\rm ee} > 0.5$ for neutrino energies $E < 0.5$ MeV. This increase in $P_{\rm ee}$ at lower solar neutrino energies can be explained only if $\Delta m^2_{\rm sol} = \Delta_{21}$ is positive. At present, there is no definite experimental evidence for either positive or negative sign of $\Delta m^2_{\rm atm} = \Delta_{31} \approx \Delta_{32}$.

The PMNS matrix is similar to the quark mixing matrix introduced by Kobayashi and Maskawa \cite{Kobayashi:1973fv}. It can be parameterized in terms of three mixing angles, $\theta_{12}$, $\theta_{13}$ and $\theta_{23}$, and one CP-violating phase $\dcp$. The following parameterization of the PMNS matrix is found to be the most convenient to analyze the neutrino data
\begin{equation}
    U = \left( \begin{array}{ccc} 1 & 0 & 0 \\
    0 & c_{23} & s_{23} \\ 0 & -s_{23} & c_{23} \end{array} \right)
    \left( \begin{array}{ccc} c_{13} & 0 & s_{13} e^{-i\dcp} \\
    0 & 1 & 0 \\ -s_{13} e^{i\dcp} & 0 & c_{13} \end{array} \right)
    \left( \begin{array}{ccc} c_{12} & s_{12} & 0 \\
    -s_{12} & c_{12} & 0 \\ 0 & 0 & 1 \end{array} \right),
\end{equation}
where $c_{ij} = \cos \theta_{ij}$ and $s_{ij} = \sin \theta_{ij}$. Among these mixing angles, CHOOZ experiment set a strong upper bound on the middle angle $\theta_{13}$ \cite{Apollonio:1997xe}
\begin{equation}
    \sin^2 2 \theta_{13} \leq 0.1.
\end{equation}
By combining this limit with the solar and atmospheric neutrino data, it can be shown that $\theta _{13} \ll 1$ \cite{Narayan:1997mk}.

The mixing angles $\theta_{ij}$ and the mass-squared  differences $\Delta m^2_{ij}$ have been determined in a series of precision experiments with man-made neutrino sources, which we briefly describe below. 
\begin{itemize}
    \item The long baseline reactor neutrino experiment KamLAND \cite{KamLAND:2004mhv} has $L \simeq 180$ km. At this long distance, it can observe the oscillations due to the small mass-squared difference $\Delta_{21}$. In the limit of neglecting $\theta_{13}$ in the three-flavour oscillation, the expression for the anti-neutrino survival probability $P( \bar{\nu}_e \to \bar{\nu}_e)$ reduces to an effective two flavour expression
    \begin{equation}
    P( \bar{\nu}_e \to \bar{\nu}_e) = 1 - \sin^2 2 \theta_{12}
    \sin^2 \left( 1.27 \frac{\Delta_{21} L}{E} \right).
    \end{equation}
    KamLAND measured the spectral distortion $P( \bar{\nu}_e \to \bar{\nu}_e)$ precisely. A combined analysis of KamLAND and solar neutrino data yields the results
    \begin{equation}
    \Delta_{21}= (7.9 \pm 0.5) \times 10^{-5}~{\rm eV}^2~~~~{\rm and}~~~~
    \tan^2 \theta_{12} = 0.4^{+0.10}_{-0.07}
    \end{equation}
    \item The short baseline reactor neutrino experiments, Daya Bay \cite{DayaBay:2018yms}, RENO \cite{RENO:2018dro} and DoubleCHOOZ \cite{DoubleChooz:2015mfm}, have baselines of the order of 1 km. At this distance, the oscillating term in $P( \bar{\nu}_e \to \bar{\nu}_e)$ containing $\Delta_{21}$ is negligibly small. In this approximation, $P( \bar{\nu}_e \to \bar{\nu}_e)$ again reduces to an effective two-flavour expression
    \begin{equation}
    P( \bar{\nu}_e \to \bar{\nu}_e) = 1 - \sin^2 2 \theta_{13}
    \sin^2 \left( 1.27 \frac{\Delta_{31} L}{E} \right).
    \end{equation}
    High statistics measurement from Daya Bay gives the measurement
    \begin{equation}
    |\Delta_{31}| = (2.47 \pm 0.07) \times 10^{-3}~{\rm eV}^2~~~~{\rm and}~~~~
    \sin^2 2 \theta_{13} = (0.0856 \pm 0.0029).
    \end{equation}
    \item The long baseline accelerator experiment MINOS \cite{MINOS:2014rjg} has a baseline of 730 km and it measured the survival probability of the accelerator $\nu_\mu$ beam. For this baseline and for accelerator neutrino energies, the oscillating term in $P(\nu_\mu \to \nu_\mu)$ due to $\Delta_{21}$ is negligibly small. Setting this term and $\theta_{13}$ to be zero in the
    three flavour expression for $P(\nu_\mu \to \nu_\mu)$, once
    again we obtain an effective two flavour expression
    \begin{equation}
    P(\nu_\mu \to \nu_\mu) = 1 - \sin^2 2 \theta_{23}
    \sin^2 \left( 1.27 \frac{\Delta_{31} L}{E} \right).
    \end{equation}
    MINOS data gives the results
    \begin{equation}
    |\Delta_{31}| = (2.3 - 2.5) \times 10^{-3}~{\rm eV}^2~~~~{\rm and}~~~~
    \sin^2 \theta_{23} = (0.35 - 0.65).
    \end{equation}
\end{itemize}
Note that both short baseline reactor neutrino experiments and long baseline accelerator experiments as well as atmospheric neutrino data determine only the magnitude of $\Delta_{31}$ but not its sign. Hence, we must consider both positive and negative sign possibilities in the data analysis. The case of positive $\Delta_{31}$ is called normal hierarchy (NH) and that of negative $\Delta_{31}$ is called inverted hierarchy (IH). Both atmospheric neutrino data and accelerator neutrino data are functions  of $\sin^2 2 \theta_{23}$ and they prefer $\sin^2 2 \theta_{23} \simeq 1$. For such values, there are two possibilities: $\theta_{23} < 45^\circ$ called lower octant (LO) and $\theta_{23} > 45^\circ$ called higher octant (HO). At present, the data is not able to make a distinction between these two cases.

A number of groups have done global analysis of neutrino oscillation data from all the available sources: solar, atmospheric, reactor and accelerator \cite{Esteban:2020cvm, deSalas:2020pgw}. In Table~\ref{nu-fit-2021}, we present the latest results obtained by the nu-fit collaboration \cite{Esteban:2020cvm}.
\begin{table}[htbp]
    \centering
    \begin{tabular}{|c|c|c|} \hline \hline
    ~~~~ & Normal Hierarchy & Inverted Hierarchy \\ \hline
    $\sin^2 \theta_{12}$ & $0.304^{+0.012}_{-0.012}$ &  $0.304^{+0.013}_{-0.012}$ \\ \hline
    $\sin^2 \theta_{13}$ & $0.02246 \pm 0.00062$ & $0.02241^{+0.00074}_{-0.00062}$ \\ \hline
    $\sin^2 \theta_{23}$ & $0.45^{+0.019}_{-0.016}$ & $0.570^{+0.016}_{-0.022}$ \\ \hline
    $\frac{\Delta_{21}}{10^{-5}~{\rm eV^2}}$ & $7.42 \pm 0.2$  & 
    $7.42 \pm 0.2$  \\ \hline
    $\frac{\Delta_{31}}{10^{-3}~{\rm eV^2}}$ & $2.51 \pm 0.027$  & 
    $-2.49^{+0.026}_{-0.028}$    \\ \hline
    \end{tabular}
    \caption{Neutrino oscillation parameters determine by nu-fit collaboration using global neutrino oscillation data \cite{Esteban:2020cvm}.}
    \label{nu-fit-2021}
\end{table}

Among the neutrino oscillation parameters, the mass-squared differences and the mixing angles (except for $\theta_{23}$) are measured to a precision of a few percent. On the other hand, the CP-violating phase $\dcp$ still eludes measurement. In addition, we also need to resolve the issues of the sign of $\Delta_{31}$ (also called the problem of neutrino mass hierarchy) and the octant of $\theta_{23}$. Thus, there are currently three main unknowns in the three-flavour neutrino oscillation paradigm.

It can be shown that the survival probability of neutrinos, $P(\nu_\alpha \to \nu_\alpha)$ is equal to that of the anti-neutrinos $P(\bar{\nu}_\alpha \to \bar{\nu}_\alpha)$ due to CPT invariance. But the oscillation probabilities $P(\nu_\alpha \to \nu_\beta)$ and $P(\bar{\nu}_\alpha \to \bar{\nu}_\beta)$ are not equal if there is CP violation. A measurement of the difference between these two probabilities will establish CP-violation in neutrino oscillations and will also determine $\dcp$. While considering oscillation probabilities, in principle, we can enumerate six possibilities, with $\alpha = e/\mu/\tau$ and $\beta$ taking two possible values other than $\alpha$. It can be shown that, in all six cases, 
the difference
\begin{equation}
    \Delta P^{\alpha \beta}_{\rm CP} = 
    P(\nu_\alpha \to \nu_\beta) - P(\bar{\nu}_\alpha \to \bar{\nu}_\beta)
\end{equation}
is proportional to 
\begin{equation}
 \frac{\Delta_{21}}{\Delta_{31}} \cos \theta_{13} \sin 2 \theta_{12} 
 \sin 2 \theta_{13} \sin 2 \theta_{23} \sin \dcp = 
 \frac{\Delta_{21}}{\Delta_{31}} J,
 \label{jarlskog}
\end{equation}
where $J$ is the Jarlskog invariant of the PMNS matrix \cite{Jarlskog:1985ht}.

Practically speaking, $\alpha = e/\tau$ at production are not possible because there are no intense sources of $\nu_e/\nu_\tau/\bar{\nu}_\tau$. Nuclear reactors do produce $\bar{\nu}_e$ copiously but their energies are in the  range of a few MeV. When such $\bar{\nu}_e$ oscillate into $\bar{\nu}_\mu$, the resultant anti-neutrinos are not energetic enough to produce $\mu^+$ by their interactions in the detector. Hence, $\alpha = e/\tau$ are not practical choices. The neutrino beams produced by accelerators yield intense fluxes of $\nu_\mu/\bar{\nu}_\mu$. In principle, it is possible to search for CP-violation by contrasting $P(\nu_\mu \to \nu_e)$ with $P(\bar{\nu}_\mu \to \bar{\nu}_e)$ or by contrasting $P(\nu_\mu \to \nu_\tau)$ with $P(\bar{\nu}_\mu \to \bar{\nu}_\tau)$. However, the second option is much more difficult compared to the first for the reasons listed below.
\begin{itemize}
    \item To produce $\tau^\mp$ in the detector, due to the interactions of $\nu_\tau/\bar{\nu}_\tau$, the neutrinos should have energies of tens of GeV. At these energies, the oscillation probabilities are quite small.
    \item Even when we have energetic-enough beams to produce $\tau^\mp$ in the detector, the efficiency of reconstructing these particles is very poor. Thus the event numbers will be very limited.
    \item From eq.~(\ref{jarlskog}), we see that the CP violating asymmetry 
    $$
    A_{CP}^{\mu \tau} = \frac{P(\nu_\mu \to \nu_\tau) - P(\bar{\nu}_\mu \to \bar{\nu}_\tau)}{P(\nu_\mu \to \nu_\tau) + P(\bar{\nu}_\mu \to \bar{\nu}_\tau)}
    $$
    will be very small because the numerator is a product of two small quantities $(\Delta_{21}/\Delta_{31})$ and $\sin 2 \theta_{13}$, whereas the denominator is close to $1$. Hence a measurement of this CP-asymmetry
    requires very high statistics. 
\end{itemize}
Thus the most feasible method to establish CP-violation in neutrino oscillations and to determine $\dcp$ is to measure the difference between $P(\nu_\mu \to \nu_e)$ and $P(\bar{\nu}_\mu \to \bar{\nu}_e)$. The neutrinos do not require large energies to produce electrons/positrons on interacting in the detector. So the neutrino beam energy can be tuned to the oscillation maximum. The produced electrons and positrons are relatively easy to identify in the detector. The dominant term in the expression for $P(\nu_\mu \to \nu_e)$ is proportional to $\sin^2 \theta_{23} \sin^2 2 \theta_{13}$ \cite{Cervera:2000kp}. The expression for the CP-asymmetry in $\nu_\mu \to \nu_e$ oscillations has the form
\begin{equation}
    A_{\rm CP}^{\mu e} \sim \frac{\Delta_{21}}{\Delta_{31}} 
    \frac{J}{\sin^2 2 \theta_{13}} \sim \frac{\Delta_{21}}{\Delta_{31}} 
     \frac{1}{\sin 2 \theta_{13}},
\end{equation}
which is much larger than $A_{\rm CP}^{\mu \tau}$. Thus, CP-violation in these oscillations can be established with moderate statistics. 

There are, however, some other difficulties to overcome before the goal of establishing CP-violation in neutrino oscillations can be achieved. The matter effect, which modifies the solar neutrino oscillation probabilities, modifies $P(\nu_\mu \to \nu_e)$ and $P(\bar{\nu}_\mu \to \bar{\nu}_e)$ \cite{Lipari:1999wy,Narayan:1999ck}. 
These modifications depend on the sign of $\Delta_{31}$. Since the dominant terms in these oscillation probabilities are proportional to $\sin^2 \theta_{23}$, they are also subject to the octant ambiguity of $\theta_{23}$. That is, the two oscillation probabilities, $P(\nu_\mu \to \nu_e)$ and $P(\bar{\nu}_\mu \to \bar{\nu}_e)$, depend on {\bf all the three unknowns} of the three flavour neutrino oscillation parameters. In such a situation, the change in the probabilities induced by changing one of the unknowns can be compensated by changing another unknown. This leads to degenerate solutions which can explain a given set of measurements. Unravelling these degeneracies and making a distinction between the degenerate solutions requires a number of careful measurements and moderately high statistics.

In this review article, we have discussed the theory of oscillation probability and parameter degeneracy in section~\ref{degeneracy}. Details of the $\chi^2$ analysis have been discussed in section \ref{chisq}. In section~\ref{evolution}, we have discussed the chronology of \nova and T2K data. In the same section, we have also tried to explain the results of the analysis of data from \nova and T2K in the past and the present ones as well with the help of parameter degeneracy and explain the cause for the tension between the data
of the two experiments. The resolution of the tension in terms of BSM physics has been discussed in section~\ref{resolution}. A summary of the article has been drawn in section~\ref{summary}.

\section{Oscillation probability and parameter degeneracy}
\label{degeneracy}
Two accelerator experiments, T2K \cite{Itow:2001ee} and \nova \cite{Ayres:2007tu},
are taking data with the aim of establishing CP-violation as well as determining
neutrino mass hierarchy and the octant of $\theta_{23}$. 
Both experiments share the following common features.
\begin{itemize}
\item They aim a beam of $\nu_\mu/\bar{\nu}_\mu$ to a far detector
a few hundred km away, which is at an off-axis location.
\item The off-axis location leads to a sharp peak in neutrino
spectrum \cite{McDonald:2001mc}, which is crucial to suppress
the $\pi^0$ events, produced in via the
neutral current reaction $\nu_\mu N \to \nu_\mu \pi^0 N$, that
form a large background for the $\nu_\mu \to \nu_e$ oscillation signal.
\item They have a near detector, a few hundred meters from the accelerator,
which measures the neutrino flux accurately.
\item  The energy of the neutrino beam is tuned to be close to the
oscillation maximum.
\item They measure the two survival probabilities, 
$P(\nu_\mu \to \nu_\mu$ and $P(\bar{\nu}_\mu \to \bar{\nu}_\mu)$, 
and the oscillation probabilities 
$P(\nu_\mu \to \nu_e$ and $P(\bar{\nu}_\mu \to \bar{\nu}_e)$.
\end{itemize}
The survival probabilities lead to further improvement in the
precision of $|\Delta_{31}|$ and $\sin 2 \theta_{23}$. A careful analysis
of the oscillation probabilities can lead to the determination of 
three unknowns of the neutrino oscillation parameters. The crucial
parameters of T2K and \nova experiments are summarised in 
table~\ref{T2K-NOvA-Info}. Note that the integrated flux of accelerator 
neutrinos is specified in units of protons on target (POT).

\begin{table}[htbp]
    \centering
    \begin{tabular}{|c|c|c|} \hline \hline
    ~~~~ & T2K & \nova \\ \hline
    Baseline & 295 km &  810 km \\ \hline
    Energy of Peak Flux & 0.7 GeV & 2.0 GeV \\ \hline
    Detector Type & Water Cerenkov & Liquid Scintillator \\ \hline
    Detector Mass & 22.5 ktons  &  14 ktons  \\ \hline
    Total Flux in $\nu$ mode & $13.6 \times 10^{20}$ POT  & 
    $16.3 \times 10^{20}$ POT    \\ \hline
    Total Flux in $\bar{\nu}$ mode & $19.7 \times 10^{20}$ POT  & 
    $12.5 \times 10^{20}$ POT    \\ \hline
    \end{tabular}
    \caption{Summary of important information of T2K and \nova experiments.}
    \label{T2K-NOvA-Info}
\end{table}



We first start with a discussion of the  oscillation probabilities, $P(\nu_\mu \to \nu_e)$
and $P(\bar{\nu}_\mu \to \bar{\nu}_e)$ and
describe how they vary with each of the 
unknown neutrino oscillation parameters. 
The three flavour $\nu_\mu \to \nu_e$ oscillation probability in the presence of matter effect with constant matter density can be written as
\cite{Cervera:2000kp}
\begin{eqnarray}
  \pme &\simeq& \sin^2 2 \ty \sin^2 \tz\frac{\sin^2\dhat(1-\ahat)}{(1-\ahat)^2}\nonumber\\
  &+& \alpha \cos \ty \sin2\tx \sin 2\ty \sin 2\tz \cos(\dhat+\dcp)
 \frac{\sin\dhat \ahat}{\ahat}
  \frac{\sin \dhat(1-\ahat)}{1-\ahat}\nonumber \\
  &+&  \alpha^2 \sin^2 2\theta_{12}\cos^2 \theta_{13}\cos^2\theta_{23}\frac{\sin^2 \dhat \ahat}{\ahat^2},
  \label{pme}
   \end{eqnarray}
where $\alpha=\frac{\ds}{\dl}$, $\dhat=\frac{\dl L}{4E}$ and $\ahat=\frac{A}{\dl}$, with $E$ being the energy of the neutrino and $L$ being the length of the baseline. The parameter $A$ is the Wolfenstein matter term \cite{msw1}, given by $A=2\sqrt{2}G_FN_eE$, where $G_F$ is the
Fermi coupling constant and $N_e$ is the number density of the electrons in the matter. Anti-neutrino oscillation probability $\pmebar$ can be obtained by changing the sign of $A$ and $\dcp$ in eq.~(\ref{pme}). The oscillation probability mainly depends on hierarchy (sign of $\dl$), octant of $\tz$ and $\dcp$, and precision in the value of $\ty$. $\pme$ is enhanced if $\dcp$ is in the lower half plane (LHP, $-180^\circ<\dcp<0$), and it is suppressed if $\dcp$ is in the upper half plane (UHP, $0<\dcp<180^\circ$), compared to the CP conserving $\dcp$ values. In the following paragraph, for the sake of discussion, we will treat $\dcp$ as a binary variable that either increases or decreases oscillation probability.

The dominant term in $\pme$ is proportional to $\sin^2 2\ty$. Therefore, the oscillation probability is rather small. It can be enhanced (suppressed), by $8\%$ for T2K and
$22\%$ for NO$\nu$A, due to the matter effect if $\dl$ is positive (negative). This dominant term is also proportional to $\sin^2 \tz$. If $\sin^2 2\tz<1$, there can be two possible cases: i) $\sin^2 \tz<0.5$ which will suppress $\pme$, and ii) $\sin^2 \tz>0.5$ which will enhance $\pme$ relative to the maximal $\tz$. Since each of the unknowns can take 2 possible values, there are 8 different combinations of three unknowns. Any given value of $\pme$ can be reproduced by any of these eight combinations of the three unknowns with the appropriate choice of $\ty$ value. Thus, if the value of $\ty$ is not known precisely, it will lead to eight fold degeneracy in $\pme$. Given that $\theta_{13}$ has been measured quite precisely, this
degeneracy becomes less severe. 

 \paragraph*{}
\textbf{a. Hierarchy-$\dcp$ degeneracy}: 
To start with, we assume that $\tz$ is maximal and the values of $\ty$ and $\tx$ are precisely known. With these
assumptions, the only two unknowns are hierarchy and
$\dcp$.
From table \ref{nu-fit-2021}, we see that, according to the current
measurements, $\sin 2 \ty \approx 0.3\pm 0.005$
whereas $|\alpha| \approx 0.03\pm 0.001$. Therefore, the first term in the
expression of $\pme$ (and in $\pmebar$) in 
eq.~(\ref{pme}) has the maximum matter effect contribution. 
This term is much larger than second term and the third
term is extremely small. We will neglect the third term in all further discussions.

\begin{figure}[htb]
\centering
\includegraphics[width=0.7\textwidth]{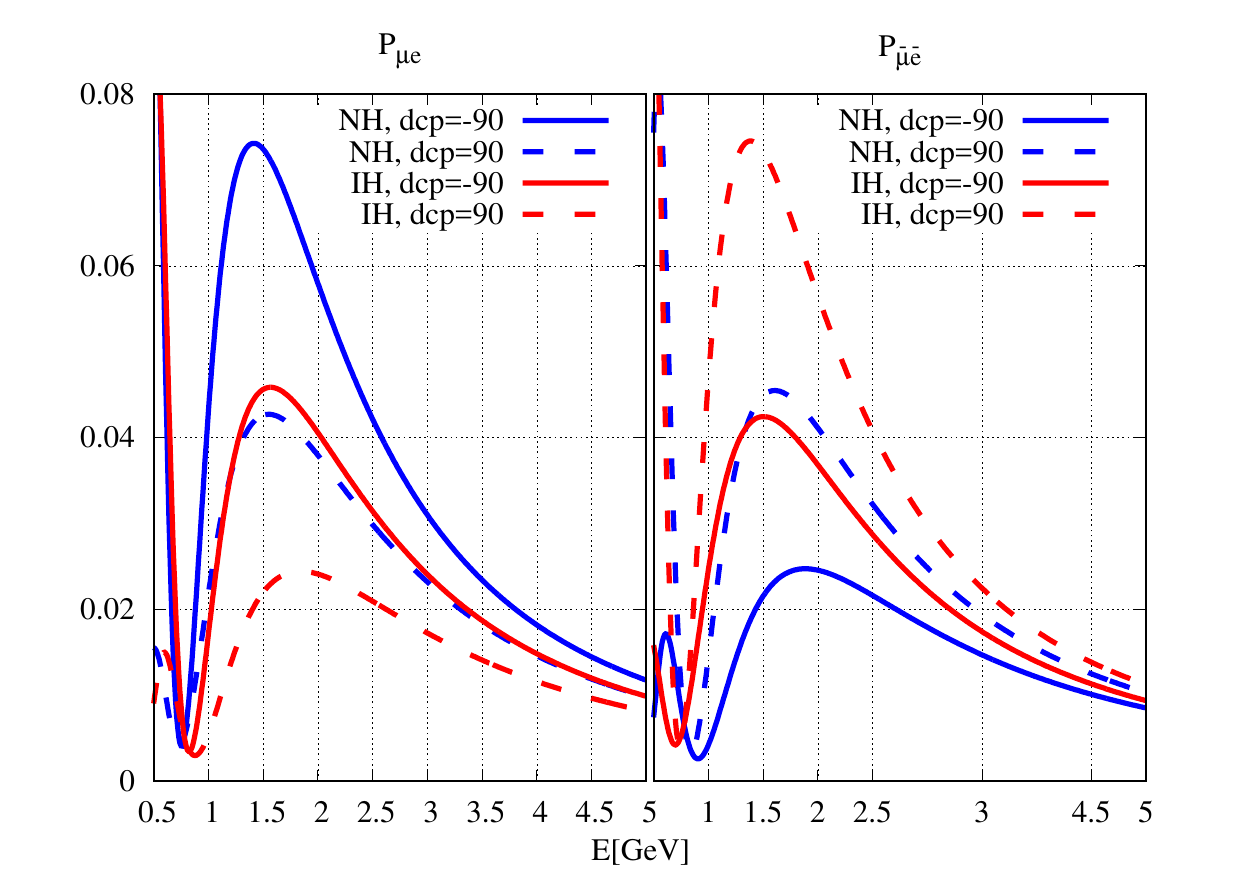}
\caption{\footnotesize{$\pme$ (left panel) and $\pmebar$ (right panel) vs. neutrino energy for the
NO$\nu$A baseline. Variation of $\dcp$ 
leads to the blue (red) bands for NH (IH). The plots are drawn for
maximal $\tz$ and other neutrino parameters given in the main text.}}
\label{prob}
\end{figure}

For NH (IH), the first term in $\pme$ becomes larger (smaller). 
For $\pmebar$, the situation is reversed. These changes in $\pme$
and in $\pmebar$ can be {\it amplified or canceled} by the second 
term, depending on the value of $\dcp$. Because of the
dependence on $\ahat$ term, $\pme$ ($\pmebar$) for NH is always larger (smaller) than
that for IH. At the oscillation maxima, $\dhat \simeq 90^\circ$. Thus for $\pme$
($\pmebar$), the term $\cos (\dhat+\dcp)$ is maximum (minimum) for $\dcp=-90^\circ$
and it is minimum (maximum) for $\dcp=90^\circ$. Therefore, for NH and $\dcp=-90^\circ$,
$\pme$ ($\pmebar$) is maximum (minimum) and for IH and $\dcp=90^\circ$, it is minimum
(maximum). 
These two hierarchy-$\dcp$ combinations, for both $\pme$ and $\pmebar$, are well
separated from each other. It can be shown that oscillation probability, for the NH and $\dcp$ in the LHP, is well separated from that for the IH and $\dcp$ in the UHP, for both neutrino
and anti-neutrino. But for the other two hierarchy-$\dcp$ combinations, NH and $\dcp$ in the UHP, and IH and
$\dcp$ in the LHP, $\pme$ and $\pmebar$ are quite close to each other, leading to 
hierarchy-$\dcp$ degeneracy.
This is illustrated in 
Fig. \ref{prob}, where $\pme$ and $\pmebar$ are plotted for 
the \nova experiment baseline. For these plots, we have used maximal $\tz$, i.e., $\sin^2 \tz=0.5$.
The other mixing angle values are $\sin^2 \tx=0.32$ and $\sin^2 2\ty=0.089$.
For the mass-squared differences, we have used $\ds=7.50 \times 10^{-5}\ {\rm eV}^2$
and $|\Delta_{\mu \mu}|=2.40 \times 10^{-3}\ {\rm eV}^2$. $\dmm$ is related with $\dl$ by the following equation \cite{Nunokawa:2005nx}:
\begin{equation}
\Delta_{\mu \mu}= \sin^2 \tz \dl + \cos^2 \tx \Delta_{32}+\cos \dcp \sin 2\tx \sin \ty \tan \tx \ds.
\end{equation}
$\Delta_{\mu \mu}$ is positive (negative) for NH (IH).
For the NH and $\dcp$ in the  
LHP, the values of $\pme$ 
($\pmebar$) are reasonably greater (lower) than the values of 
$\pme$ ($\pmebar$) for the IH and any value of $\dcp$. Similarly,  
for the IH and $\dcp$ in the UHP the values of $\pme$ ($\pmebar$) 
are reasonably lower (greater) than the values of 
$\pme$ ($\pmebar$) for the NH and any value of $\dcp$. 
Hence, for these {\it favourable combinations}, \nova is capable 
of determining the hierarchy at a confidence level (C.L.) of 
$2\sigma$ or better, with 3 years run each for $\nu$ and $\anu$. However, as mentioned above, the change in the first term in eq.~(\ref{pme}) can be 
canceled by the second term for unfavourable values of
$\dcp$. This leads to hierarchy-$\dcp$ 
degeneracy \cite{Barger:2001yr, Minakata:2003wq, Mena:2004sa}. 
From Fig.~(\ref{prob}),
 it can be seen that    
$\pme$ and $\pmebar$ for NH and $\dcp$ in the UHP 
are very close to or degenerate with 
those of IH and $\dcp$ in the LHP.
For these {\it unfavourable combinations},
\nova has {\bf no} hierarchy sensitivity \cite{Prakash:2012az}. 
Since the neutrino energy of T2K is only one third of the energy of
NO$\nu$A, the matter effect of T2K is correspondingly smaller.
Therefore, T2K has very little hierarchy sensitivity. The cancellation of
change due to matter effect occurs for different values of $\dcp$
in the case of \nova and T2K. Therefore, combining the data of
\nova and T2K leads to a small hierarchy discrimination capability for
unfavourable hierarchy-$\dcp$ combinations \cite{Mena:2004sa, Prakash:2012az, Agarwalla:2012bv}.

\begin{figure}[htb]
\centering
\includegraphics[width=0.7\textwidth]{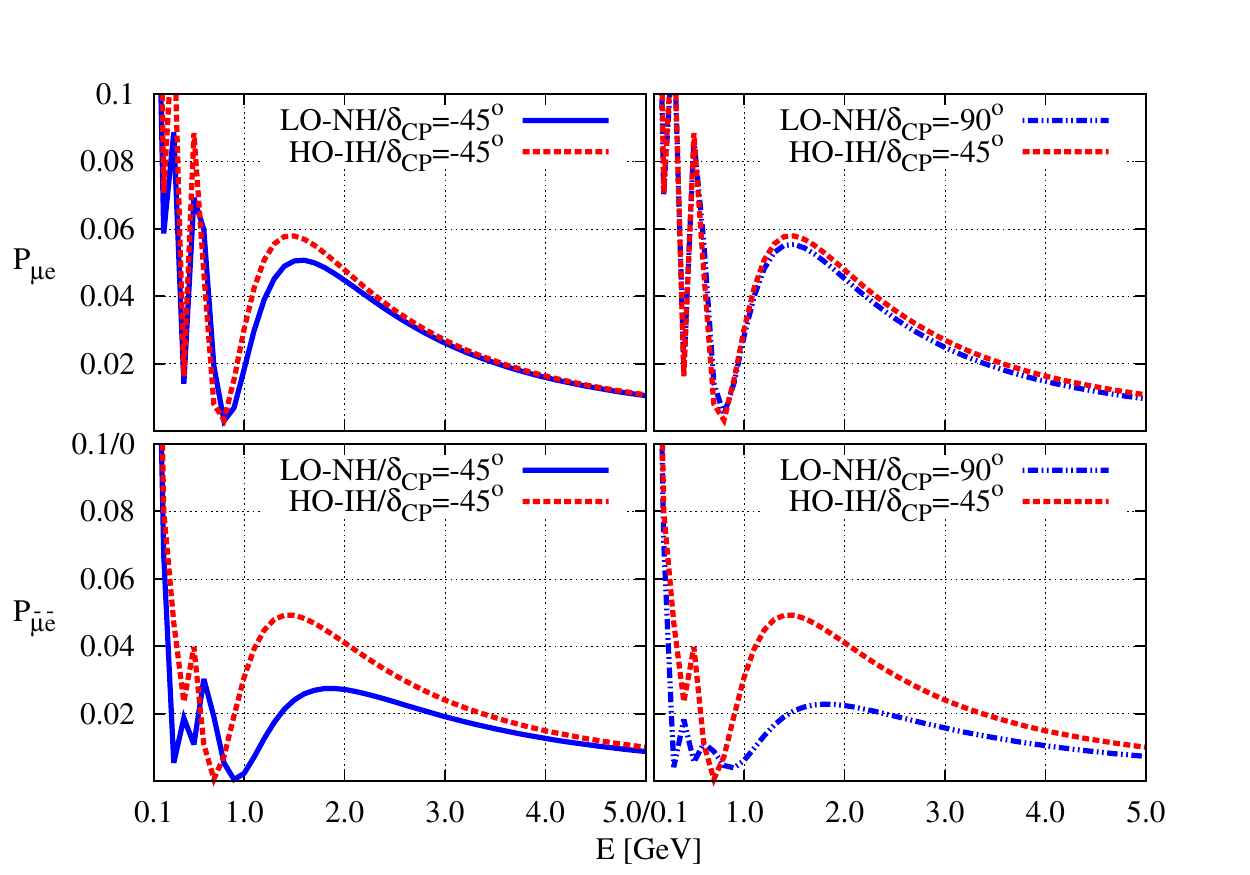}
\caption{\footnotesize{Illustration of degenerate $\pme$ and non-degenerate $\pmebar$ for the following two cases.
Left: (LO-NH, $\dcp=-45^\circ$) and 
(HO-IH, $\dcp=-45^\circ$), 
Right: (LO-NH, $\dcp=-90^\circ$) and 
(HO-IH, $\dcp=-45^\circ$). 
}}
\label{deg-prob}
\end{figure}

\paragraph*{}

\textbf{b. Octant-hierarchy degeneracy}: Even though the atmospheric neutrino experiments prefer maximal $\tz$ ($\sin^2 2\tz=1$), MINOS experiment prefers non-maximal values, $\sin^22\tz=0.96$ \cite{Kyoto2012MINOS}.
The global fits, before the \nova and T2K experiment started taking data, also favour a non-maximal value of $\tz$ 
\cite{Tortola:2012te,Fogli:2012ua,GonzalezGarcia:2012sz}, 
leading to two degenerate solutions: $\tz$
in the lower octant (LO) ($\sin^2 \tz =0.41$) and 
$\tz$ in the higher octant (HO) ($\sin^2 \tz = 0.59$). Given the two hierarchy and two octant possibilities, there are four possible octant-hierarchy combinations:
LO-NH, HO-NH, LO-IH and HO-IH. The 
first term of $\pme$ in eq.~(\ref{pme}) becomes larger (smaller) for NH (IH).
The same term also becomes smaller (larger) for LO (HO).
If HO-NH (LO-IH) is the true octant-hierarchy combination,
then the values of $\pme$ 
are significantly higher (smaller) than those for IH (NH) and 
for any octant. For these two cases, only $\nu$ data has good
hierarchy determination capability. But the situation is 
very different for the two cases LO-NH and HO-IH.
The increase (decrease) in the first term of $\pme$ due to NH 
(IH) is canceled (compensated) by the decrease (increase) for LO (HO).
Therefore the two octant-hierarchy combinations, LO-NH and HO-IH, have degenerate values 
for $\pme$. However, this degeneracy is not present in
$\pmebar$, which receives a double boost (suppression) for the 
case of HO-IH (LO-NH). Thus the octant-hierarchy degeneracy in 
$\pme$ is broken by $\pmebar$ (and vice-verse). Therefore 
$\nu$-only data has 
{\bf no} hierarchy sensitivity if the cases LO-NH or HO-IH
are true, but a combination of $\nu$ and $\anu$ data will
have a good sensitivity.

This has been illustrated in Figure 
\ref{deg-prob}. From the figure, we can
see that $\pme$ has degeneracy
for the octant-hierarchy combinations
LO-NH and HO-IH. This degeneracy does not exist
in the case of $\pmebar$ \cite{Prakash:2013dua}.

\begin{figure}[htb]
\centering
\includegraphics[width=0.7\textwidth]{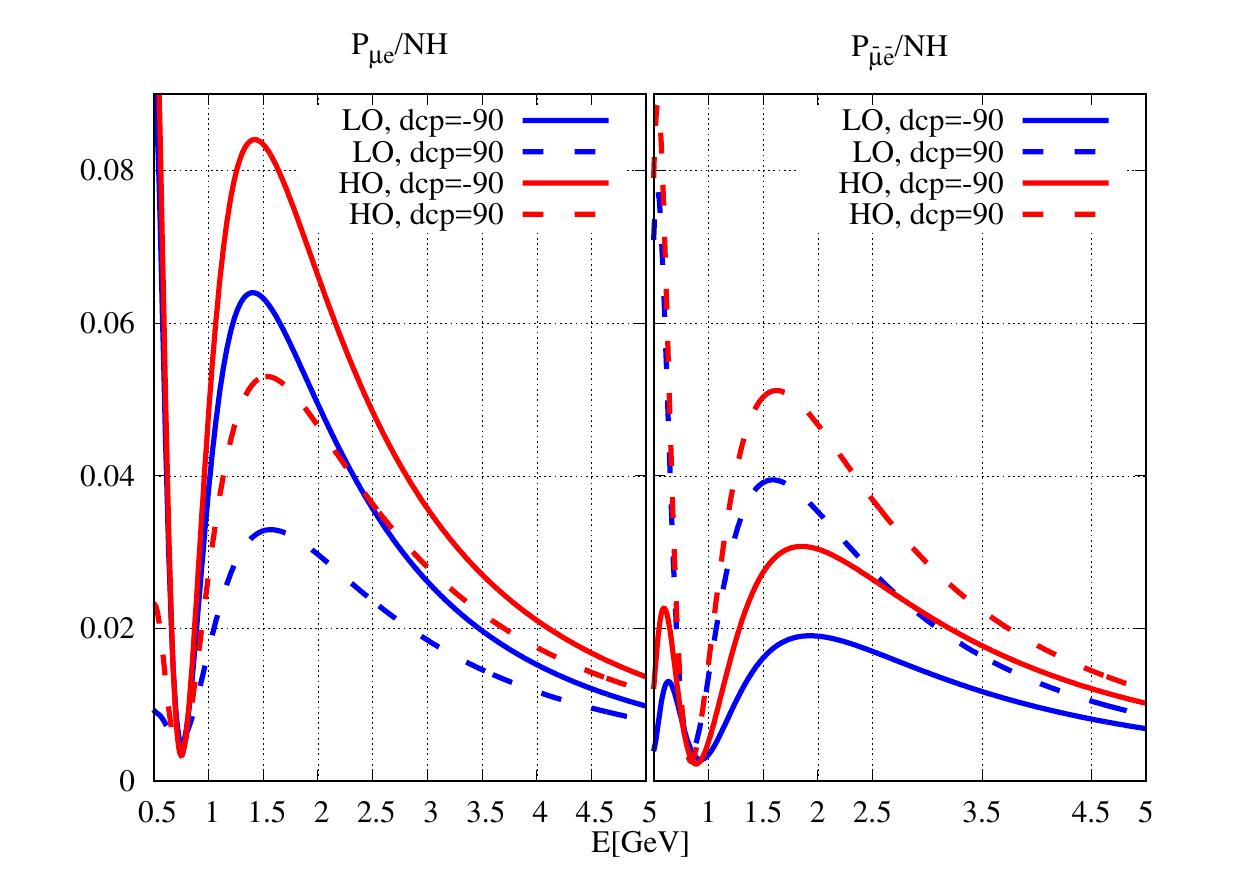}
\caption{\footnotesize{$\pme$ as a function of neutrino energy for the \nova baseline. The left 
(right) panel is for $\nu$ ($\anu$). The plots have been drawn for
different possible 
$\dcp$ values between $-180^\circ$ and 
$180^\circ$. The $\sin^2 2\ty$ value is 0.089. The value
of $\sin^2\tz$ is 0.41 (0.59) for LO (HO).}}
\label{oct-dcp}
\end{figure}

\paragraph*{}
\textbf{c. Octant-$\dcp$ degeneracy}: The possibility of two octants of $\tz$ also leads to octant-$\dcp$ degeneracy.
To highlight this degeneracy, we rewrite the expression
for $\pme$ in
eq.~(\ref{pme}) as \cite{Agarwalla:2013ju}
\begin{equation}
 \pme=\beta_1 \sin^2 \tz +\beta_2 \cos(\hat{\Delta}+\dcp)+\beta_3 \cos^2\tz,
 \label{pmebeta}
\end{equation}
where
\begin{eqnarray}
 &&\beta_1=\sin^2 2\ty_{13}\frac{\sin^2 \hat{\Delta}(1-\hat{A})}{(1-\hat{A})^2}, \nonumber \\
 &&\beta_2=\alpha \cos \ty \sin 2\tx \sin 2\ty \sin 2\tz \frac{\sin (\dhat \ahat)}{\ahat}\nonumber \\
 &&\beta_3=\alpha^2 \sin^2 2\tx \cos^2\ty\frac{\sin^2 (\dhat \ahat)}{\ahat^2}.
\end{eqnarray}
In Figure~\ref{oct-dcp} we have plotted $\pme$ ($\pmebar$) for
the \nova experiment
as a function of neutrino energy $E_\nu$, for normal hierarchy 
and for different values of $\dcp$. In our calculation, $\sin^2 \tz=0.41$, when 
$\tz$ is in the LO and $\sin^2 \tz=0.59$, when 
$\tz$ is in the HO.
The $\sin^2 2\ty$ has been taken equal to 0.089.
From the left panel of the figure, we can see that when $\tz$ is in  the LO and $\dcp$ is $90^\circ$, $\pme$ is 
quite distinctive from probability values with other
$\tz$ and $\dcp$ combinations. Similar arguments hold for
$\pme$ with $\tz$ in 
the HO and $\dcp=-90^\circ$.
Therefore, $\tz$ in the LO (HO) and $\dcp=90^\circ$ ($-90^\circ$)
is a favourable octant-$\dcp$ combination to determine the octant of
$\tz$. However $\pme$ for $\tz$ in LO and
$\dcp=-90^\circ$ overlaps with that for
$\tz$ in HO and $\dcp=90^\circ$. Therefore, 
$\tz$ in the LO (HO) and $\dcp=-90^\circ$ ($90^\circ$)
is an unfavourable combination to determine the octant.

But these unfavourable combinations, become favourable
for $\pmebar$ and vice-versa, as can be seen from the right panel of
Figure \ref{oct-dcp}. Thus the octant-$\dcp$ degeneracy,
present in neutrino data, can be removed by anti-neutrino data
and vice-versa. This aspect is different from the 
hierarchy-$\dcp$ degeneracy.

The above features can also be understood
by following algebraic analysis of
eq.~(\ref{pmebeta}). From that equation, we see that
$\pme$ increases with the increase
in $\tz$. But $\pme$ can increase or
decrease with change in $\dcp$. In the
case of octant-$\dcp$ degeneracy,
for different $\dcp^{\rm LO}$ and 
$\dcp^{\rm HO}$, we can have 
$\pme(\rm LO, \dcp^{\rm LO})=\pme(\rm HO, \dcp^{\rm HO})$.
It leads to
\begin{equation}
 \cos(\dhat+\dcp^{\rm LO})-\cos(\dhat+\dcp^{\rm HO})=\frac{\beta_1-\beta_3}{\beta_2}(\sin^2 \tz^{\rm HO}-\sin^2 \tz^{\rm LO}).
\end{equation}
The \nova experiment has a baseline of
810 km and the flux peaks at an energy
2 GeV. Now for the NH and neutrino,
\begin{equation}
 \cos(\dhat+\dcp^{\rm LO})-\cos(\dhat+\dcp^{\rm HO})=1.7.
\end{equation}
 This equation can have solutions only if
  \begin{eqnarray}
  &&0.7 \leq \cos(\dhat+\dcp^{\rm LO}) \leq 1.0, \nonumber \\
  &&-1.0 \leq \cos(\dhat+\dcp^{\rm HO}) \leq -0.7.
 \end{eqnarray}
From the above equation, we have the 
ranges of $\dcp$ as
\begin{eqnarray}
 &&-116^\circ \leq \dcp^{\rm LO}\leq -26^\circ \nonumber \\
 &&64^\circ \leq \dcp^{\rm HO}\leq 161^\circ.
\end{eqnarray}
 Therefore, for the \nova experiment, for NH and neutrino,
 $\pme(\rm LO, -116^\circ \leq \dcp\leq -26^\circ)$ is
 close to $\pme(\rm HO, 64^\circ \leq \dcp\leq 161^\circ)$ \cite{Agarwalla:2013ju}.
 Similar equations for $\pmebar$ show that this degeneracy can be removed by anti-neutrino run.
 
\section{Details of data analysis}
\label{chisq}

In this section, we describe the methodology by which we have
done data analyses of T2K and \nova data in Sections \ref{evolution} and 
\ref{resolution}. We have computed $\chi^2$ and $\Delta \chi^2$
between the data and a given theoretical model. 
In section \ref{evolution}, we have discussed the evolution of \nova and T2K data with time. To do so, we have presented the analysis of the latest data from both the experiments in standard 3-flavour oscillation scenario. In section \ref{resolution}, we have discussed how different BSM scenarios alleviate the tension between the two experiments.
This is done by analyzing the latest data from \nova and T2K in each of the different BSM frameworks. In both Sections \ref{evolution} and \ref{resolution}, the results have been presented in the form of $\dchsq$. 

The observed and the expected number of events in the $i$-th energy bin of a given experiment are denoted by $N_{i}^{\rm obs}$ and $N_{i}^{\rm th}$ respectively. The $\chi^2$ between these two distributions is calculated as 
\begin{eqnarray}
%
\chi^2 &=& 2\sum_i \left\{
(1+z) N_i^{\rm th} - N_i^{\rm exp} + N_i^{\rm exp} 
\ln\left[ \frac{N_i^{\rm exp}}{(1+z) N_i^{\rm th}} \right]
\right\} + 2 \sum_j (1+z) N_j^{\rm th} + z^2, \,\, \nonumber, \\
\label{poisionian}
\end{eqnarray}
where $i$ stands for the bins for which the observed event numbers are non-zero, $j$ stands for bins for which the observed event numbers are zero and $z$ is the parameter defining systematic uncertainties. 

The theoretical expected event numbers for each energy bin and the $\chi^2$ between theory and experiment have been calculated using the software GLoBES \cite{Huber:2004ka, Huber:2007ji}. To do so, we fixed the signal and background efficiencies of each energy bin according to the Monte-Carlo simulations provided by the experimental collaborations \cite{Himmel:2020, NOvA:2021nfi, Dunne:2020, T2K:2021xwb}. We kept $\sin^2\tx$ and $\ds$ at their best-fit values $0.304$ and $7.42\times 10^{-5}\, {\rm eV}^2$, respectively. We varied $\sin^2 2\ty$ in its $3\, \sigma$ range around its central value $0.084$ with $3.5\%$ uncertainty~\cite{Dohnal:2021rcr}. $\sin^2 \tz$ has been varied in its $3\, \sigma$ range $[0.41:0.62]$ (with $2\%$ uncertainty on $\sin^2 2\tz$ \cite{Esteban:2018azc}). We varied $|\Delta_{\mu \mu}|$ in its $3\, \sigma$ range around the MINOS best-fit value $2.32\times 10^{-3}\, {\rm eV}^2$ with $3\%$ uncertainty \cite{Nichol:2012}. 
The CP-violating phase $\dcp$ has been varied in its complete range $[-180^\circ:180^\circ]$. In case of BSM physics, we modified the software to include new physics. The ranges of different new parameters for each of the BSM scenarios have been discussed in section \ref{resolution}. 

Automatic bin-based energy smearing for the generated theoretical events has been implemented within GLoBES~\cite{Huber:2004ka, Huber:2007ji} using a Gaussian smearing function
\begin{equation}
R^c (E,E^\prime)=\frac{1}{\sqrt{2\pi}}e^{-\frac{(E-E^\prime)^2}{2\sigma^2(E)}},
\end{equation}
where $E^\prime$ is the reconstructed energy. The energy resolution function is given by 
\begin{equation}
\sigma(E)=\alpha E+\beta \sqrt{E}+\gamma.
\end{equation}
For NO$\nu$A, we used $\alpha=0.11\, (0.09)$, $\beta=\gamma=0$ for electron (muon) like events \cite{NOvA:2018gge, NOvA:2019cyt}. For T2K, we used $\alpha=0$, $\beta=0.075$, $\gamma=0.05$ for both electron and muon like events. For T2K, the relevant systematic uncertainties are
\begin{itemize}
    \item $5\%$ normalisation and $5\%$ energy calibration systematics uncertainty for $e$-like events, and
    \item $5\%$ normalisation and $0.01\%$ energy calibration systematics uncertainty for $\mu$-like events.
\end{itemize}
For NO$\nu$A, we used $8.5\%$ normalization and $5\%$ energy calibration systematic uncertainties for both the $e$ like and $\mu$ like events \cite{NOvA:2018gge}. Details of systematic uncertainties have been discussed in the GLoBES manual \cite{Huber:2004ka, Huber:2007ji}. 

During the calculation of $\chi^2$ we added (for the older, pre-2020 data) priors on $\sin^2 2\ty$, $\sin^2 2\tz$, and $|\Delta_{\mu \mu}|$, in cases where we have not included muon disappearance data. In all other cases, priors have been added on $\sin^2 2\ty$ only (to account for electron disappearance data from reactor neutrino experiments). 

We calculated the $\chi^2$ for both the hierarchies. Once the $\chi^2$s had been calculated, we subtracted the minimum $\chi^2$ from them to calculate the $\dchsq$. The parameter values and hierarchy, for which the $\dchsq=0$, is called the best-fit point.

\section{Evolution of the tension between \nova and T2K data}
\label{evolution}

In this section, we consider the appearance data of T2K and \nova in both $\nu/\bar{\nu}$ modes and discuss how they give rise to degenerate solutions. We will highlight the growing tension between these appearance data when they are interpreted within the three flavour oscillation paradigm. In the next section, we will consider
new physics scenarios which can reduce that tension. 

\subsection{Evolution of the \nova data}
In 2017, \nova published 
first result with combined analysis of $\nu_e$ appearance and $\nu_\mu$ disappearance data, corresponding to $6.05 \times 10^{20}$ POT \cite{NOvA:2017abs}. This analysis gave three almost degenerate solutions
\begin{itemize}
    \item NH, $\sin^2\tz=0.4$, $\dcp=-90^\circ$ (NH, LO, $-90^\circ$),
    \item NH, $\sin^2\tz=0.62$, $\dcp=135^\circ$ (NH, HO, $135^\circ$), and
    \item IH, $\sin^2\tz=0.62$, $\dcp=-90^\circ$ (NH, HO, $-90^\circ$).
\end{itemize}
In ref.~\cite{Bharti:2018eyj}, an effort 
was made to understand this degeneracy in the \nova data from 2017. To do so, the authors of ref.~\cite{Bharti:2018eyj} first calculated the expected $\nu_e$ appearance event numbers for vacuum oscillation, maximal $\tz=45^\circ$, and $\dcp=0$ for $6.05\times 10^{20}$ POT. This case was labelled as $'000'$.  Then they considered changes in this number due to a) matter effect, b) octant of non-maximal $\tz$, and c) large value of $\dcp$. The parameter value for which $\pme$ is increased (decreased) was labelled as $'+'$ ($'-'$). First, one change at a time was introduced in the following manner:
\begin{itemize}
    \item NH which increases $\pme$ (labelled as $'+'$) or IH which decreases it (labelled as $'-'$),
    \item HO which increases $\pme$ (labelled as $'+'$) or LO which decreases it (labelled as $'-'$),
    \item $\dcp=-90^\circ$ which increases $\pme$ (labelled as $'+'$) or $\dcp=+90^\circ$ which decreases it (labelled as $'-'$).
\end{itemize}
The event numbers for $6.05\times 10^{20}$ POT were calculated using the software GLoBES \cite{Huber:2004ka, Huber:2007ji}. Other parameters were fixed at following constant values: $\ds=7.5\times 10^{-5}\, {\rm eV}^2$, $\sin^2\tx=0.306$, $\dl(\rm NH)=2.74\times 10^{-3}\, {\rm eV}^2$, $\dl(\rm IH)=-2.65\times 10^{-3}\, {\rm eV}^2$, and $\sin^2 2\ty=0.085$. The values of $\dl$(NH) and $\dl$(IH) were taken from the analysis of \nova $\nu_\mu$ disappearance data. The results 
are presented in table~\ref{events-1}. From this table, it is obvious that the increase (decrease) in $\pme$ for any single $'+'$ ($'-'$) change in the unknown parameters is essentially same. 
\begin{table}
\begin{center}
\begin{tabular}{|c|c|c|}
  \hline
  Hierarchy-$\sin^2\tz$-$\dcp$ & Label&
  $\nu_e$ Appearance events\\
  
  \hline
   Vacuum-$0.5$-$0$ & $000$ & $26.49$ \\
  \hline
  NH-$0.5$-$0$ & $+00$ & $31.28$\\
  \hline
  IH-$0.5$-$0$  & $-00$ & $21.08$ \\
  \hline
 Vac-$0.5$-$-90^\circ$ & $00+$ & $31.04$ \\
 \hline
  Vac-$0.5$-$+90^\circ$ & $00-$ & $21.18$\\
 \hline
 Vac-$0.62$-$0$  & $0+0$ & $32.88$ \\
 \hline
  Vac-$0.4$-$0$  & $0-0$ & $24.59$  \\
 \hline

\end{tabular}
\end{center}
 \caption{Expected $\nu_e$ appearance events of \nova for $6.05\times10^{20}$ POT. They are listed for the reference point and for change of one unknown parameter at a time. \nova observed $33$ $\nu_e$ appearance events in 2017.}
  \label{events-1}
\end{table}

Next all eight possible combinations in the changes of the three unknown parameters were considered. All three unknown parameters can shift in a way that each change leads to an increase in $\pme$. This can be lebelled as $'+++'$. In this case, one gets the maximum number of $\nu_e$ appearance events. Another case is that two of the unknown parameters change to increase $\pme$, whereas the third one decreases it. This can happen in three possible ways, labelled as $'+-+'$, $'++-'$, and $'-++'$. These three combinations lead to a moderate increase in $\nu_e$ appearance events compared to the reference $'000'$ case. Similarly a moderate decrease in $\nu_e$ appearance event numbers compared to the reference case can occur due to increase in $\pme$ by one unknown parameter and decrease by the other two.
This can occur in three possible ways: $'+--'$, $'-+-'$ and $'--+'$. Finally there is a possible case where each of the three changes lowers $\pme$ and we get the minimum number of $\nu_e$ appearance events. This can be labelled as $'---'$. In table~\ref{events-2}, the number of $\nu_e$ appearance events for $6.05\times 10^{20}$ POT and for all the eight combinations, mentioned above, have been listed. From the table, we can see that the number of events for $'++-'$, $'+-+'$ and $'-++'$ are nearly the same. A similar statement can be made about $'+--'$, $'-++'$ and $'--+'$. The predicted number of events for $'+++'$, and $'---'$ are totally unique. Thus, the eight-fold degeneracy, present before the $\ty$ was measured, breaks itself down to $1+3+3+1$ pattern after the precise measurement of $\ty$. The 2017 \nova data saw a moderate increase in $\nu_e$ appearance events, compared to the expected event numbers for the reference case $'000'$. Hence, there was a three-fold degeneracy in the \nova solutions. In table~\ref{events-bestfit-2017}, the expected number of $\nu_e$ appearance events for $6.05\times 10^{20}$ POT at each of the \nova solutions have been listed. The predictions for two NH solutions matched the experimentally observed event numbers $33$. The prediction of the IH solution (which was $0.5\, \sigma$ away from the NH solutions) was lower by $3$ (half the statistical uncertainty in the expected number). The occurrence of three-fold degeneracy in the NO$\nu$A data, based on the inherent degeneracy in $\pme$ was also discussed in ref.~\cite{Lindner:2017nmb}. 

\begin{table}
  \begin{center}
\begin{tabular}{|c|c|c|}
  \hline
  Hierarchy-$\sin^2\tz$-$\dcp$ & Label&
  $\nu_e$ Appearance events\\
  
  \hline
   NH-$0.62$-$-90^\circ$ & $+++$ & $43.67$ \\
  \hline
  NH-$0.4$-$-90^\circ$ & $+-+$ & $33.54$\\
  \hline
  NH-$0.62$-$+90^\circ$  & $++-$ & $33.04$ \\
  \hline
 IH-$0.62$-$-90^\circ$ & $-++$ & $30.94$ \\
 \hline
  NH-$0.4$-$+90^\circ$ & $+--$ & $22.79$\\
 \hline
 IH-$0.4$-$-90^\circ$  & $--+$ & $24.47$ \\
 \hline
  IH-$0.62$-$+90^\circ$  & $-+-$ & $22.63$  \\
  \hline
IH-$0.4$-$+90^\circ$  & $---$ & $16.07$  \\
  \hline
\end{tabular}
\end{center}
 \caption{Expected $\nu_e$ appearance events of \nova for $6.05\times10^{20}$ POT, and for eight different combinations of unknown parameters.}
  \label{events-2}
\end{table}

\begin{table}
  \begin{center}
\begin{tabular}{|c|c|c|}
  \hline
  Hierarchy-$\sin^2\tz$-$\dcp$ & Label&
  $\nu_e$ Appearance events\\
  
  \hline
   NH-$0.404$-$-86^\circ$ & $+-+$ & $33.55$ \\
  \hline
  NH-$0.62$-$+135^\circ$ & $++-$ & $34.36$\\
  \hline
 IH-$0.62$-$-90^\circ$ & $-++$ & $30.94$ \\
 \hline
\end{tabular}
\end{center}
 \caption{Expected $\nu_e$ appearance events of \nova for $6.05\times10^{20}$ POT, and for the three solutions in ref.~\cite{NOvA:2017abs}}
  \label{events-bestfit-2017}
\end{table}

In ref.~\cite{Bharti:2018eyj} the authors also considered the possible resolution of the three-fold degeneracy with the anti-neutrino run. One can obtain the anti-neutrino oscillation probability $\pmebar$ by reversing the signs of matter term $A$, and $\dcp$ in eq.~(\ref{pme}). $\pmebar$ decreases (increases) for NH (IH). Similarly, $\dcp$ in the UHP (LHP) increases (decreases) $\pmebar$. But, we will continue to label the NH (IH) as $'+'$ ($'-'$). In the same way, $\dcp$ in the LHP (UHP) will be labelled as $'+'$ ($'-'$). However, it should be noted that $'+'$ ($'-'$) sign in hierarchy and $\dcp$ leads to an decrease (increase) in $\pmebar$. But for the octant of $\tz$, $'+'$ ($'-'$) sign leads to an increase (decrease) in $\pmebar$.
Now, for the $'++-'$ solution, $\pmebar$ decreases due to the hierarchy, and increases due to the octant of $\tz$ and $\dcp$. Similarly, for $'-++'$ solution, $\pmebar$ increases due to the hierarchy and the octant of $\tz$, and decreases due to $\dcp$. Hence these two solutions are degenerate for anti-neutrino data also, and the \nova $\bar{\nu}_e$ appearance data would not be able to distinguish between them. However, for the third solution $'+-+'$, $\pmebar$ decreases due to all three unknown parameters. The expected $\bar{\nu}_e$ appearance events for this case would be the smallest. In principle, the \nova $\bar{\nu}_e$ appearance data for $6.05\times 10^{20}$ POT would be able to distinguish this solution for the other two. Since the expected number of events for this particular scale is rather small, as can be seen from table~\ref{events-bestfit_anu_2017}, the statistical uncertainty would be large, and \nova $\bar{\nu}_e$ appearance data with $6.05\times 10^{20}$ POT would not be able to distinguish this solution from the other two at the $3\, \sigma$ level.

\begin{table}
  \begin{center}
\begin{tabular}{|c|c|c|}
  \hline
  Hierarchy-$\sin^2\tz$-$\dcp$ & Label&
  $\bar{\nu}_e$ Appearance events\\
  
  \hline
   NH-$0.404$-$-86^\circ$ & $+-+$ & $6.85$ \\
  \hline
  NH-$0.62$-$+135^\circ$ & $++-$ & $11.78$\\
  \hline
 IH-$0.62$-$-90^\circ$ & $-++$ & $12.86$ \\
 \hline
\end{tabular}
\end{center}
 \caption{Expected $\bar{\nu}_e$ appearance events of \nova for $6.05\times10^{20}$ POT, and for the three solutions in ref.~\cite{NOvA:2017abs}}
  \label{events-bestfit_anu_2017}
\end{table}

In the Neutrino 2018 conference, \nova published results after analysing data corresponding to $8.85\times 10^{20}$ ($6.9\times 10^{20}$) POTs in the neutrino (anti-neutrino) mode \cite{sanchez_mayly_2018_1286758, NOvA:2018gge}. They found out the best-fit point at $\dcp=30.6^\circ$ ($-95.4^\circ$), $\sin^2\tz=0.58\pm 0.03$ ($0.58\pm 0.04$), $\Delta_{32}=2.51^{+0.12}_{-0.08}\times 10^{-3}\, {\rm eV}^2$ ($-2.56\times 10^{-3}\, {\rm eV}^2$) for NH (IH). NH was preferred over IH at $1.8\, \sigma$ C.L. Also $\dcp=90^\circ$ in the IH was excluded at more than $3\, \sigma$ C.L. The neutrino disappearance data were consistent with the maximal $\tz$, whereas the $\bar{\nu}_\mu$ disappearance data preferred a non-maximal mixing \cite{Nizam:2018got}. Therefore, there was a mild tension between the two different data sets of the same experiment. However, since the statistics from the anti-neutrino disappearance data were very low, this tension was not statistically significant. In the case of appearance data, the expected number of $\nu_e$ ($\bar{\nu}_e$) appearance events at the reference point $'000'$ was $39$ ($15.5$) \cite{Nizam:2018got}. The observed $\nu_e$ ($\bar{\nu}_e$) appearance events were $58$ ($18$). Hence, there was a moderate excess in both these channels. As we have already seen, the moderate excess in both $\nu_e$ and $\bar{\nu}_e$ appearance channels is possible when the $20\%$ change due to the hierarchy and $\dcp$ cancels each other, and there is an increase in both channels induced by $\tz$ in the HO. We have already labelled these possibilities by A) $'++-'$ and B) $-++$. These are two of the three degenerate solutions of the previous 2017 data. The other degenerate solution of 2017 data, namely $'+-+'$ was ruled out by 2018 data because although this solution leads to moderate excess in the $\nu_e$ appearance events, it causes minimum number of $\bar{\nu}_e$ appearance events. In ref.~\cite{Nizam:2018got}, an analysis of the $\nu_e$ and $\bar{\nu}_e$ appearance data from 2018 was performed. It was found out that there were two degenerate best-fit solutions: i) NH, $\sin^2\tz=0.65$, $\dcp=120^\circ$, and ii) IH, $\sin^2\tz=0.67$, $\dcp=-50^\circ$. The first solution is in the form of A) and the second solution is in the form of B). The best-fit points given by the \nova collaboration were also in these two forms. However, because of the inclusion of the disappearance data in the analysis of the \nova collaboration, a smaller value, compared to those mentioned above, of $\sin^2\tz$ was obtained.

\begin{figure}[htb]
\centering
\vskip -1.cm
\includegraphics[width=0.85\textwidth]{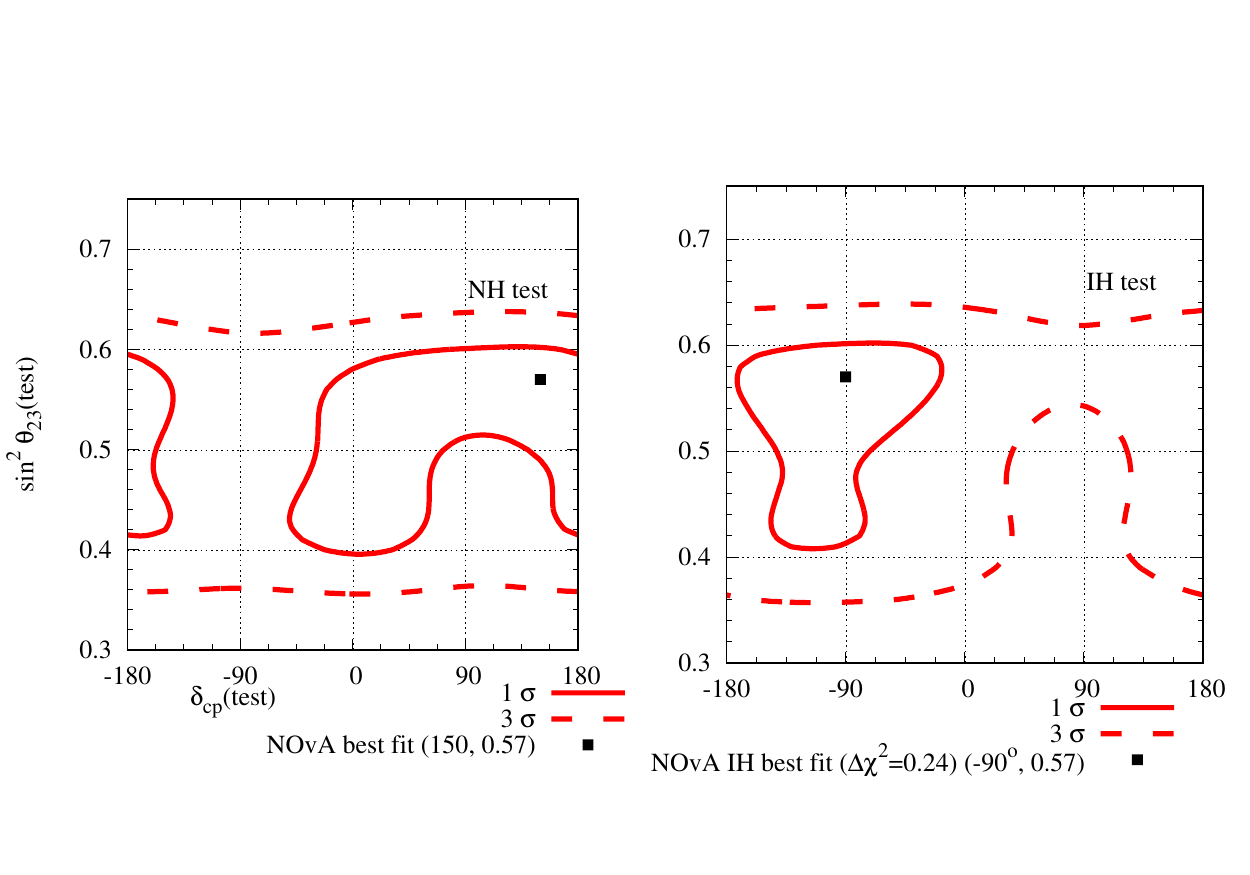}
\caption{\footnotesize{Expected allowed regions in the $\sin^2\tz-\dcp$ plane from the \nova appearance data, as given in ref.~\cite{Himmel:2020, NOvA:2021nfi}, for both the $\nu_e$ and $\bar{\nu}_e$ channels. In the left (right) panel, the hierarchy is assumed to be NH (IH). The best-fit point is at NH with a minimum $\chi^2=6.77$ for 12 energy bins. IH has a minimum $\dchsq=0.2$.}}
\label{appearance-2020-nova}
\end{figure}

In 2020, \nova published the analysis of their data corresponding to $1.36 \times 10^{21}$ ($1.25 \times 10^{21}$) POT in the $\nu$ ($\bar{\nu}$) mode \cite{Himmel:2020, NOvA:2021nfi}. The best-fit points were $\Delta_{32}=+(2.41\pm 0.07)\times 10^{-3}\, {\rm eV}^2$ ($-2.45\times 10^{-3}\, {\rm eV}^2$), $\sin^2\tz=0.57^{+0.03}_{-0.04}$ ($0.56$), $\dcp/\pi=0.82^{+0.27}_{-0.87}$ ($1.52$). In ref.~\cite{Rahaman:2021zzm}, a detailed analysis of the \nova data has been done. The expected $\nu_e$ and $\bar{\nu}_e$ appearance event numbers at the reference point $'000'$ are $76.14$ and $32.93,$ respectively. The observed event numbers in these channels are $82$ and $33$, respectively. Thus, there is a moderate excess in the $\nu_e$ appearance channel, and this excess can be explained by, as explained before, three possible solutions: A) $'+-+'$, B) $'++-'$ and C) $'-++'$. As for the $\bar{\nu}_e$ appearance data, the observed event number $33$ is consistent with the $'000'$ solution. However, due to the lack of statistics in the $\bar{\nu}_e$ appearance channel, other solutions are also allowed at the $1\, \sigma$ C.L.. Exceptions are the cases labelled as $'-+-'$, and $'+-+'$, since these cases lead to the maximum and minimum number of expected $\bar{\nu}_e$ appearance events, respectively. Thus the solution in the form of B) is excluded when $\nu_e$ and $\bar{\nu}_e$ appearance data are analysed together. The result is shown in Fig.~\ref{appearance-2020-nova}. Best-fit points are of the forms A) and C). The best-fit points obtained by the \nova collaboration after analysing appearance and disappearance data together are of the same forms as well.

\subsection{Evolution of the T2K data}
In 2013 T2K published their first analysis of $\nu_e$ appearance data corresponding to $6.57\times 10^{20}$ POT \cite{T2Kapp, T2Kdisapp}. They found out their best-fit point at NH and $\dcp=-90^\circ$. Both the hierarchies with $\dcp$ in the LHP were allowed at $2\, \sigma$ C.L., whereas $\dcp$ values in the UHP were disfavoured at $2\, \sigma$ C.L.\ for both the hierarchies. From the $\nu_\mu$ disappearance data published in 2014 \cite{T2Kdisapp}, $\tz$ was found to be close to the maximal mixing. In ref.~\cite{Bharti:2016hfb}, a detailed analysis of the physics potential of \nova in the presence of the information from T2K data were made. It was shown that if the hierarchy and $\dcp$ are in favourable combinations, T2K data have no effect on the hierarchy determination potential of NO$\nu$A. Among the unfavourable hierarchy-$\dcp$ combinations, T2K data picked out the correct (incorrect) solution from the two degenerate solutions allowed by the \nova data for the hierarchy being IH (NH) and $\dcp$ in the LHP (UHP). Therefore it was concluded that if the combination of \nova and T2K prefer IH and $\dcp$ in the LHP as the correct solution, one needs to be careful, because the actual combination might be NH and $\dcp$ in the LHP. We will find out that this prediction in ref.~\cite{Bharti:2016hfb} was quite accurate in the context of the latest published data from \nova and T2K.

In 2018, T2K published the analysis of their data with $14.7 \times 10^{20}$ ($7.6\times 10^{20}$) POT in the neutrino (anti-neutrino) mode \cite{Abe:2018wpn}.  In ref.~\cite{Nizam:2018got} a detailed analysis of T2K disappearance and appearance data was done separately. It was found out that the analysis of T2K disappearance data gave the best-fit point at $\sin^2 \tz=0.51$.  The $3\, \sigma$ C.L. on $\sin^2\tz$ was constrained in the range $[0.43:0.6]$. The constraint on $\sin^2\tz$ was valid for all values of $\dcp$, since the disappearance data do not have $\dcp$ sensitivity. As for the appearance data it was found out that the expected $\nu_e$ appearance event number at the reference point $'000'$ for the given neutrino run was found out to be $60$. Inclusion of matter effect changed the number by $4$, and inclusion of maximal CP violation changed the number by $11$. Therefore, for NH and $\dcp=-90^\circ$, the expected number was increased to 80 \cite{Nizam:2018got}. T2K observed $89$ $\nu_e$ events. Hence, the $\nu_e$ appearance data of T2K pulled the $\sin^2\tz$ to a value larger than $0.5$. An analysis of T2K $\nu_e$ appearance data was done in ref.~\cite{Nizam:2018got}. The $\bar{\nu}_e$ data were not included because the observed number of events in this channel was too small to have any statistical significance. The best-fit point was found to be at $\sin^2\tz=0.63$, although $\sin^2\tz=0.5$ was allowed at $1\, \sigma$ C.L. Therefore there was a mild tension between the T2K appearance and disappearance data. The T2K collaboration, after analysing the appearance and disappearance data together got the best-fit point at $\sin^2\tz=0.53$ \cite{Abe:2018wpn}. Because of the larger statistical weight of disappearance data, the final value of $\sin^2 \tz$ was determined by the disappearance data and the final value of $\sin^2\tz$ was found to be close to the maximal value. Because of the large excess in the observed $\nu_e$ appearance events in T2K, the $\dcp$ was found to be in the vicinity of $-90^\circ$. For $\dcp$ in the UHP, the expected number of $\nu_e$ events was smaller than that at the reference point. Thus $\dcp$ in the UHP was highly disfavoured. This data also disfavoured IH, because the corresponding matter effect reduced the number of expected $\nu_e$ events. IH with $\dcp=-90^\circ$ was barely allowed at $2\,\sigma$ C.L.\ \cite{Abe:2018wpn}. 

In 2019, T2K published data corresponding to $14.9\times 10^{20}$ ($16.4\times 10^{20}$) POT in the neutrino (anti-neutrino) mode. The best-fit point was at $\dcp/\pi=-1.89^{+0.70}_{-0.58}$, ($-1.38^{+0.48}_{-0.54}$) $\sin^2 \tz=0.53^{+0.03}_{-0.04}$ for NH (IH) \cite{Abe:2019vii}. They also found out that $\dcp=0$ was excluded at $99\%$ C.L.

\begin{figure}[htb]
\centering
\vskip -1.cm
\includegraphics[width=0.85\textwidth]{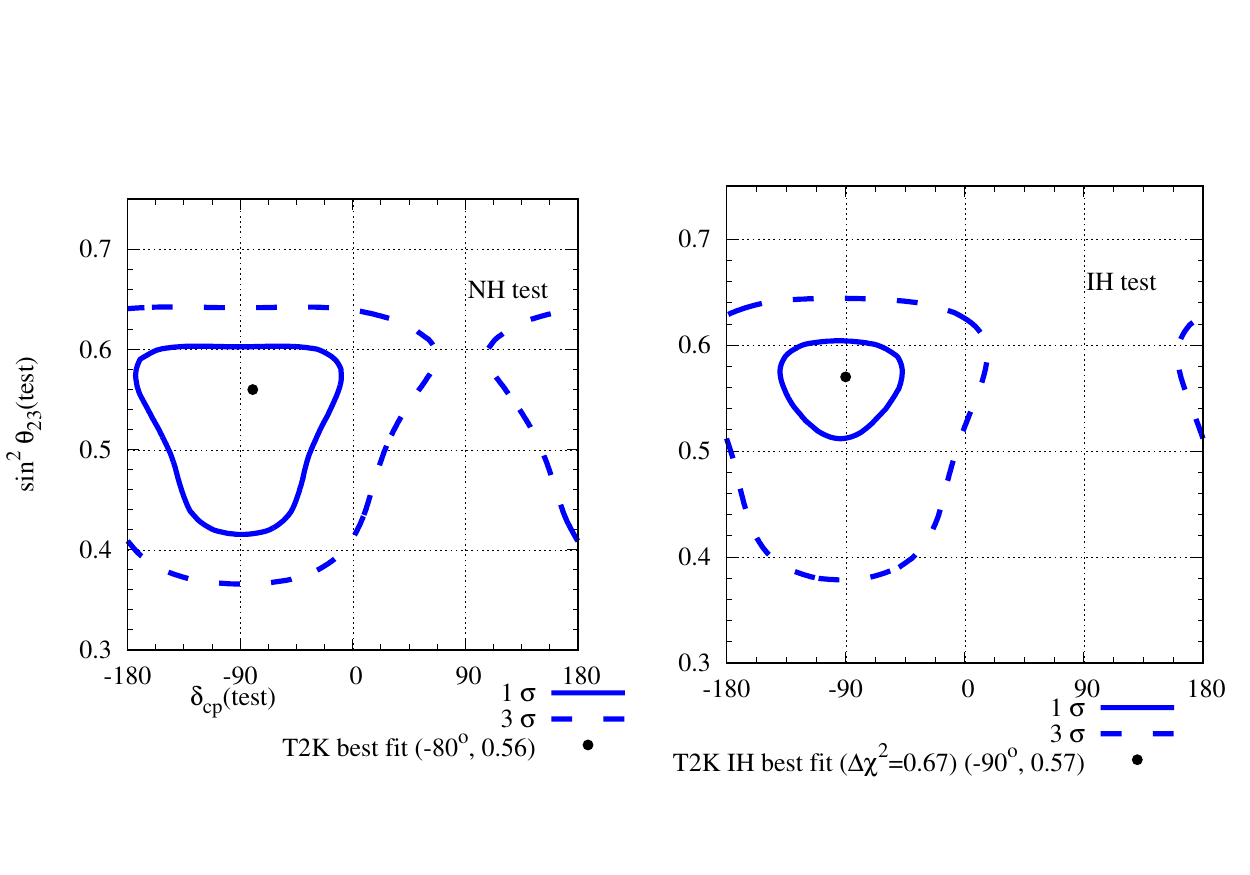}
\caption{\footnotesize{Expected allowed regions in the $\sin^2\tz-\dcp$ plane from the appearance data from T2K, as given in ref.~\cite{Dunne:2020, T2K:2021xwb}, in both the $\nu_e$ and $\bar{\nu}_e$ channels. In the left (right) panel, the hierarchy is assumed to be NH (IH). The best-fit point is at NH with a minimum $\chi^2=20.23$ for 18 energy bins. }}
\label{appearance-2020-T2K}
\end{figure}

In 2020, T2K published their latest data with $1.97\times 10^{21}$ ($1.63\times 10^{21}$) POT in the neutrino (anti-neutrino) mode \cite{Dunne:2020, T2K:2021xwb}. The best-fit point is at $\dcp/\pi=-2.14^{+0.90}_{-0.69}$ ($-1.26^{+0.61}_{-0.69}$), $\sin^2\tz=0.512^{+0.045}_{-0.042}$ for NH (IH). This result can be explained with the change in event number due to the change in unknown parameter values from their reference point values \cite{Rahaman:2021zzm}. At the reference point $'000'$, the expected number of $\nu_e$ ($\bar{\nu}_e$) appearance event is $78$ ($19$). T2K observed $113$ ($15$) $\nu_e$ ($\bar{\nu}_e$) appearance events. The large excess in the $\nu_e$ appearance channel observed by T2K can only be explained by making the choice of unknowns to be $'+++'$. Choosing NH leads to only an $8\%$ boost and we need $\tz$ to have a large value in HO to explain the large excess $\nu_e$ events. But the disappearance data limits $\sin^2\tz \leq 0.59$. Given that only about $20\%$ boost is possible from the hierarchy and octant, $\dcp$ has to be firmly anchored around $\dcp=-90^\circ$ to accommodate the large excess in the $\nu_e$ appearance channel. The $\bar{\nu}_e$ appearance data see a reduction in the observed events. This reduction is consistent with the event numbers expected from 
$'+++'$ choice of the unknowns. However, the number of events in this channel is too small to have any statistical significance.  In ref.~\cite{Rahaman:2021zzm}, an analysis of the T2K $\nu_e$ and $\bar{\nu}_e$ appearance data has been done and the result is presented in Fig.~\ref{appearance-2020-T2K}. It is obvious that the large excess in the T2K $\nu_e$ appearance channel is responsible for $\dcp$ being close to $-90^\circ$, and this is what leads to the tension between the \nova and T2K data.

\subsection{Combined analysis of \nova and T2K data}
\begin{figure}[htb]
\centering
\vskip -1.5cm
\includegraphics[width=1.0\textwidth]{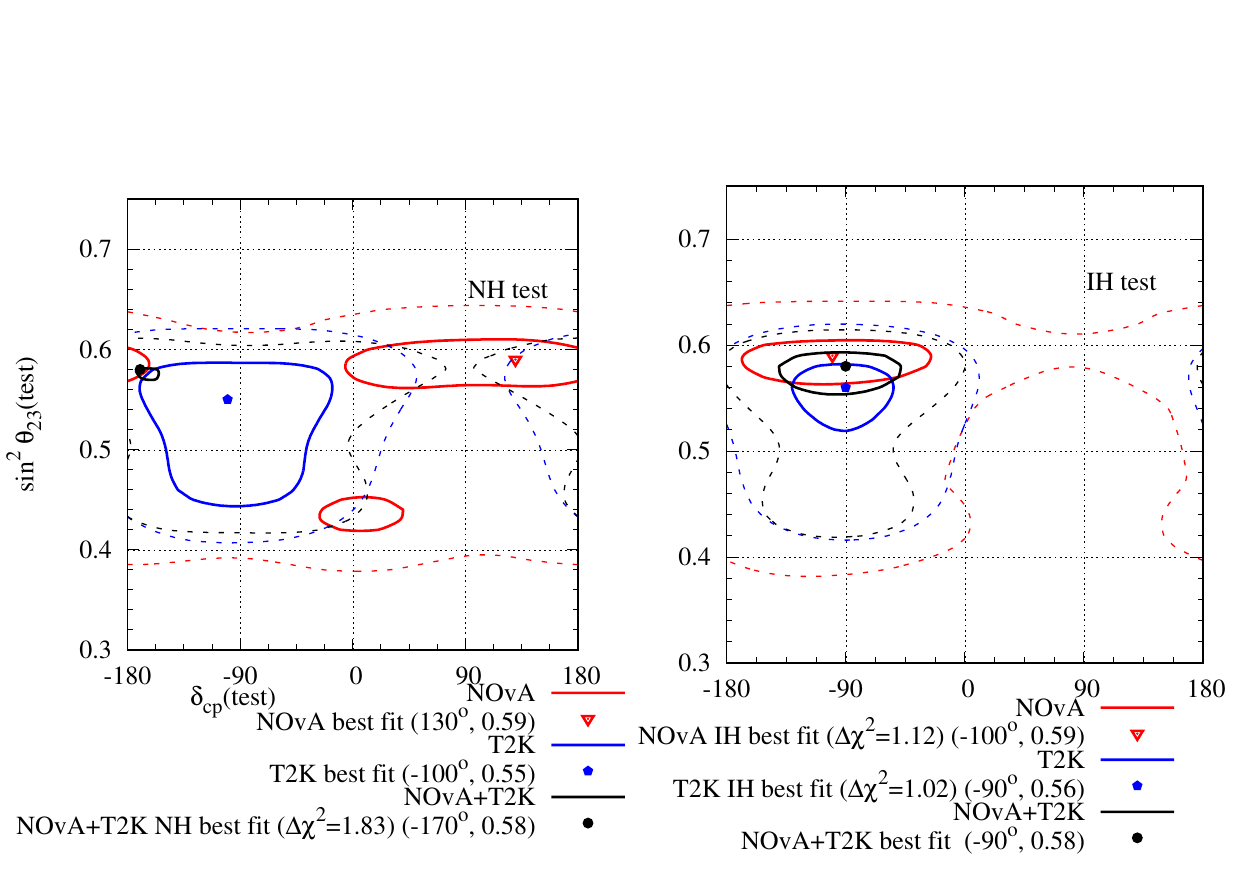}
\caption{\footnotesize{Allowed region in the $\sin^2 \tz-\dcp$ plane after analysing \nova and T2K complete data sets. The left (right) panel represents the test hierarchy to be NH (IH). The red (blue) lines indicate the results for \nova (T2K)
and the black line indicates the combined analysis of both. The solid (dashed) lines indicate the $1\, \sigma$ ($3\, \sigma$)
allowed regions. The minimum $\chi^2$ for \nova
(T2K) with 50 (88) bins is 48.65 (95.85) and it occurs at NH. For the combined analysis, the minimum $\chi^2$ with 138 bins is 147.14, which occurs for IH.}}
\label{nova+t2k-2020}
\end{figure}

Joint analyses have been planned between the \nova and T2K collaborations with the aim of obtaining better constraints on the oscillation parameters due to resolved degeneracy and to understand the non-trivial systematic correlations between them~\cite{Berns:2021iss}. However, independent joint analyses of the two experiments have been done in ref.~\cite{Kelly:2020fkv}, and it has been found out that the combined analysis prefers IH over NH. The result has been shown in Fig.~\ref{nova+t2k-2020}. Note that there is {\bf no} overlap between the
$1\,\sigma$ individual allowed regions of T2K and NO$\nu$A. This also leads to an extremely tiny
$1\,\sigma$ allowed region for the combined analysis.

\section{Resolution of the \nova-T2K tension with new physics}
\label{resolution}
One of the possible reasons for the tension between \nova and T2K is the existence of new 
physics in the neutrino oscillation. Effect of new physics on the determination of unknown oscillation parameters in the long-baseline accelerator neutrino experiments have been studied in details in the literature \cite{Arguelles:2019xgp}. In recent times, efforts have been made to resolve the tension between \nova and T2K experiments with the help of non-standard physics \cite{Miranda:2019ynh, Rahaman:2021leu, Chatterjee:2020kkm, Denton:2020uda}. In this section, we will discuss the present status of different non-standard physics to resolve this tension.

\subsection{Non-unitary mixing}
The anomalies observed in LSND \cite{Aguilar:2001ty} and MiniBooNE \cite{AguilarArevalo:2007it} experiments can be explained with the existence of one or more ``sterile'' neutrino states with mass at or below a few eV \cite{Abazajian:2012ys}\footnote{A recent results from the MicroBooNE experiment \cite{MicroBooNE:2021zai} rules out any excess electron-like events at $94.8\%$ C.L. However, they do not rule out the complete parameter space suggested by the MiniBooNE experiment and other data, nor do they probe the $\nu_e$ interpretation of MiniBooNE result in a model independent way \cite{Arguelles:2021meu}.}. The effect of sterile neutrino on the long-baseline accelerator neutrino experiments have been discussed in detail in Ref.~\cite{Gupta:2018qsv, Chatla:2018sos, Choubey:2017ppj, Choubey:2017cba, Berryman:2015nua}. If the sterile neutrinos exist as iso-singlet neutral heavy leptons (NHL), then in the minimum extension of the standard model, they do not take part in neutrino oscillation. However, their admixture in the charged current weak interactions affects neutrino oscillation. In such scenarios, neutrino oscillation will be governed by a non-unitary mixing matrix. The mixing matrix, in this case, can be parameterised as \cite{Escrihuela:2015wra}
\begin{equation}
    N=N_{\rm NP}U_{3\times 3}= \left[ {\begin{array}{ccc}
   \alpha_{00} & 0 & 0 \\
   \alpha_{10} & \alpha_{11} & 0 \\
   \alpha_{20} & \alpha_{21} & \alpha_{22}
  \end{array} } \right] U_{\rm PMNS} \,.
\end{equation}
Here $U_{\rm PMNS}$ is the standard unitary PMNS mixing matrix. The diagonal (off-diagonal) terms of $N_{\rm NP}$ matrix are real (complex). To allow the effect of non-unitarity, the diagonal terms of $N_{\rm NP}$ matrix must deviate from unity, and/or the off-diagonal terms must deviate from $0$. The present constraints on non-unitary parameters at $3\, \sigma$ C.L.\ are \cite{Escrihuela:2016ube}:
\begin{eqnarray}
\alpha_{00} > 0.93 \,;\, \alpha_{11} > 0.95 \,;\, \alpha_{22} > 0.61 \nonumber \\ 
|\alpha_{10}| < 3.6\times 10^{-2} \,;\, |\alpha_{20}| < 1.3\times 10^{-1} 
\,;\, |\alpha_{21}| < 2.1\times 10^{-2}\,.
\label{eq:nubounds}
\end{eqnarray}

The calculation of oscillation probability in the presence of matter effect in the case of non-unitary mixing has been discussed in Ref.~\cite{Escrihuela:2015wra, Miranda:2019ynh}. The effects of non-unitary mixing on the determination of the unknown oscillation parameters in the present and future long-baseline accelerator neutrino experiments have been discussed in literature \cite{Ge:2016xya, Escrihuela:2015wra, Soumya:2018nkw, Fong:2017gke, Verma:2016nfi}. 

In the context of non-unitary mixing, the mixing between flavour states and mass eigenstates can be written as
\begin{equation}
    |\nu_{\beta} \rangle = \sum_{i=1}^{3}N_{\beta i}|\nu_i \rangle ,
\end{equation}
where $\beta$ denotes the flavour states and $i$ denotes the mass eigenstates. The evolution of neutrino mass eigenstates during the propagation, can be written as
\begin{equation}
    i\frac{d}{dt}|\nu_i \rangle = H_{\rm vac}|\nu_i \rangle,
\end{equation}
where $H_{\rm vac}$ is the Hamiltonian in vacuum and it is defined as
\begin{equation}
    H_{\rm vac}=\frac{1}{2E}\left[ {\begin{array}{ccc}
   0 & 0 & 0 \\
   0 & \ds & 0 \\
   0 & 0 & \dl
  \end{array} } \right]
\end{equation}
The non-unitary neutrino oscillation probability in vacuum can be written as
\begin{eqnarray}
    \pme^{\rm NU}({\rm vac})&=& \sum_{i,j}^{3}N_{\mu i}^{*}N_{ei}N_{\mu j}N_{ej}^{*}-4\sum_{j>i}^{3}Re\left[N_{\mu j}N_{ej}N_{\mu i}N_{ei}^{*}\right]\sin^2 \left(\frac{\Delta_{ji}L}{4E}\right)\nonumber \\
    &&+2\sum_{j>i}^{3}Im\left[N_{\mu j}^{*}N_{ej}N_{\mu i}N_{ei}\right]\sin \left(\frac{\Delta_{ji}L}{2E}\right).
    \label{pme-nu}
\end{eqnarray}
If written explicitly, neglecting the cubic products of $\alpha_{10}$, $\sin \ty$, and $\ds$, the oscillation probability takes the form \cite{Escrihuela:2015wra}
\begin{equation}
    \pme^{\rm NU}({\rm vac})= (\alpha_{00}\alpha_{11})^2\pme^{\rm SO}+\alpha_{00}^2\alpha_{11}|\alpha_{10}|\pme^{\rm I} +\alpha_{00}^2|\alpha_{10}|^2.
\end{equation}
$\pme^{\rm SO}$ is the standard three-flavour unitary neutrino oscillation probability in vacuum and can be written as
\begin{eqnarray}
    \pme^{\rm SO}&=& 4\cos^2 \tx \cos^2 \tz \sin^2 \tx\sin^2\left(\frac{\ds L}{4E}\right)\nonumber \\
    &&+ 4\cos^2 \ty \sin^2 \ty \sin^2 \tz\sin^2\left(\frac{\dl L}{4E}\right)\nonumber \\
    &&+ \sin (2\tx)\sin \ty \sin (2\tz)\sin\left(\frac{\ds L}{2E}\right)\sin\left(\frac{\dl L}{4E}\right)\cos\left(\frac{\dl L}{4E}-I_{123}\right),
   \end{eqnarray}
 and
\begin{eqnarray}
    \pme^{\rm I} &=&-2\left[\sin (2\ty)\sin\tz\sin \left(\frac{\dl L}{4E}\right)\sin\left(\frac{\dl L}{4E}+I_{\rm NP}-I_{012}\right)\right]\nonumber \\
    &&-\cos \ty \cos \tz \sin (2\tx)\sin\left(\frac{\ds L}{2E}\right)\sin (I_{\rm NP}),
\end{eqnarray}
where $I_{012}=-\dcp=\phi_{10}-\phi_{20}+\phi_{21}$, $I_{123}=\phi_{21}-\phi_{31}+\phi_{32}$ and $I_{\rm NP}=\phi_{10}-{\rm Arg}(\alpha_{10})$. $\phi_{ij}$'s are the phases associated with the off-diagonal terms $\alpha_{ij}=|\alpha_{ij}|e^{i \phi_{ij}}$. It should be noted that non-unitary parameters $\alpha_{00}$, $\alpha_{11}$, and $\alpha_{10}$ have the most significant effects on $\pme^{\rm NU}({\rm vac})$.

While propagating through matter, the neutrinos undergo forward scattering and the neutrino oscillation probability gets modified due to interaction potential between neutrino and matter. In case of non-unitary mixing, the CC and NC terms in the interaction Lagrangian is given as
\begin{equation}
    \mathcal{L}=V_{\rm CC}\sum_{i,j}N_{ei}^{*}N_{ej}\bar{\nu}_i\gamma^0\nu_j+V_{\rm NC}\sum_{\alpha,i,j}N_{\alpha i}^{*}N_{\alpha j}\bar{\nu}_i\gamma^0\nu_j,
\end{equation}
where $V_{\rm CC}=\sqrt{2}G_FN_e$, and $V_{\rm NC}=-G_F N_n/\sqrt{2}$ are the potentials for CC and NC interactions, respectively. Therefore, the effective Hamiltonian becomes
\begin{equation}
    H_{\rm matter}^{\rm NU}= \frac{1}{2E}\left[ {\begin{array}{ccc}
   0 & 0 & 0 \\
   0 & \ds & 0 \\
   0 & 0 & \dl
  \end{array} } \right]+ \frac{1}{2E} N^\dagger\left[ {\begin{array}{ccc}
   V_{\rm CC}+V_{\rm NC} & 0 & 0 \\
   0 & V_{\rm NC} & 0 \\
   0 & 0 & V_{\rm NC}
  \end{array} } \right]N \,.
\end{equation}
The non-unitary neutrino oscillation probability, after neutrinos travel through a distance $L$, is given as
\begin{equation}
    P_{\alpha \beta} (E,L)= |\langle\nu_\beta|\nu_\alpha (L)\rangle|^2=\left|\left(Ne^{-iH_{\rm matter}^{\rm NU}L}N^\dagger\right)_{\beta \alpha}\right|^2 \,.
\end{equation}

\begin{figure}[H]
\centering
\vskip -1.5cm
\includegraphics[width=1.0\textwidth]{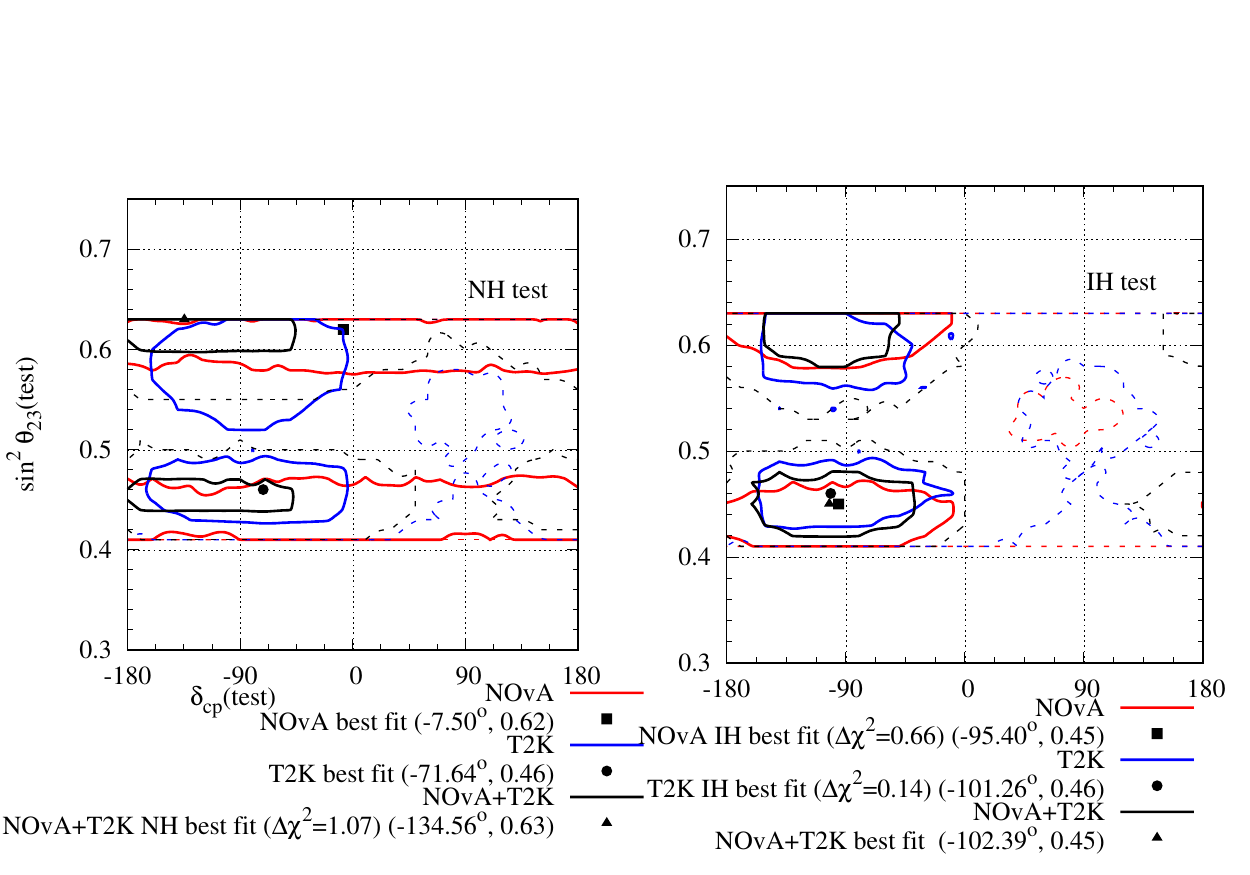}
\caption{\footnotesize{Allowed region in the $\sin^2 \tz-\dcp$ plane after analysing \nova and T2K complete data set with non-unitary hypothesis. The left (right) panel represents test hierarchy NH (IH). The red (blue) lines indicate the results for \nova (T2K)
and the black line indicates the combined analysis of both. The solid (dashed) lines indicate the $1\, \sigma$ ($3\, \sigma$)
allowed regions. The minimum $\chi^2$ for \nova
(T2K) with 50 (88) bins is 45.88 (93.36) and it occurs at NH. For the combined analysis, the minimum $\chi^2$ with 138 bins is 142.72.}}
\label{nova+t2k-2020-non-uni}
\end{figure}

A detailed description of non-unitary neutrino oscillation probability in the presence of matter effect has been discussed in Ref.~\cite{Miranda:2019ynh}, 
where an effort has been made to resolve the tension between the two experiments with non-unitary mixing. To do so, the authors first analysed the present data from \nova and T2K with standard unitary oscillation hypothesis. We have already presented the result in Fig.~\ref{nova+t2k-2020} using the 2020 data. As mentioned in the caption of Fig.~\ref{nova+t2k-2020}, the minimum $\chi^2$ for \nova (T2K) is $48.65$ ($95.85$) for $50$ ($88$) energy bins. For the combined analysis, the minimum $\chi^2$ was $147.14$ for $138$ energy bins. In the next step, data were analysed with non-unitary mixing hypothesis. To do so, only the effects of $\alpha_{00}$, $|\alpha_{10}|$, and $\alpha_{11}$ were considered, as these three parameters have the most significant effects on the oscillation probabilities $\pme$, and $\pmebar$. Only those values were chosen for which the condition $|\alpha_{10}|\leq \sqrt{(1-\alpha_{00}^2)(1-\alpha_{11}^{2})}$  \cite{Antusch_2014, Escrihuela:2016ube} is satisfied. All other non-unitary parameters have been kept fixed at their unitary values. The software GLoBES \cite{Huber:2004ka, Huber:2007ji} was used to analyse the data for both standard and non-unitary oscillations. In the later case, the software was modified to include non-unitary mixing. We have already talked about the standard unitary parameter values in section \ref{chisq}. After analysing the data with non-unitary mixing hypothesis, the minimum $\chi^2$ was found to be $45.88$ ($93.36$) for \nova (T2K). For the combined analysis, the minimum $\chi^2$ was $142.72$. Thus it can be said that each of the two experiments individually prefer non-unitary mixing over unitary mixing by $1\, \sigma$ C.L., and the combined analysis rules out unitary mixing at $2\, \sigma$ C.L. The result has been presented in Fig.~\ref{nova+t2k-2020-non-uni}. It can be observed from the figure that there is a large overlap between the $1\,\sigma$ allowed regions of the two experiments for both the hierarchies. Both the experiments lose their hierarchy sensitivity when the data are analysed with non-unitary mixing. For the NH best-fit point, \nova (T2K) prefers $\tz$ to be in LO (HO), but a nearly degenerate best-fit point occurs at HO (LO). Thus the experiments lose their octant determination sensitivity as well. Best-fit points at IH coincide for both the experiments. 

\begin{figure}[H]
\includegraphics[width=85 mm,scale=2.0]{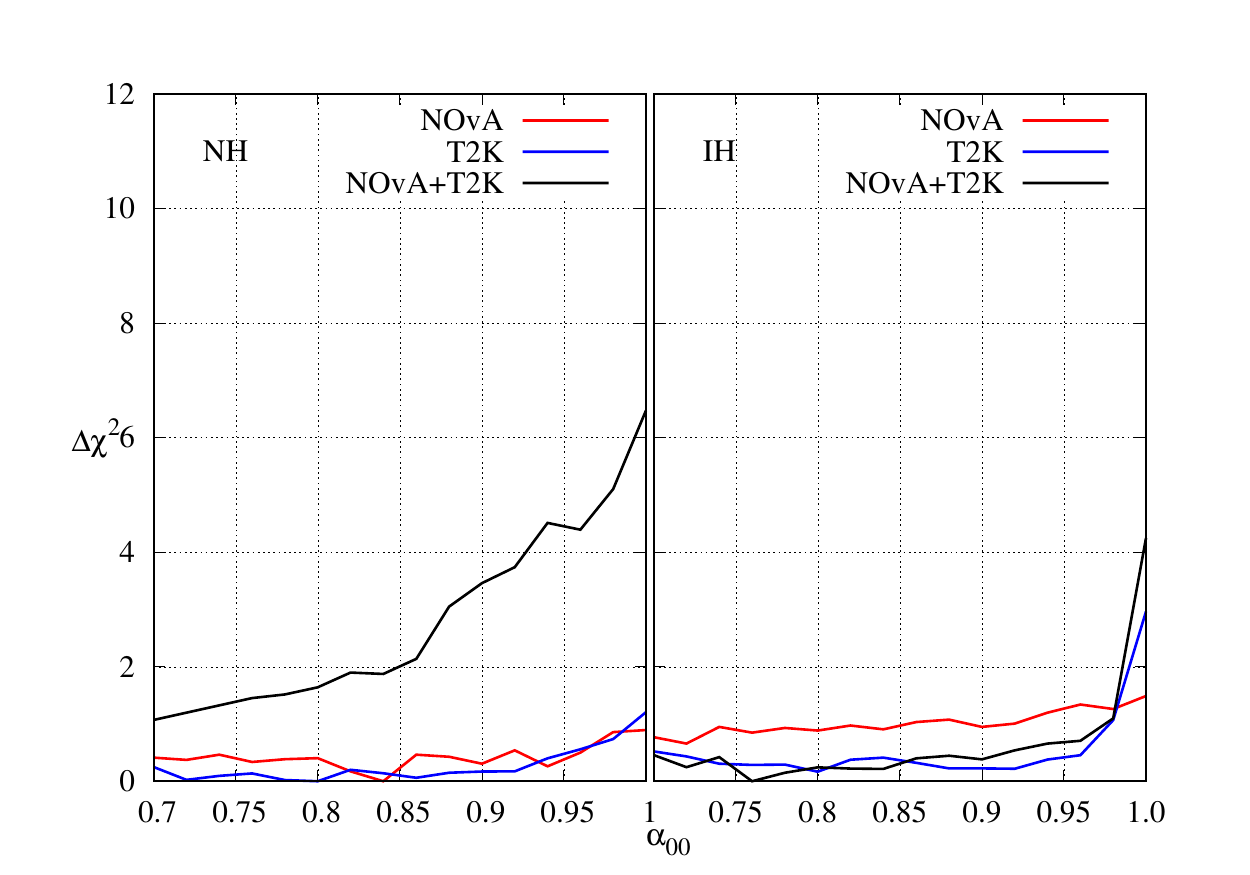}
\includegraphics[width=85 mm,scale=2.0]{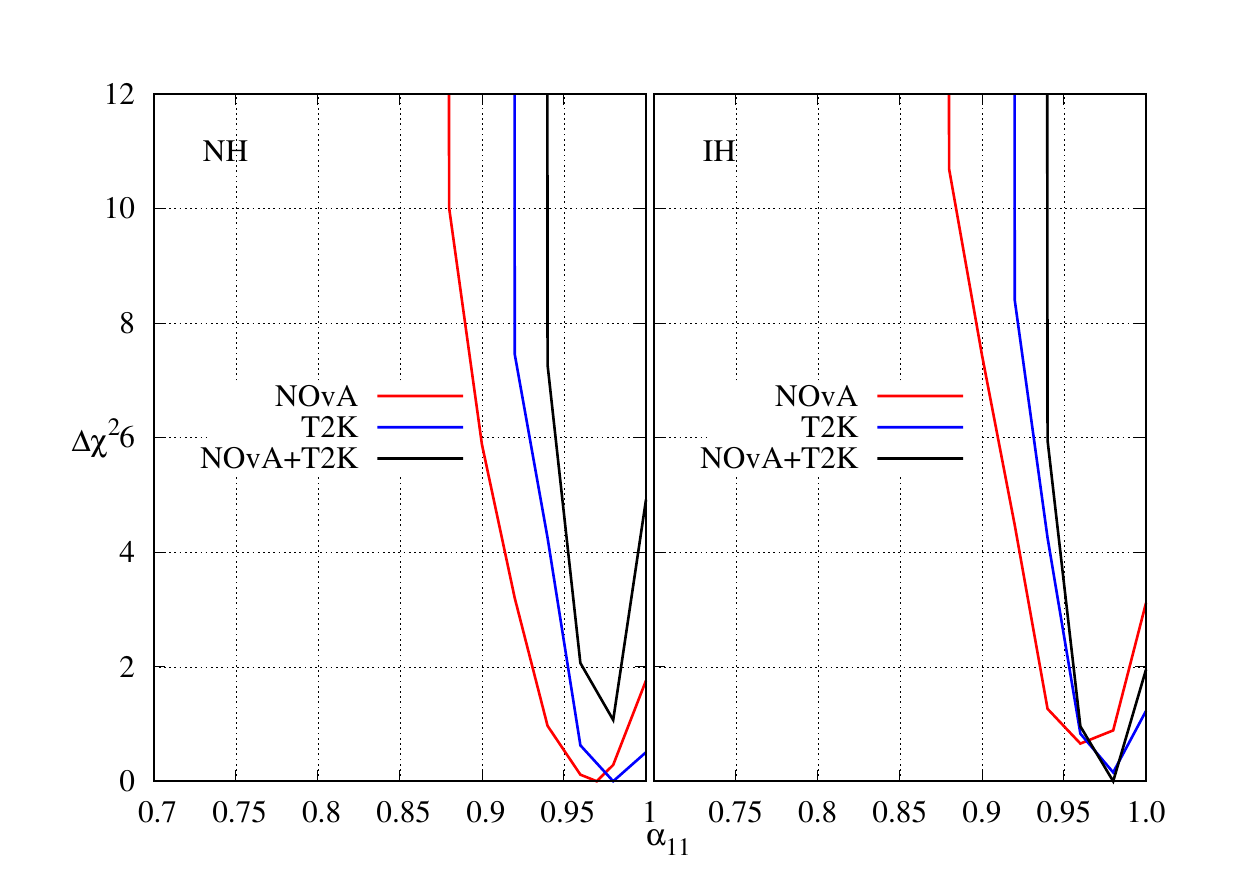}
\vskip -0.5cm
\includegraphics[width=85 mm, scale=2.0]{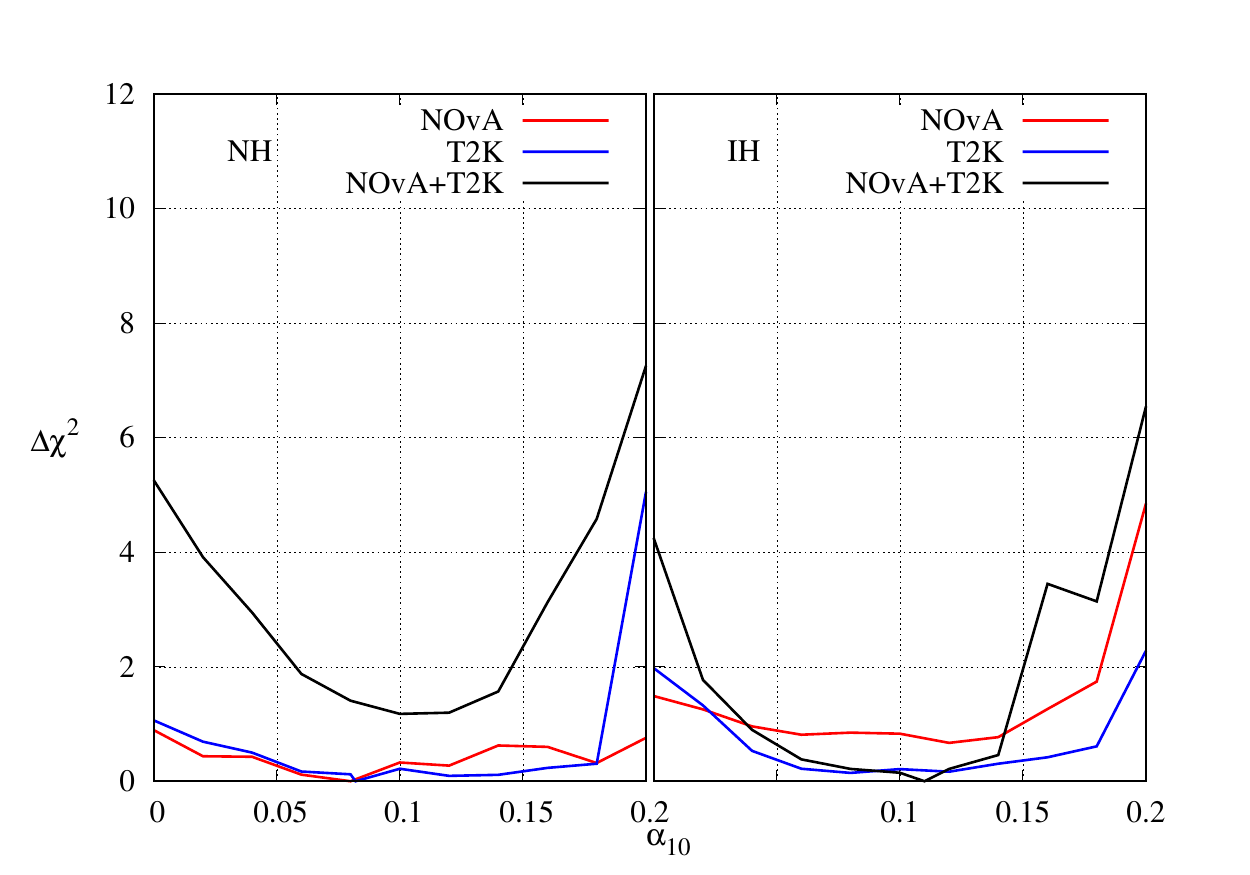}
\includegraphics[width=85 mm,scale=2.0]{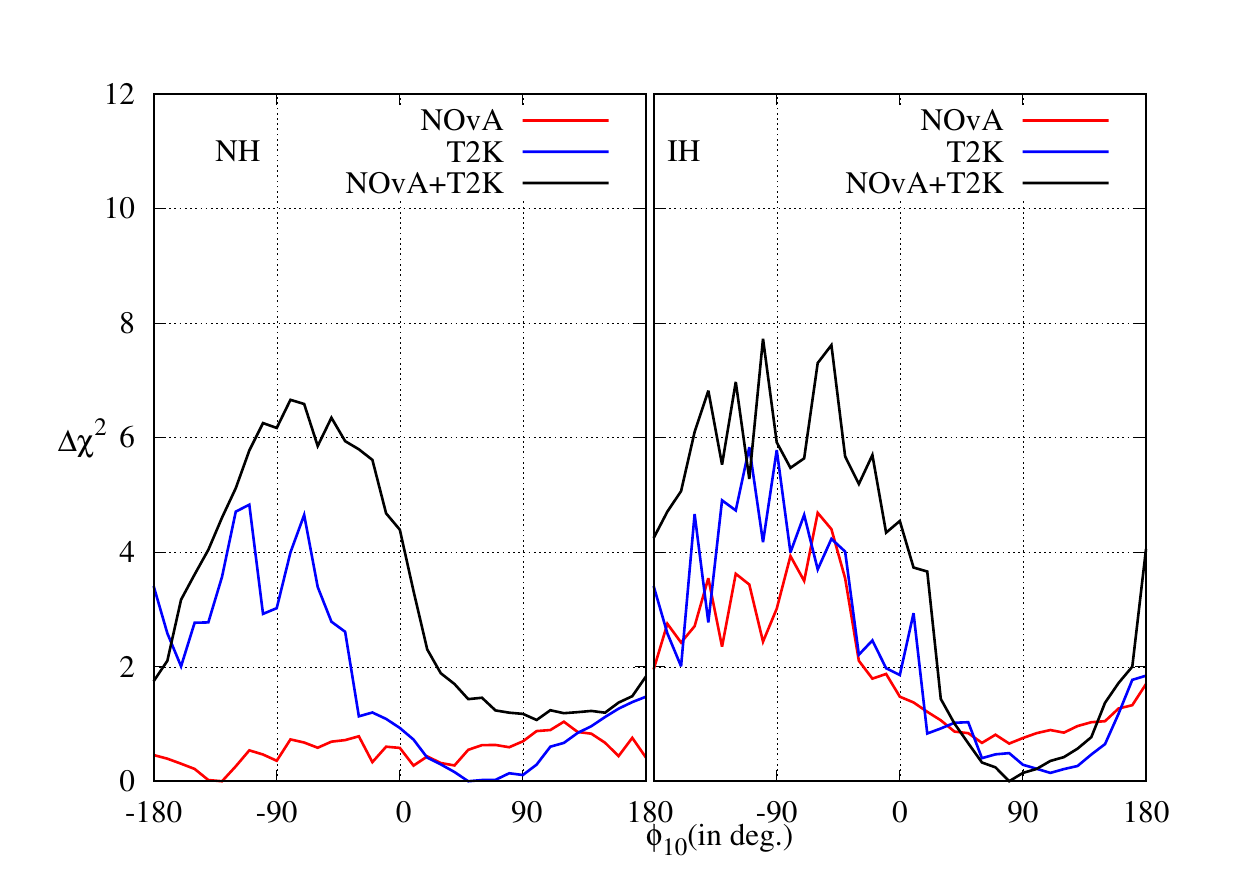}
\caption{\footnotesize{$\dchsq$ as a function of individual non-unitary parameters after analysing 2020 data.}}
\label{alpha-precision}
\end{figure}

In Fig.~\ref{alpha-precision}, $\dchsq$ as a function of individual non-unitary parameters has been represented. It is obvious that each experiment rules out unitary mixing at $1\, \sigma$ C.L., and the combined analysis does the same at $2\, \sigma$ C.L. 

\begin{figure}[H]
\centering
\includegraphics[width=0.75\textwidth]{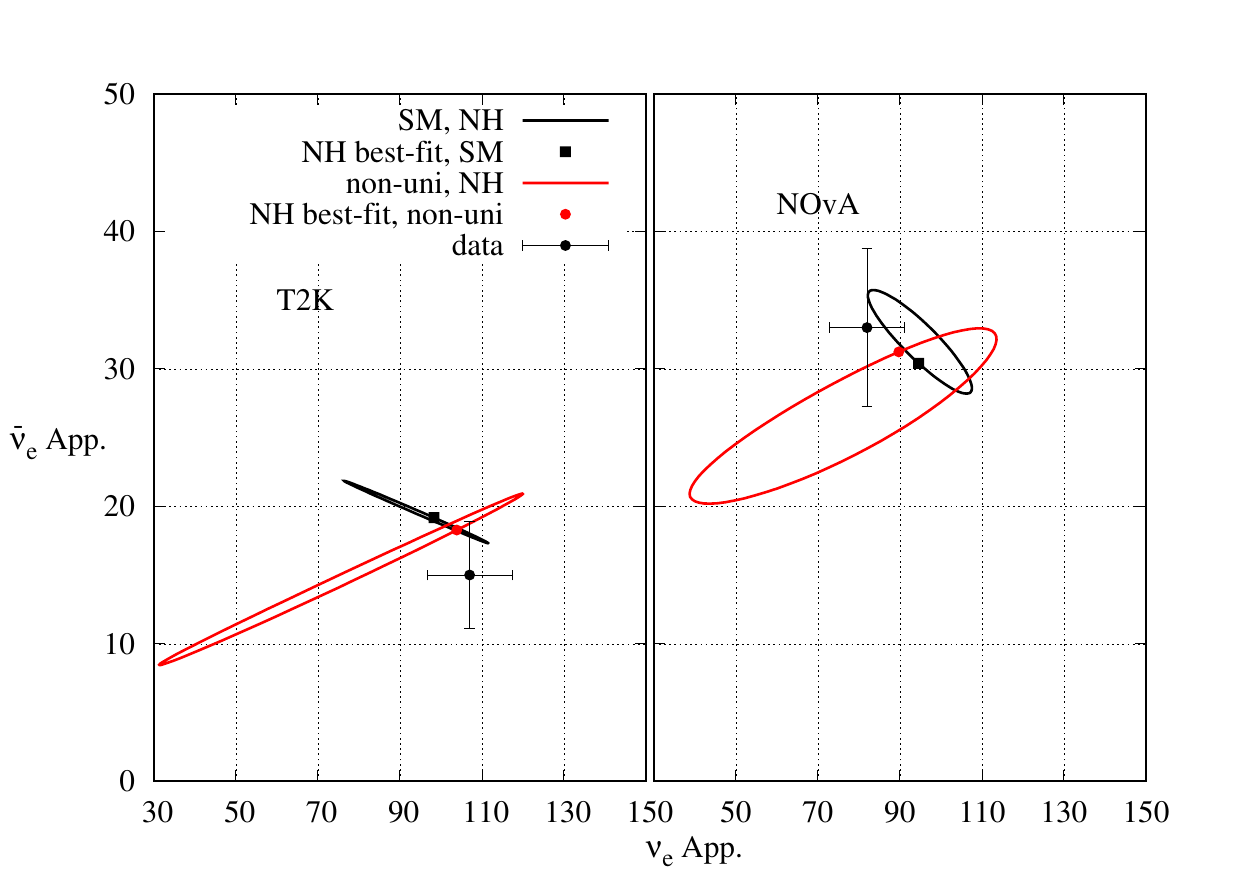}
\includegraphics[width=0.75\textwidth]{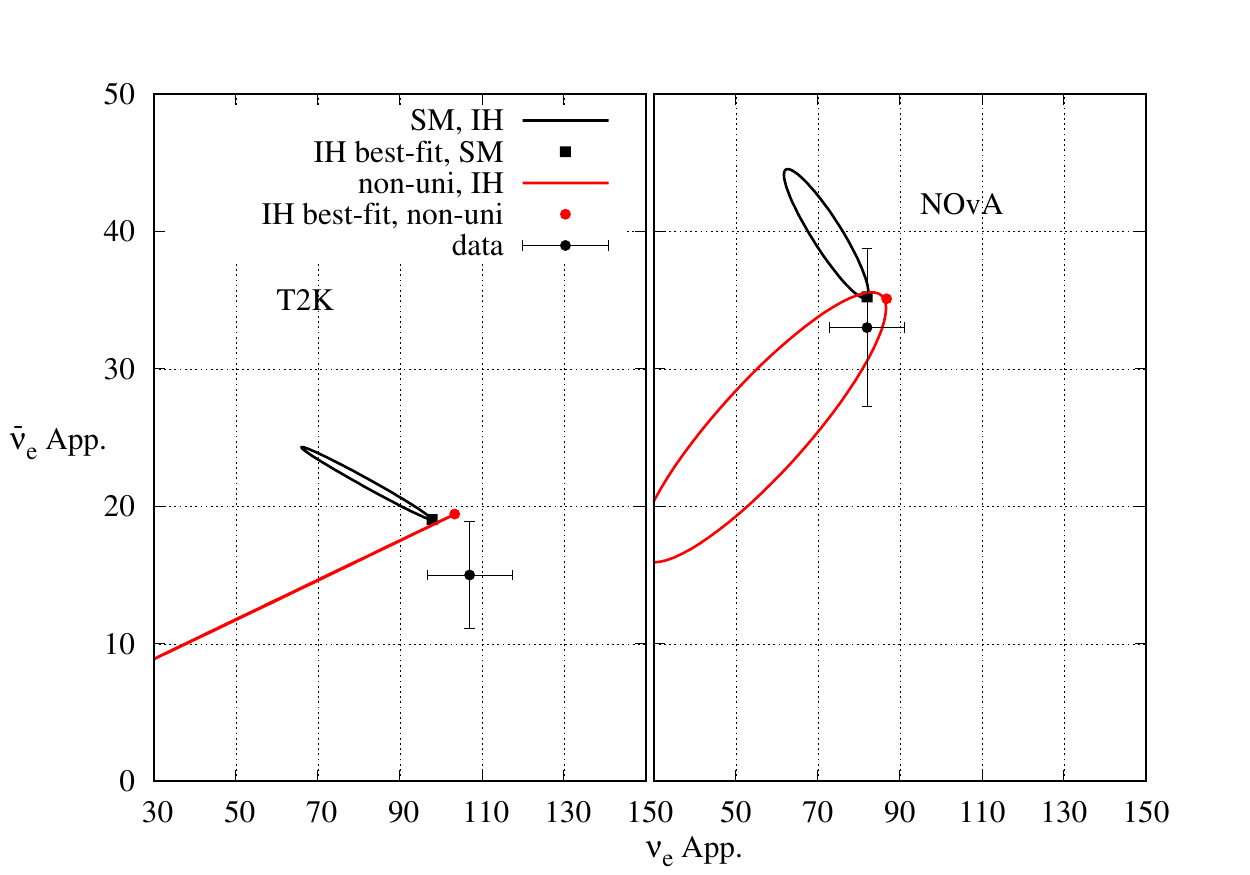}
\caption{\footnotesize{Bi-event plot for T2K (NO$\nu$A) in left (right) panel. The upper (lower) panel is for NH (IH). To generate the ellipses, $\dcp$ has been varied in the range $[-180^\circ:180^\circ]$ while keeping all other parameters at their combined best-fit values. The black (red) ellipses represent the SM (non-unitary) case with the best-fit points indicated by black square (red circle). The ellipses and the best-fit points have been determined by fitting the combined data from \nova and T2K. The black circle with error bars represent the experimental data.}}
\label{bievents-nonuni}
\end{figure}

In Fig.~\ref{bievents-nonuni}, we have represented the bi-event plots with $x$-axis ($y$-axis) denoting the $\nu_e$ ($\bar{\nu}_e$) appearance events. These plots are helpful in understanding the origin of the tension and their resolution with the help of new physics. As we have already seen, in \nova and T2K, all the information from appearance channels can be represented by total number of events, because of the limited statistics, the information extracted from the shape of the energy spectrum is limited. The ellipses have been generated by fitting the combined data from \nova and T2K. The black ellipses, corresponding to the standard unitary oscillation, have been generated for the combined best-fit values of $|\dl|$, $\sin^2 2\ty$, and $\sin^2 \tz$. The red ellipses, corresponding to the non-unitary oscillation, have been generated for the combined best-fit values of $|\dl|$, $\sin^2 2\ty$, and $\sin^2 \tz$, $\alpha_{00}$, $|\alpha_{10}|$, $\alpha_{11}$, and $\phi_{10}$. In both cases, only the $\dcp$ has been varied in its complete range $[-180^\circ:180^\circ]$ while keeping all other parameters fixed at their best-fit values from the combined analysis. For the black (red) ellipse, the square (circle) represent the best-fit event numbers for $\dcp=-170^\circ$ ($-134.56^\circ$) for NH, and $\dcp=-90^\circ$ ($-102.39^\circ$) for IH. These best-fit values come as a compromise between T2K and \nova. However, for the standard case, T2K appearance events strongly prefer NH and $\dcp=-80^\circ$, and \nova prefers NH-$\dcp=150^\circ$ or IH-$\dcp=-90^\circ$ (Table II of Ref.~\cite{Rahaman:2021zzm}). Thus we see that for NH, neither of the two experiments can give a good fit to the data at the combined best-fit point. Non-unitary mixing hypothesis, brings the expected appearance event numbers closer to the observed event numbers. For IH, however, T2K alone cannot give a good fit to the data at the combined best-fit point for standard oscillation, and non-unitary hypothesis brings the expected event number 
closer to the observed event number. But \nova can give a good fit to the data at the combined best-fit point for standard oscillation, and including non-unitary mixing does not affect it in any significant way. Hence, the tension between the two experiments gets reduced by the non-unitary mixing.

\subsection{Lorentz invariance violation}
Neutrino oscillation is the first experimental signature towards BSM physics, as it requires neutrinos to be massive, albeit extremely light. Without loss of any generality, SM can be considered as the low-energy effective theory derived from a more general theory -- governed by the Planck mass ($M_P\simeq 10^{19}\, {\rm GeV}$) -- which unify the  gravitational interactions along with the weak, strong, and electromagnetic interactions. There are models which include spontaneous Lorentz invariance violation (LIV) and CPT violations in that more complete framework at the Planck scale \cite{Kostelecky:1988zi, Kostelecky:1989jp, Kostelecky:1991ak, Kostelecky:1994rn, Kostelecky:1995qk}. At the observable low energy, these violations can give rise to minimal extension of SM through perturbative terms suppressed by $M_P$. CPT invariance imposes that particles and anti-particles have the same mass and lifetime. Observation of difference between masses and lifetimes of particles and anti-particles would be a hint of CPT violation. The present upper limit on CPT violation from the kaon system is $|m_{K^0}-m_{\bar{K}_0}|/m_K<6 \times 10^{-18}$ \cite{Tanabashi:2018oca}. Since kaons are bosons and the natural mass term appearing in the Lagrangian is mass squared term, the above constraints can be rewritten as $|m_{K^0}^{2}-m_{\bar{K}_0}^{2}|<0.25\, {\rm eV}^2$. Current neutrino oscillation data provide the bounds $|\ds-\bar{\Delta}_{21}|<5.9 \times 10^{-5}\, {\rm eV}^2$ and $|\dl-\bar{\Delta}_{31}|<1.1 \times 10^{-3}\, {\rm eV}^2$ \cite{Ohlsson:2014cha}. The non-zero differences are manifestations of some kind of CPT violation and this can change neutrino oscillation probability \cite{Kostelecky:2003cr, Diaz:2011ia, Kostelecky:2004hg, Katori:2006mz}. 

Several studies have been done about the LIV/CPT violation with neutrinos \cite{Dighe:2008bu, Barenboim:2009ts, Rebel:2013vc, deGouvea:2017yvn, Barenboim:2017ewj, Barenboim:2018ctx, Majhi:2019tfi, Giunti:2010zs, Datta:2003dg, Chatterjee:2014oda, Koranga:2014dua, Diaz:2016fqd, Hooper:2005jp, Tomar:2015fha, Liao:2017yuy, Agarwalla:2019rgv}. Different neutrino oscillation experiments have looked for LIV/CPT violations and put on constraints on the LIV/CPT violating parameters \cite{Auerbach:2005tq, Adamson:2008aa, Adamson:2012hp, Aguilar-Arevalo:2018gpe, Abe:2012gw, Abe:2014wla, Aartsen:2017ibm, Abe:2017eot}. Constraints on all the relevant LIV/CPT violating parameters have been listed in Ref.~\cite{Kostelecky:2008ts}. In Ref.~\cite{Rahaman:2021leu}, an effort has been made to resolve the tension between \nova and T2K by considering the changes in neutrino oscillation probability due to CPT-violating LIV.

The effective Lagrangian for the Lorentz invariance violating neutrinos and antineutrinos can be written as \cite{Kostelecky:2003cr, Kostelecky:2011gq} 
\begin{equation}
    \mathcal{L}=\bar{\Psi}_A\left(i \gamma_\mu \partial_\mu \delta_{AB}-M_{AB}+\hat{\mathcal{Q}}_{AB}\right)\Psi_B+{\rm h.c.}.
    \label{LIV-lag}
\end{equation}
$\Psi_{A(B)}$ is a $2N$ dimensional spinor containing $\psi_{\alpha(\beta)}$, which is a spinor field with $\alpha(\beta)$ ranging over $N$ spinor flavours, and their charge conjugates given by $\psi_{\alpha(\beta)}^{C}=C\bar{\psi}_{\alpha(\beta)}^T$. Therefore, $\Psi_{A(B)}$ can be expressed as
\begin{equation}
    \Psi_{A(B)}=\left(\psi_{\alpha(\beta)}, \psi_{\alpha(\beta)}^{C}\right)^T.
\end{equation}
$\hat{\mathcal{Q}}$ in eq.~(\ref{LIV-lag}) is a generic Lorentz invariance violating operator. The first term in the right side of eq.~(\ref{LIV-lag}) is the kinetic term, the second term is the mass term involving the mass matrix $M$ and the third term gives rise to the LIV effect. $\hat{\mathcal{Q}}$ is small and perturbative in nature.

Restricting ourselves only to the renormalizable Dirac couplings in the theory (terms only with mass dimension $\leq 4$ will be incorporated), the LIV  Lagrangian in the flavour basis can be written as \cite{Kostelecky:2003cr}
\begin{equation}
    \mathcal{L}_{\rm LIV}= -\frac{1}{2}\left[a^{\mu}_{\alpha \beta}\bar{\psi}_\alpha \gamma_\mu \psi_\beta+b^{\mu}_{\alpha \beta}\bar{\psi}_\alpha \gamma_5 \gamma_\mu \psi_\beta-i c_{\alpha \beta}^{\mu \nu}\bar{\psi}_\alpha \gamma_\mu \partial_\nu \psi_\beta-i d_{\alpha \beta}^{\mu \nu}\bar{\psi}_\alpha \gamma_\mu \gamma_5 \partial_\nu \psi_\beta\right],
\end{equation}
where $a^{\mu}_{\alpha \beta}$, $b^{\mu}_{\alpha \beta}$, $c^{\mu \nu}_{\alpha \beta}$ and $d^{\mu \nu}_{\alpha \beta}$ are Lorentz invariance violating parameters. Considering that only left handed neutrinos are present in the SM, these terms can be written as
\begin{equation}
    \left(a_L\right)^{\mu}_{\alpha \beta}=\left(a+b\right)^{\mu}_{\alpha \beta}, \left(c_L\right)^{\mu \nu}_{\alpha \beta}=\left(c+d\right)^{\mu \nu}_{\alpha \beta}.
\end{equation}
$\left(a_L\right)^{\mu}_{\alpha \beta}$, and $ \left(c_L\right)^{\mu}_{\alpha \beta}$ are constant Hermitian matrices which can modify the standard Hamiltonian in vacuum. In Ref.~\cite{Rahaman:2021leu}, only direction-independent isotropic terms were considered where $\mu=\nu=0$. From now on, for simplicity, we will call $a_{\alpha \beta}^{0}$ terms as $a_{\alpha \beta}$ and $c_{\alpha \beta}^{00}$ term as $c_{\alpha \beta}$. $a_{\alpha \beta}$ involves CPT violating terms, and $c_{\alpha \beta}$ involves CPT conserving Lorentz invariance violating terms. Taking into account only these isotropic LIV terms, the neutrino Hamiltonian with LIV effect becomes:
\begin{equation}
    H=H_{\rm vac}+H_{\rm mat}+H_{\rm LIV},
    \label{Ham}
\end{equation}
where 
\begin{equation}
    H_{\rm vac}=\frac{1}{2E}U\left[
\begin{array}{ccc}
m_{1}^{2} & 0 & 0\\
0 & m_{2}^{2} & 0\\
0 & 0 & m_{3}^{2}\\
\end{array}
\right]U^\dagger; H_{\rm mat}=\sqrt{2}G_FN_e \left[
\begin{array}{ccc}
1 & 0 & 0\\
0 & 0 & 0\\
0 & 0 & 0\\
\end{array}
\right];
\end{equation}
\begin{equation}
    H_{\rm LIV}=\left[
\begin{array}{ccc}
a_{ee} & a_{e\mu} & a_{e\tau}\\
a_{e\mu}^{*} & a_{\mu \mu} & a_{\mu \tau}\\
a_{e\tau}^{*} & a_{\mu\tau}^{*} & a_{\tau \tau}\\
\end{array}
\right] -\frac{4}{3}E\left[
\begin{array}{ccc}
c_{ee} & c_{e\mu} & c_{e\tau}\\
c_{e\mu}^{*} & c_{\mu \mu} & c_{\mu \tau}\\
c_{e\tau}^{*} & c_{\mu\tau}^{*} & c_{\tau \tau}\\
\end{array}
\right].
\label{Ham-LIV3}
\end{equation}
Here $G_F$ is the Fermi coupling constant and $N_e$ is the electron density along the neutrino path. The $-4/3$ in front of the second term arises due to non observability of the Minkowski trace of the CPT conserving LIV term $c_L$ which relates the $xx$, $yy$, and $zz$ component to the $00$ component \cite{Kostelecky:2003cr}. The effects of $a_{\alpha \beta}$ are proportional to the baseline $L$ and the effects of $c_{\alpha \beta}$ are proportional to the product of energy and baseline $LE$. In Ref.~\cite{} only the CPT violating LIV was considered. More specifically, the authors restricted themselves to the effects of $a_{e\mu}=|a_{e\mu}|e^{i\phi_{e\mu}}$, and $a_{e\tau} = |a_{e_\tau} |e^{i\phi_{e\tau}}$, because these two terms have the maximum effects on the $\nu_\mu \to \nu_e$ oscillation probability \cite{Agarwalla:2019rgv}. The current constraint on these parameters from Super-kamiokande experiment at $95\%$ C.L.\ is \cite{Abe:2014wla} 
\begin{equation}
    |a_{e\mu}|<2.5\times 10^{-23}\, {\rm GeV};\, |a_{e\tau}|<5\times 10^{-23}\, {\rm GeV}
\end{equation}

The $\nu_\mu \to \nu_e$ oscillation probability in matter after inclusion of LIV can be approximately written as \cite{Agarwalla:2019rgv}
\begin{equation}
    P_{\mu e}^{\rm SM+LIV}\simeq P_{\mu e} ({\rm SM})+P_{\mu e} (a_{e\mu})+P_{\mu e}(a_{e\tau}).
    \label{pmue-LIV}
\end{equation}
The $P_{\mu e} (\rm SM)$ term in eq.~(\ref{pmue-LIV}) has been given in eq.~(\ref{pme}). The other two terms can be written as \cite{Agarwalla:2019rgv}
\begin{eqnarray}
    P_{\mu e}(a_{e\beta})&=&\frac{4|a_{e\beta}|\ahat \dhat\sin \theta_{13}\sin 2\theta_{23}\sin \dhat}{\sqrt{2}G_F N_e}\left[Z_{e\beta}\sin(\dcp+\phi_{e\beta})+W_{e\beta}\cos(\dcp+\phi_{e\beta}) \right] \nonumber \\
    &=& \frac{4|a_{e\beta}|L \sin \theta_{13}\sin 2\theta_{23}\sin \dhat}{2}\left[Z_{e\beta}\sin(\dcp+\phi_{e\beta})+W_{e\beta}\cos(\dcp+\phi_{e\beta}) \right]\nonumber\\
    \label{pmue-a}
\end{eqnarray}
where $\beta=\mu,\, \tau$; 
\begin{eqnarray}
 Z_{e\beta}&=&-\cos \theta_{23} \sin \dhat,\, {\rm if}\, \beta=\mu \nonumber\\
 &=& \sin \theta_{23}\sin \dhat,\, {\rm if}\, \beta=\tau
\end{eqnarray}
and
\begin{eqnarray}
 W_{e\beta}&=&\cos \theta_{23} \left(\frac{\sin^2\tz \sin \dhat}{\cos^2 \tx \dhat}+\cos \dhat\right),\, {\rm if}\, \beta=\mu \nonumber\\
 &=& \sin \theta_{23} \left(\frac{\sin\dhat}{\dhat}-\cos \dhat\right),\, {\rm if}\, \beta=\tau.
 \label{W}
\end{eqnarray}
From, eq.~(\ref{pmue-a}), it can be concluded that the LIV effects considered in this paper are matter independent.
The oscillation probability $\pmuebar$ for antineutrino can be calculated from eqs.~(\ref{pme}) and (\ref{pmue-a}) by substituting $A\to -A$, $\dcp \to -\dcp$, $|a_{e\beta}|\to -|a_{e\beta}|$ and $\phi_{e\beta}\to -\phi_{e\beta}$, where $\beta=\mu,\, \tau$.

Since the effects of the CPT-violating LIV terms are proportional to $L$, \nova will be more sensitive to CPT-violating LIV than T2K. Thus, \nova might be sensitive to CPT-violating LIV which T2K is insensitive to. Hence, LIV can be a possible explanation of the disagreement between the two experiments.

\begin{figure}[htbp]
\centering
\includegraphics[width=1.0\textwidth]{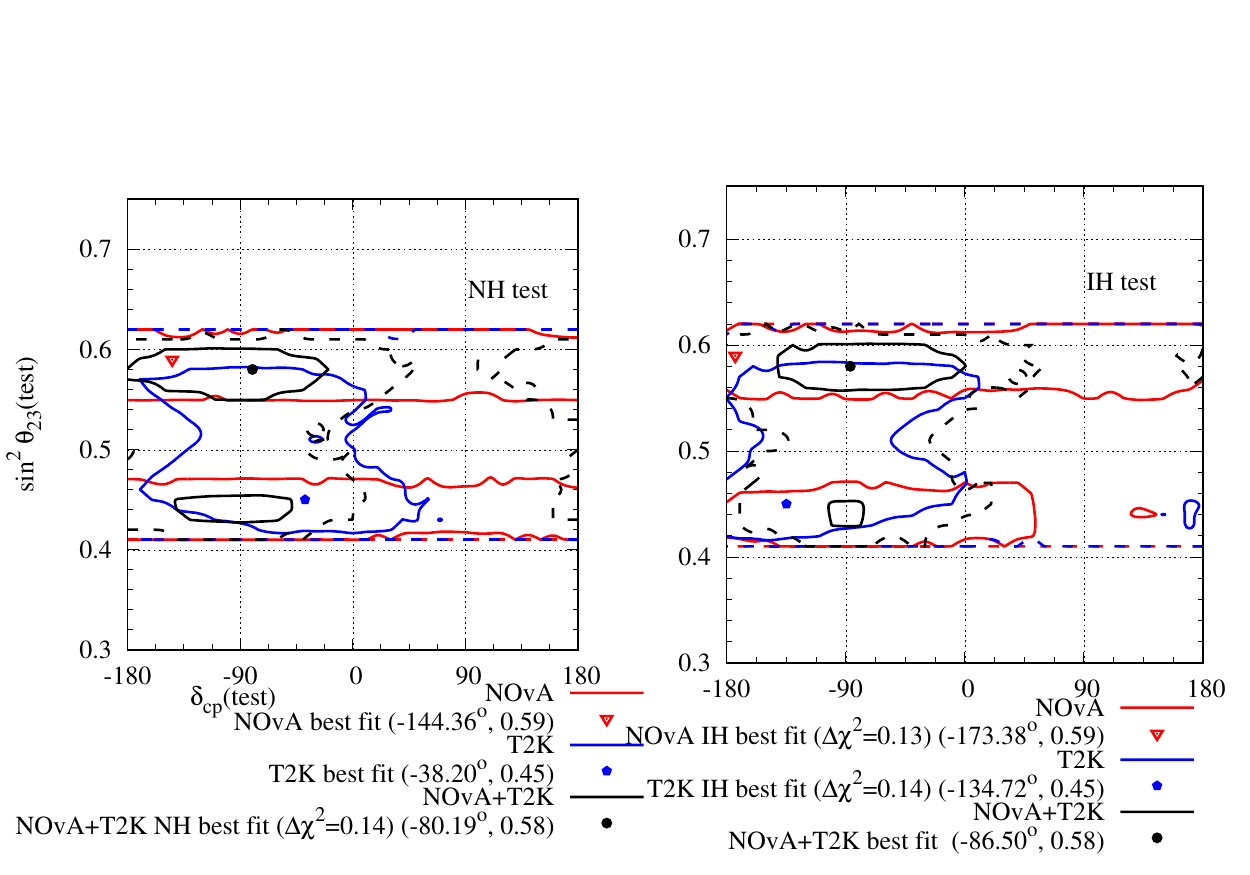}
\caption{\footnotesize{Allowed region in the $\sin^2 \tz-\dcp$ plane after analysing \nova and T2K complete data set with LIV hypothesis. The left (right) panel represents test hierarchy to be NH (IH). The red (blue) lines indicate the results for \nova (T2K)
and the black line indicates the combined analysis of both. The solid (dashed) lines indicate the $1\, \sigma$ ($3\, \sigma$)
allowed regions. The minimum $\chi^2$ for \nova (T2K) with 50 (88) bins is 47.71 (93.14) and it occurs at NH. For the combined analysis, the minimum $\chi^2$ with 138 bins is 145.09.}}
\label{LIV}
\end{figure}

In Ref.~\cite{Rahaman:2021leu}, the individual data from \nova and T2K, as well as the combination of data from both the experiments have been analysed with LIV. As before, GLoBES \cite{Huber:2004ka, Huber:2007ji} software was used to analyse the data. The software was modified to include LIV, and the oscillation probabilities and the event numbers were calculated in case of LIV without the approximations required to calculate oscillation probabilities given in eqs.~(\ref{pmue-LIV})-(\ref{W}). As we have mentioned before, among the LIV parameters, only $|a_{e\mu}|$ ($[0:20\times 10^{-23}]$ GeV), $|a_{e\tau}|$ ($[0:20\times 10^{-23}]$ GeV), $\phi_{e\mu}$ ($-180^\circ:180^\circ$), and $\phi_{e\tau}$ ($-180^\circ:180^\circ$) have been varied. All other LIV parameters have been kept fixed to zero. The standarad parameter values are same as  described in the subsection \ref{chisq}. We have represented the result on the $\sin^2\tz-\dcp$ plane in Fig.~\ref{LIV}. The minimum $\chi^2$ for \nova (T2K) is $47.71$ ($93.14$). For the combined analysis, the minimum $\chi^2$ is $145.09$. Both the experiments individually prefer NH as the best-fit hierarchy, although there are degenerate best-fit points at IH for both of them. The experiments lose their hierarchy sensitivity when analysed with LIV. The best-fit values of $\dcp$ at NH are close to each other. Moreover, there is a large overlap between the $1\, \sigma$ allowed regions of the two experiments. Thus, it can be concluded that the tension between the two experiments for NH has been reduced. Although, there is a new mild tension between the best-fit values of $\sin^2\tz$ from \nova and T2K as the former prefers $\tz$ to be in the HO while the later prefers it to be in the LO, both of them have a degenerate best-fit point at the other octant (LO for NO$\nu$A, and HO for T2K) as well. The combined experiment prefers IH over NH. However, just like the individual data, the combined data lose hierarchy (as well as octant) sensitivity and has a degenerate best-fit point at NH.

\begin{figure}[H]
\includegraphics[width=85mm,scale=2.0]{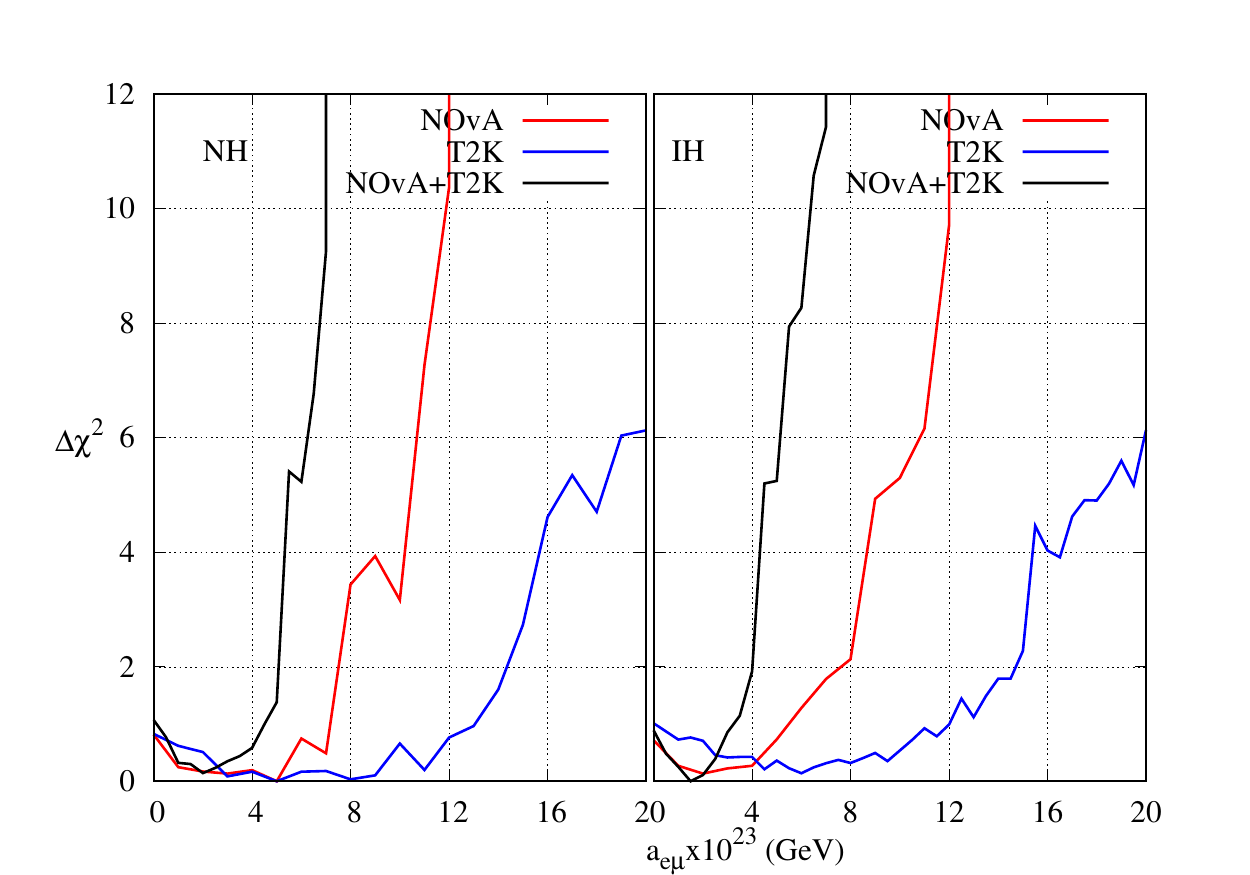}
\includegraphics[width=85 mm,scale=2.0]{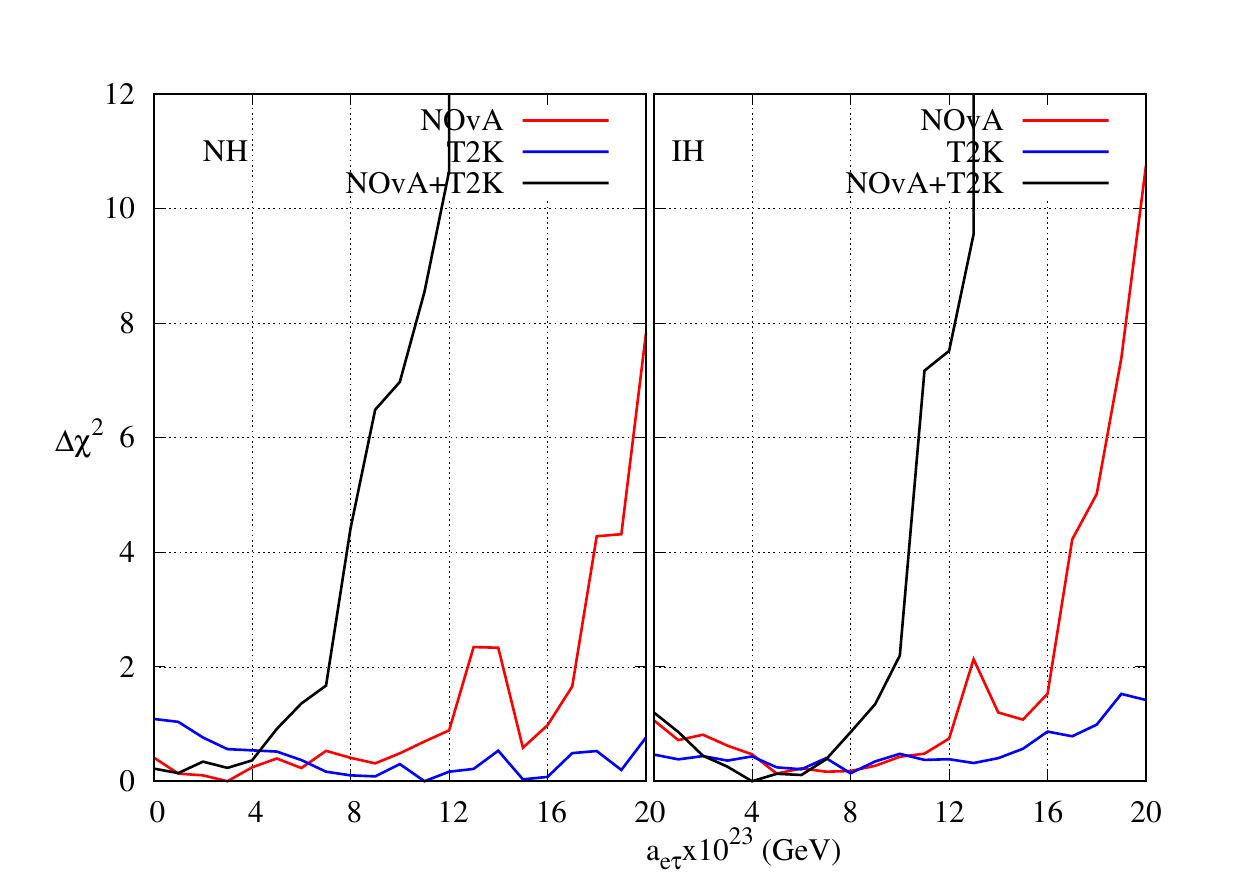}
\vskip -0.5cm
\includegraphics[width=85 mm,scale=2.0]{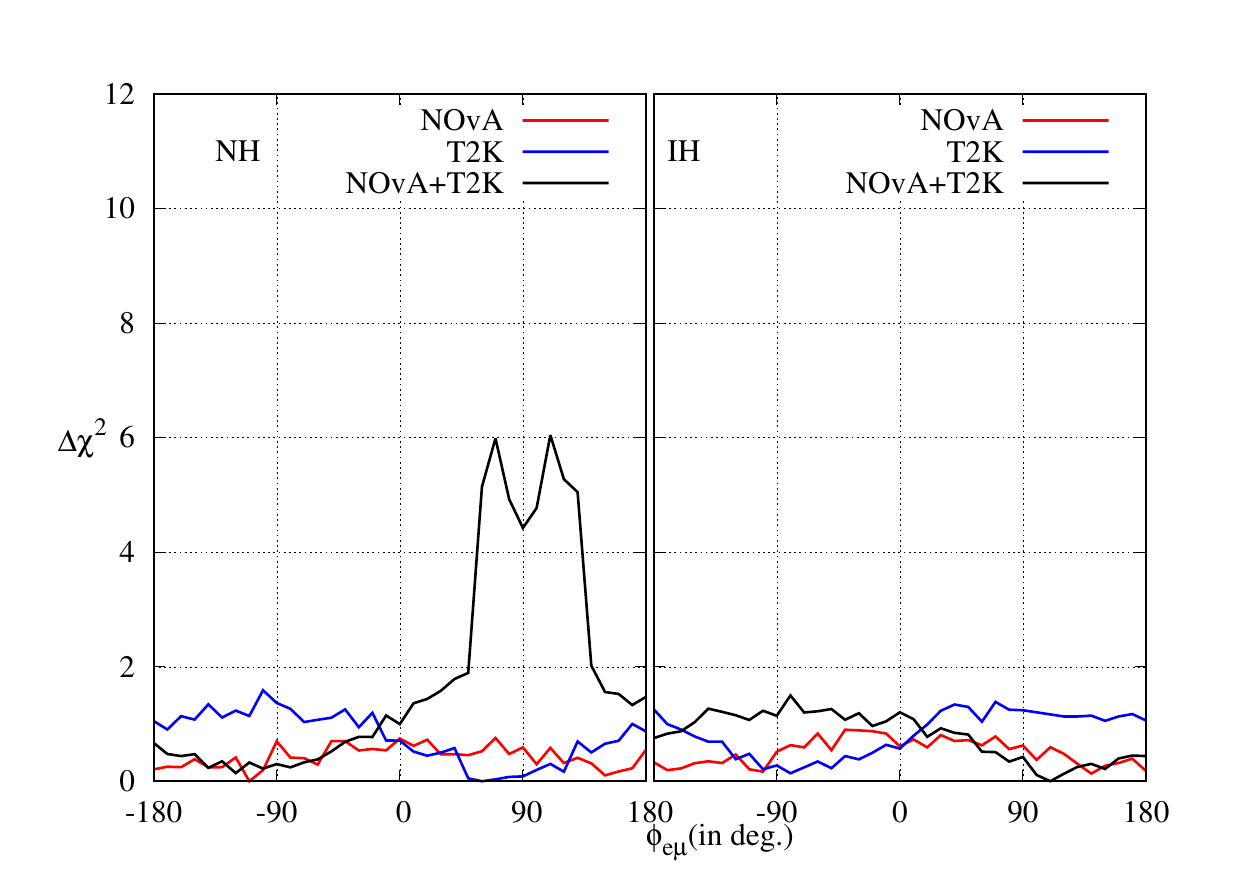}
\includegraphics[width=85 mm,scale=2.0]{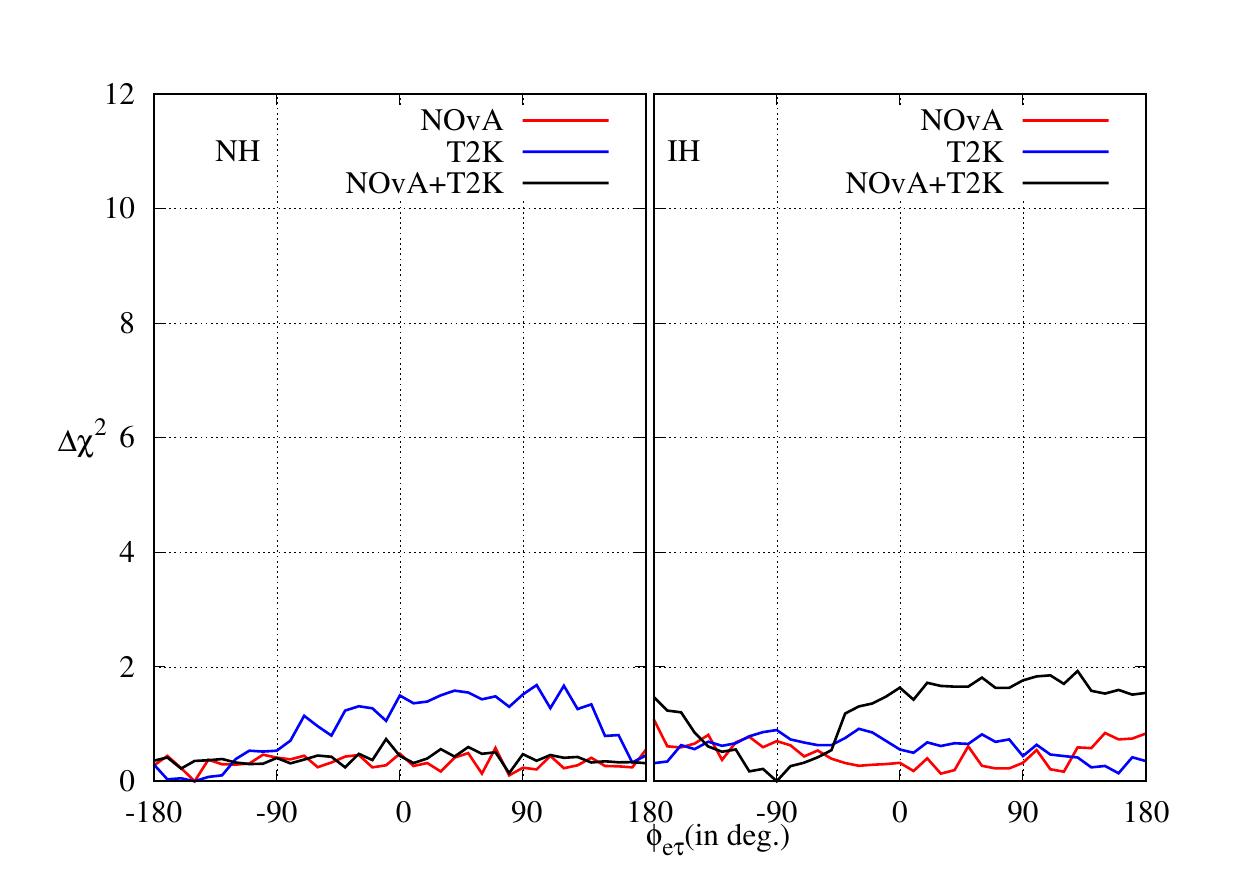}
\caption{\footnotesize{$\dchsq$ as a function of individual LIV parameters.}}
\label{param-LIV}
\end{figure}

\begin{figure}[H]
\centering
\includegraphics[width=0.75\textwidth]{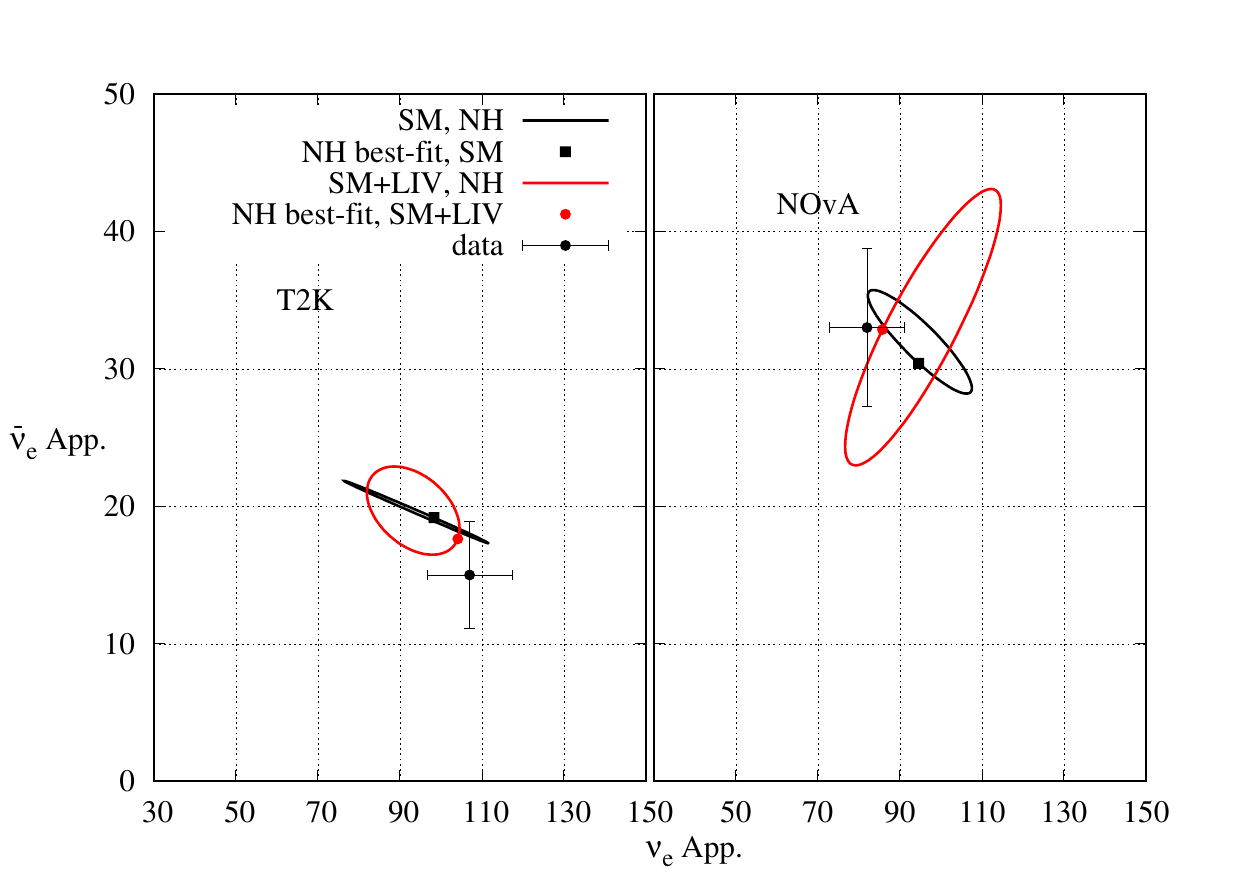}
\includegraphics[width=0.75\textwidth]{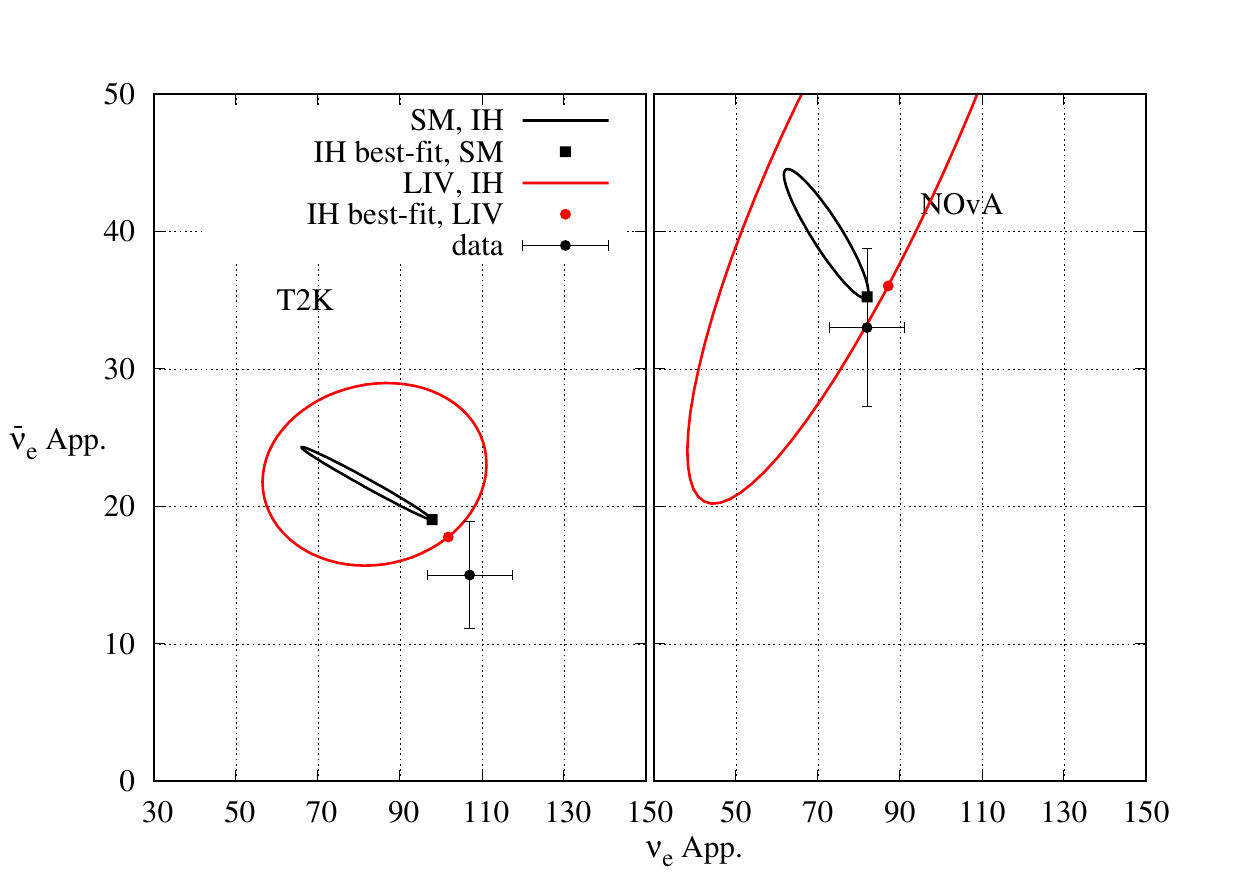}
\caption{\footnotesize{Bi-event plot for T2K (NO$\nu$A) in left (right) panel. The upper (lower) panel is for NH (IH). To generate the ellipses, $\dcp$ has been varied in the range $[-180^\circ:180^\circ]$ while keeping all other parameters fixed at the combined best-fit values. The black (red) ellipses represent the SM (LIV) case with the best-fit points indicated by black square (red circle). The ellipses and the best-fit points have been determined by fitting the combined data from \nova and T2K. The black circle with error bars represent the experimental data.}}
\label{bievents-LIV}
\end{figure}

In Fig.~\ref{param-LIV}, we have expressed $\dchsq$ as a function of the LIV parameters. It can be seen from the figure that the present T2K data disfavours standard oscillation at $1\, \sigma$ C.L., whereas the \nova data do not have any preference. The combined analysis disfavours standard oscillation at $1\, \sigma$ C.L.\ as well.
To emphasize the result, we have presented a similar bi-event plot like Fig.~\ref{bievents-nonuni} for the non-unitary oscillation analysis in Fig.~\ref{bievents-LIV} for the LIV analysis. It is obvious that inclusion of LIV in the theory brings the expected $\nu_e$, and $\bar{\nu}_e$ event numbers, of both the experiments, at the combined best-fit point at NH closer to the observed event numbers, and thus reduces the tension between the two experiments. For IH, inclusion of LIV brings the observed $\nu_e$, and $\bar{\nu}_e$ event numbers of T2K at the combined best-fit point closer to the observed event numbers. Although, LIV takes the observed $\nu_e$, and $\bar{\nu}_e$ event numbers of \nova at the combined IH best-fit point farther away from the observed event numbers, the change is minuscule. 

\subsection{Non-standard interaction (NSI)}
Non-standard interactions can arise as a low-energy manifestation of new heavy states of a more complete model at high energy \cite{Farzan:2017xzy, Biggio:2009nt, Ohlsson:2012kf, Miranda:2015dra, Proceedings:2019qno} or it can arise due to light mediators \cite{Farzan:2015doa, Farzan:2015hkd}. NSI can modify the neutrino and antineutrino flavour conversion in matter \cite{msw1, Mikheev:1986gs, Mikheev:1986wj}. The effect of NSI on the present and future long-baseline accelerator neutrino experiments have been discussed in details in literature \cite{Friedland:2012tq, Rahman:2015vqa, Coelho:2012bp, Masud:2018pig, Deepthi:2017gxg, Blennow:2016etl}. Ref.~\cite{Chatterjee:2020kkm,Denton:2020uda} have tried to invoke NSI to resolve the tension between \nova and T2K data. 

Neutral current NSI during neutrino propagation can be represented by a dimension 6 operator \cite{msw1}: 
\begin{equation}
    \mathcal{L}_{\rm NC-NSI}= -2\sqrt{2} G_F \epsilon_{\alpha \beta}^{fC}\left( \bar{\nu}_{\alpha}\gamma^\mu P_L \nu_\beta  \right)\left( \bar{f}\gamma_\mu P_C f \right)
    \label{lag-nsi},
\end{equation}
where $\alpha,\, \beta=e,\, \mu,\, \tau$ denote the neutrino flavour, $f=e,\, \mu,\, \tau$ denotes the fermions inside matter, $P$ is the projection operator with the superscript $C$ referring to the $L$ or $R$ chirality of the $ff$ current, and $\epsilon_{\alpha \beta}^{fC}$ denotes the strength of the NSI. From the hermiticity of the interaction,
\begin{equation}
    \epsilon_{\beta \alpha}^{fC}=\left( \epsilon_{\alpha \beta}^{fC} \right)^* \,.
\end{equation}
For neutrino propagation through earth matter, the relevant expression is
\begin{equation}
    \epsilon_{\alpha \beta} \equiv \sum_{f=e,u,d} \epsilon_{\alpha \beta}^{f}\frac{N_f}{N_e} \equiv \sum_{f=e,u,d} \left(\epsilon_{\alpha \beta}^{fL}+\epsilon_{\alpha \beta}^{fR}\right)\frac{N_f}{N_e},
\end{equation}
where $N_f$ is the density of $f$ fermion. If we consider earth matter to be neutral and isoscalar, then $N_n \simeq N_p=N_e$. Thus, 
\begin{equation}
    \epsilon_{\alpha \beta}\simeq \epsilon_{\alpha \beta}^{e}+3 \epsilon_{\alpha \beta}^{u}+ 3 \epsilon_{\alpha \beta}^{d} \,.
\end{equation}
The effective Hamiltonian for neutrino propagation in matter in presence of NSI can be written in the flavour basis as 
\begin{equation}
    H=H_{\rm vac}+H_{\rm mat}+H_{\rm NSI},
    \label{Ham-NSI}
\end{equation}
where 
\begin{equation}
    H_{\rm vac}=\frac{1}{2E}U\left[
\begin{array}{ccc}
m_{1}^{2} & 0 & 0\\
0 & m_{2}^{2} & 0\\
0 & 0 & m_{3}^{2}\\
\end{array}
\right]U^\dagger; H_{\rm mat}=\sqrt{2}G_FN_e \left[
\begin{array}{ccc}
1 & 0 & 0\\
0 & 0 & 0\\
0 & 0 & 0\\
\end{array}
\right];
\end{equation}
\begin{equation}
    H_{\rm NSI}=\sqrt{2}G_FN_e \left[
\begin{array}{ccc}
\epsilon_{ee} & \epsilon_{e\mu} & \epsilon_{e\tau}\\
\epsilon_{e\mu}^{*} & \epsilon_{\mu \mu} & \epsilon_{\mu \tau}\\
\epsilon_{e\tau}^{*} & \epsilon_{\mu\tau}^{*} & \epsilon_{\tau \tau}\\
\end{array}
\right].
\label{Ham-NSI3}
\end{equation}
It is important to note that the NC NSI Hamiltonian during propagation presented in eqs.~(\ref{Ham-NSI})-(\ref{Ham-NSI3}) are analogous to the CPT violating LIV Hamiltonian given in eqs.~(\ref{Ham})-(\ref{Ham-LIV3}). A relationship between CPT-violating LIV and NSI can be found by the following relation \cite{Diaz:2015dxa}:
\begin{equation}
    \epsilon_{\alpha \beta}=\frac{a_{\alpha \beta}}{\sqrt{2}G_F N_e}.
    \label{eps-a}
\end{equation}
The $\nu_\mu \to \nu_e$ oscillation probability with matter effect in presence of NSI during propagation can be written in the similar way as in eq.~(\ref{pmue-LIV}) \cite{kikuchi:2008vq, Agarwalla:2016fkh, Masud:2018pig}
    \begin{equation}
    P_{\mu e}^{\rm SM+NSI}\simeq P_{\mu e} (\rm SM)+P_{\mu e} (\epsilon_{e\mu})+P_{\mu e}(\epsilon_{e\tau}).
    \label{pmue-NSI}
\end{equation}
Just like the LIV case, we can write the second and third terms in eq.~(\ref{pmue-NSI}) as,
\begin{eqnarray}
    P_{\mu e}(\epsilon_{e\beta})&=& 4|\epsilon_{e\beta}|\ahat \dhat\sin \theta_{13}\sin 2\theta_{23}\sin \dhat\left[Z_{e\beta}\sin(\dcp+\phi_{e\beta})+W_{e\beta}\cos(\dcp+\phi_{e\beta}) \right] 
    \label{pmue-eps}
\end{eqnarray}
where $\beta=\mu,\, \tau$; 
\begin{eqnarray}
 Z_{e\beta}&=&-\cos \theta_{23} \sin \dhat,\, {\rm if}\, \beta=\mu \nonumber\\
 &=& \sin \theta_{23}\sin \dhat,\, {\rm if}\, \beta=\tau \,,
\end{eqnarray}
and
\begin{eqnarray}
 W_{e\beta}&=&\cos \theta_{23} \left(\frac{\sin^2\tz \sin \dhat}{\cos^2 \tx \dhat}+\cos \dhat\right),\, {\rm if}\, \beta=\mu \nonumber\\
 &=& \sin \theta_{23} \left(\frac{\sin\dhat}{\dhat}-\cos \dhat\right),\, {\rm if}\, \beta=\tau.
 \label{W-nsi}
\end{eqnarray}
Because of the $\ahat$ term in eq.~(\ref{pmue-eps}), the oscillation probability after inclusion of NSI is dependent on matter effect unlike LIV. The first oscillation maximum of \nova (T2K) peaks at $1.4$ GeV ($0.6$ GeV). Therefore, the matter effect is almost 3 times larger at \nova ($\ahat\simeq 0.14$) than at T2K ($\ahat \simeq 0.05$). Hence, \nova can observe NSI effects which can be remain unseen by T2K. Therefore, NSI can be a possible explanation behind the tension between the two experiments.

\begin{figure}[htb]
\centering
\vskip -1.5cm
\includegraphics[width=1.0\textwidth]{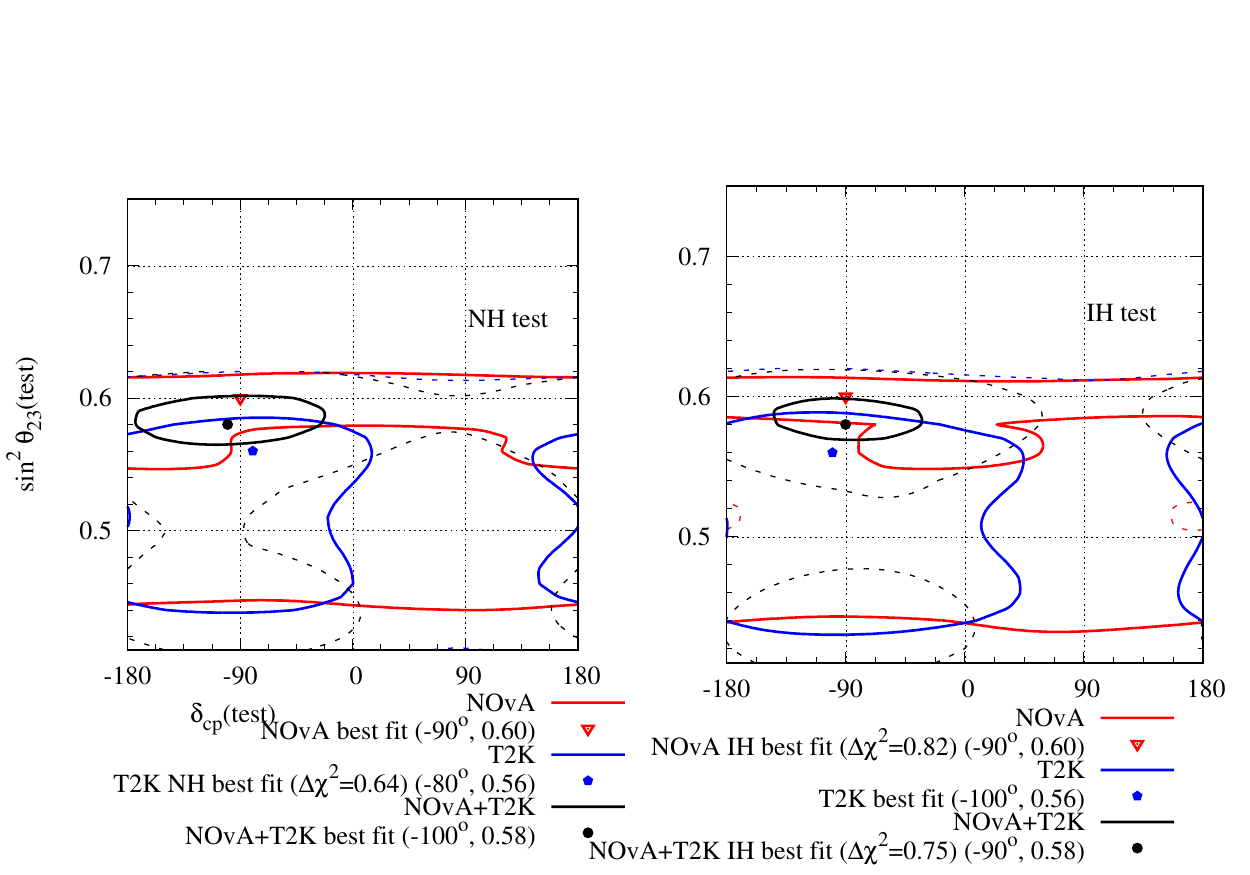}
\caption{\footnotesize{Allowed region in the $\sin^2 \tz-\dcp$ plane after analysing \nova and T2K complete data set with NSI hypothesis. Only the effect of $\epsilon_{e\mu}=|\epsilon_{e\mu}|e^{i\phi_{e\mu}^{\rm NSI}}$ has been considered. The left (right) panel represents test hierarchy to be NH (IH). The red (blue) lines indicate the results for \nova (T2K)
and the black line indicates the combined analysis of both. The solid (dashed) lines indicate the $1\, \sigma$ ($3\, \sigma$) allowed regions. The minimum $\chi^2$ for \nova (T2K) with 50 (88) bins is 49.08 (93.64) and it occurs at NH. For the combined analysis, the minimum $\chi^2$ with 138 bins is 146.26.}}
\label{NSI-1}
\end{figure}

In Ref.~\cite{Chatterjee:2020kkm, Denton:2020uda} an effort to resolve the tension with the NSI has been made. To do so, the authors first analysed the combined data of \nova and T2K. Finding the best-fit values of the NSI parameters after the combined analysis, they analysed the individual data from \nova and T2K with the NSI parameter values fixed at the combined best-fit values, and showed that the two experiments agree on their results on $\sin^2\tz-\dcp$ plane. In this review article, we have analysed the individual \nova and T2K data as well as their combined data independent of each other with NSI. To do so, at first we considered only $\epsilon_{e\mu}=|\epsilon_{e\mu}|e^{i\phi_{e\mu}^{\rm NSI}}$, and fixed all other NSI parameters to be $0$. We have presented the result on the $\sin^2\tz-\dcp$ plane in Fig.~\ref{NSI-1}. The minimum $\chi^2$ for \nova (T2K) with 50 (88) bins is 49.08 (93.64) and it occurs at NH (IH). For the combined analysis, the minimum $\chi^2$ with 138 bins is 146.26 and it is at NH. Both the experiments lose hierarchy sensitivity after analysing with NSI, and as a result \nova (T2K) has a degenerate best-fit point at IH (NH). The best-fit points of the two experiments are close to each other for both the hierarchies. However, they exclude each other's best-fit point at the $1\, \sigma$ C.L.\ for both the hierarchies. There is a significant overlap between the $1\, \sigma$ allowed regions of the two experiments. The experiments lose their octant sensitivity as well after the inclusion of NSI. At the best-fit points, the $\dcp$ values are close to $-90^\circ$ for both the experiments and both the hierarchies. The combined analysis prefers NH, $\tz$ in HO and $\dcp \sim -90^\circ$ as the best-fit point. However, there is a nearly degenerate best-fit point at IH, $\tz$ in HO and $\dcp \sim -90^\circ$. 

\begin{figure}[H]
\centering
\includegraphics[width=0.75\textwidth]{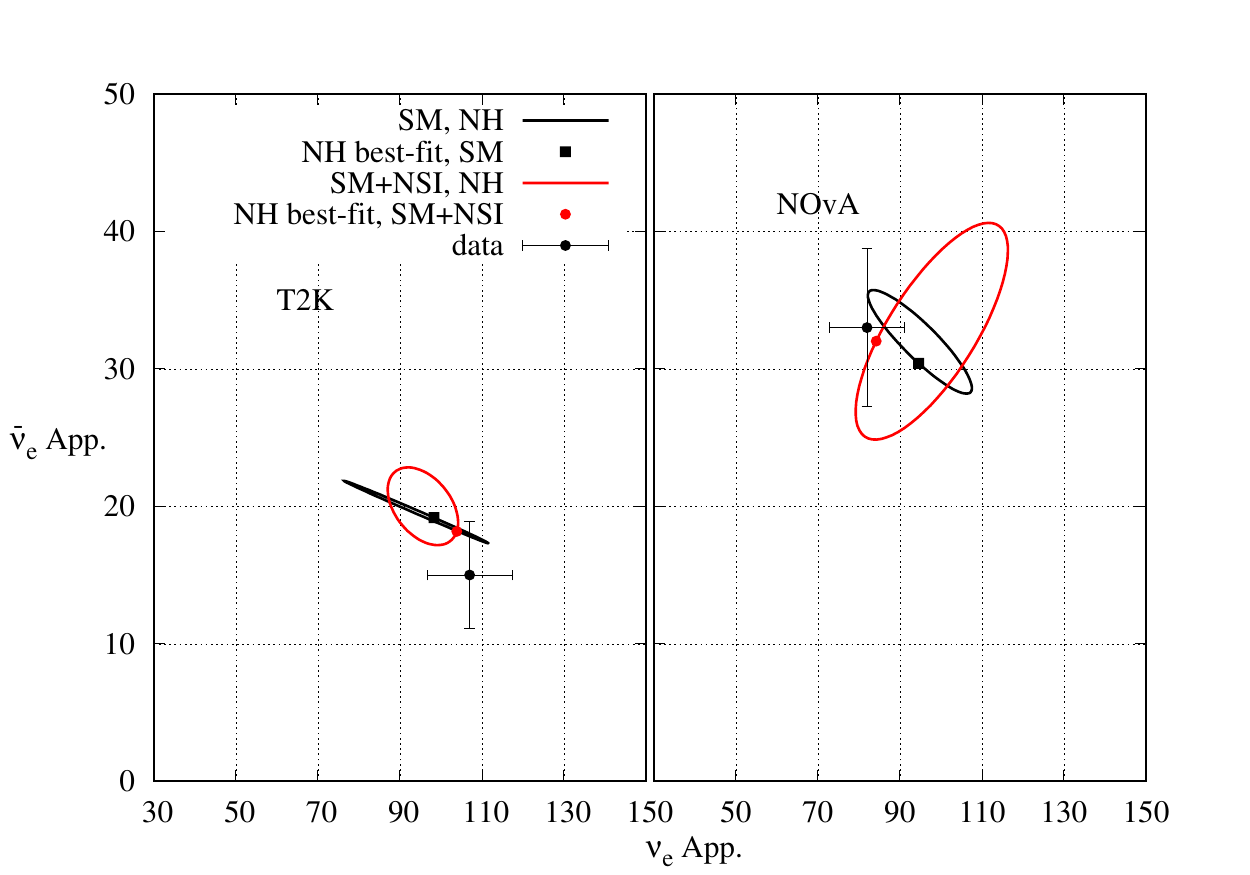}
\includegraphics[width=0.75\textwidth]{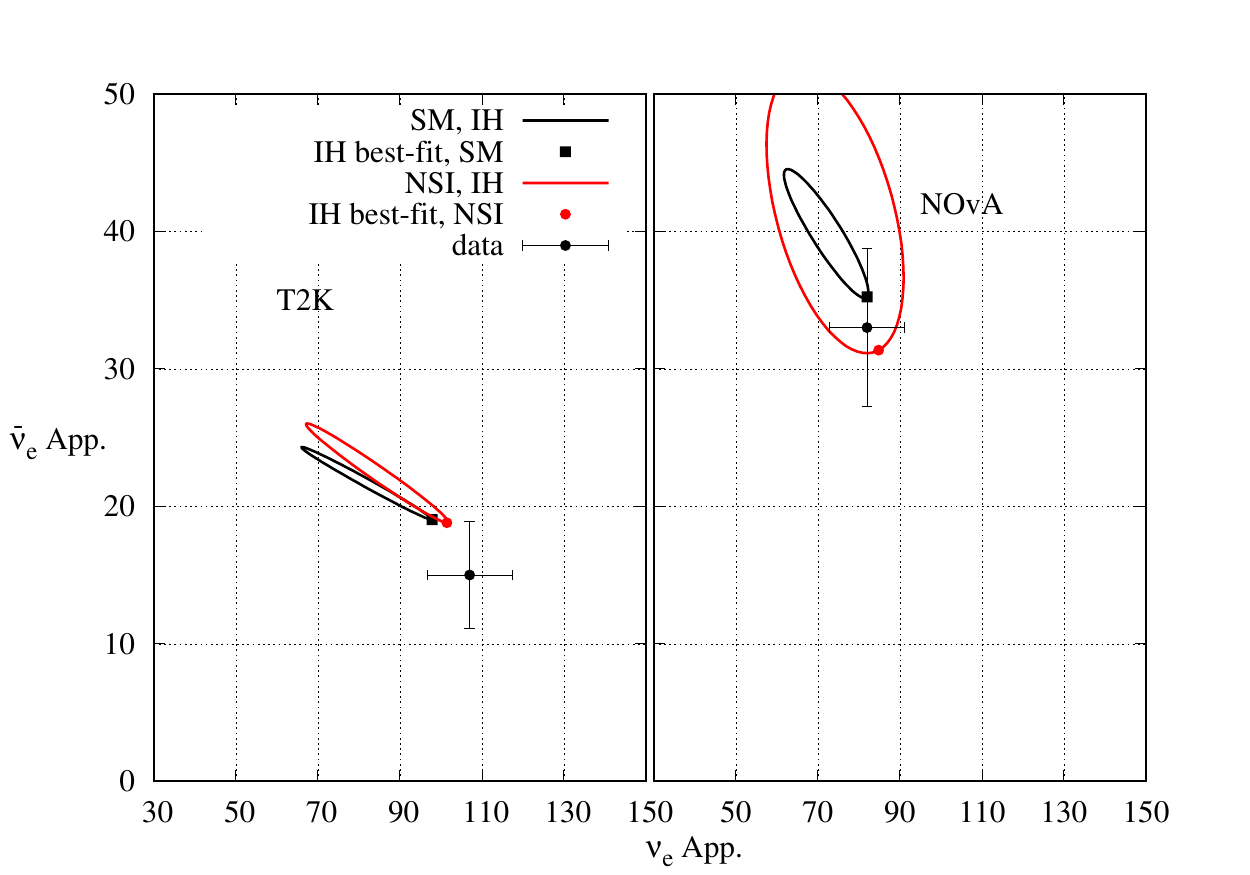}
\caption{\footnotesize{Bi-event plot for T2K (NO$\nu$A) in the left (right) panel. The upper (lower) panel is for NH (IH). To generate the ellipses, $\dcp$ has been varied in the range  $[-180^\circ:180^\circ]$ while keeping all other parameters fixed. Among the NSI parameters only the effect of $\epsilon_{e\mu}$ has been considered. The black (red) ellipses represent the SM (NSI) case with the best-fit points indicated by black square (red circle). The ellipses and the best-fit points have been determined by fitting the combined data from \nova and T2K. The black circle with error bars represent the experimental data.}}
\label{bievents-em}
\end{figure}

Just like the non-unitary mixing and LIV, we have emphasized our argument about resolution of the tension with the inclusion of NSI due to $\epsilon_{e\mu}$ through bi-event plots presented in Fig.~\ref{bievents-em}. It is clear that for both the experiments, inclusion of NSI due to $\epsilon_{e\mu}$ brings the expected $\nu_e$ and $\bar{\nu}_e$ event numbers at the combined best-fit points closer to the observed event numbers for NH, and thus resolves the tension between the two experiments for NH. For IH, the change in expected event numbers for T2K at the combined best-fit point after inclusion of NSI due to $\epsilon_{e\mu}$ is negligible. For NO$\nu$A, expected $\bar{\nu}_e$ event numbers at the combined IH best-fit point comes closer to the observed event number after inclusion of NSI due to $\epsilon_{e\mu}$, whereas change in the expected $\nu_e$ appearance event number due to the same is negligible.

\begin{figure}[H]
\centering
\vskip -1.5cm
\includegraphics[width=1.0\textwidth]{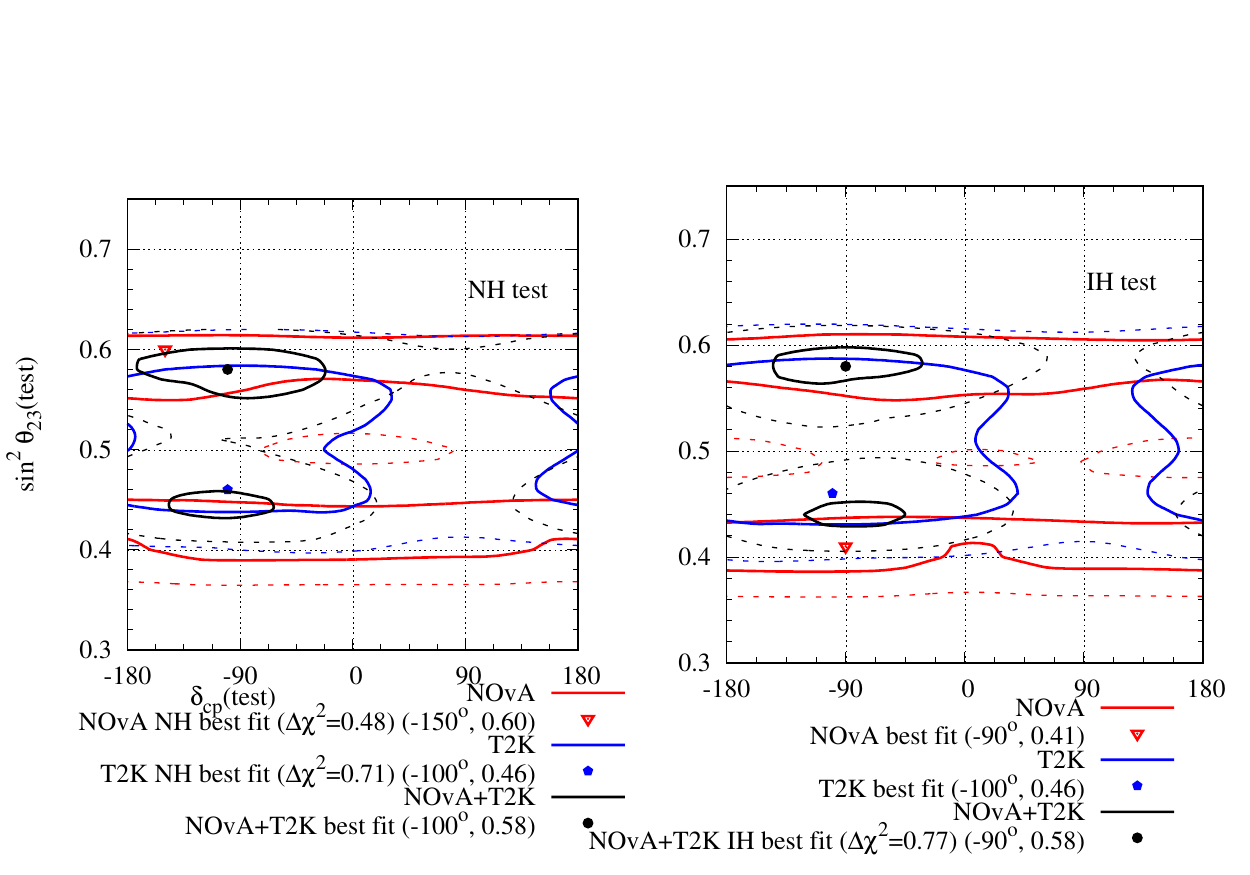}
\caption{\footnotesize{Allowed region in the $\sin^2 \tz-\dcp$ plane after analysing \nova and T2K complete data set with the NSI hypothesis. Only the effect of $\epsilon_{e\tau}=|\epsilon_{e\tau}|e^{i\phi_{e\tau}^{\rm NSI}}$ has been considered. The left (right) panel represents test hierarchy to be NH (IH). The red (blue) lines indicate the results for \nova (T2K)
and the black line indicates the combined analysis of both. The solid (dashed) lines indicate the $1\, \sigma$ ($3\, \sigma$)
allowed regions. The minimum $\chi^2$ for \nova (T2K) with 50 (88) bins is 48.59 (93.73) and it occurs at IH. For the combined analysis, the minimum $\chi^2$ with 138 bins is 146.38 and it is at NH.}}
\label{NSI-2}
\end{figure}

In the next step, we have considered the effects of $\epsilon_{e\tau}=|\epsilon_{e\tau}|e^{i\phi_{e\tau}^{\rm NSI}}$. All other NSI parameters have been kept fixed at $0$. The result has been displayed on the $\sin^2\tz-\dcp$ plane in Fig.~\ref{NSI-2}. The minimum $\chi^2$ for \nova (T2K) with 50 (88) bins is 48.59 (93.73) and it occurs at IH. For the combined analysis, the minimum $\chi^2$ with 138 bins is 146.38 and it is at NH. Although \nova and T2K both individually prefer IH and $\tz$ in LO as their best-fit point, both of them lose hierarchy and octant sensitivity after consideration of NSI parameter $\epsilon_{e\tau}$, and therefore both of them have a degenerate best-fit point at NH as well as $\tz$ in HO. The best-fit points of the two experiments are close to each other. However, both of them exclude each other's best-fit point at $1\, \sigma$ C.L.\ for both the hierarchies. Nonetheless, there is a significant overlap between the $1\, \sigma$ allowed region of the two experiments after including the effect of the NSI parameter $\epsilon_{e\tau}$. At the best-fit points the $\dcp$ values are close to $-90^\circ$ for both the experiments and both the hierarchies. The combined analysis prefers NH, $\tz$ in HO and $\dcp \sim -90^\circ$ as the best-fit point. However, there is a nearly degenerate best-fit point at IH, $\tz$ in HO and $\dcp \sim -90^\circ$.

\begin{figure}[H]
\centering
\includegraphics[width=0.75\textwidth]{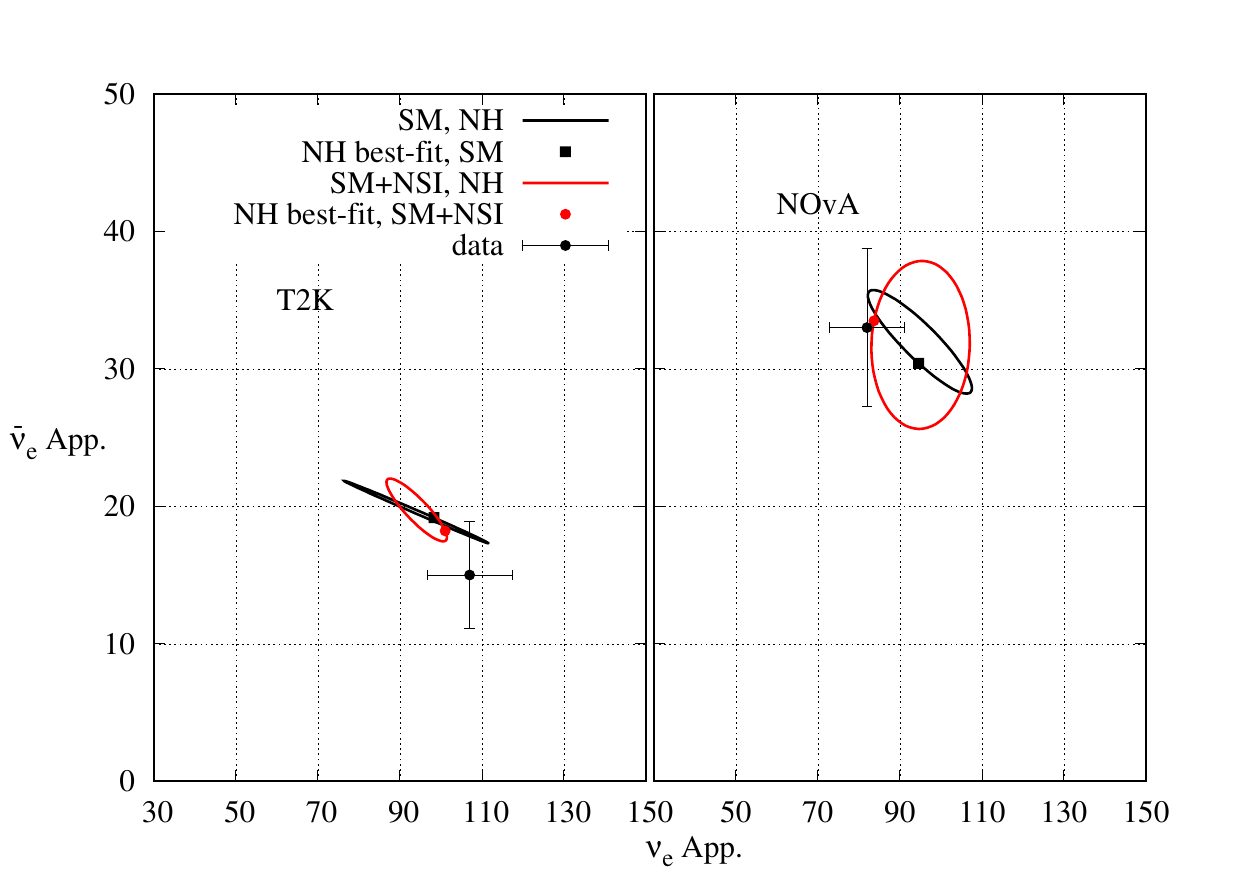}
\includegraphics[width=0.75\textwidth]{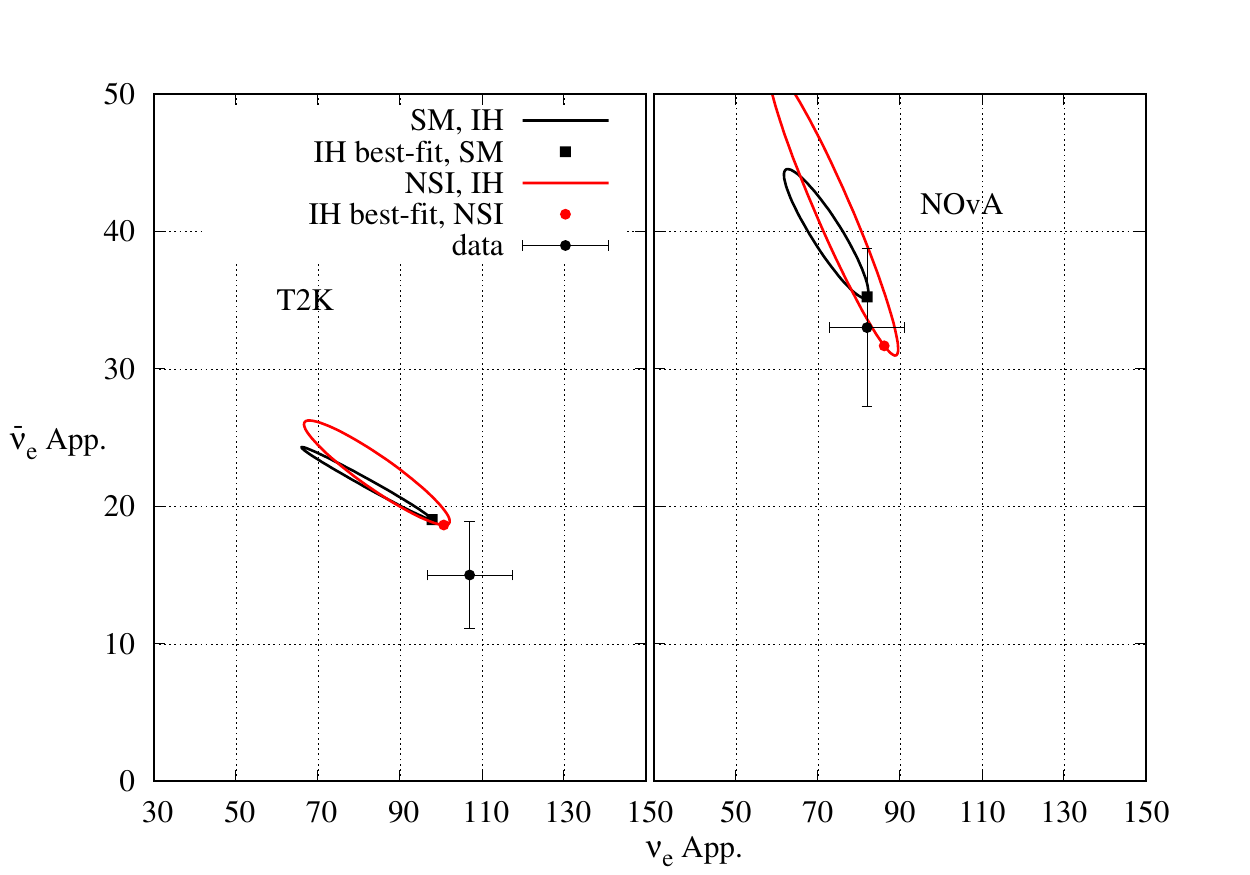}
\caption{\footnotesize{Bi-event plot for T2K (NO$\nu$A) in left (right) panel. The upper (lower) panel is for NH (IH). To generate the ellipses, $\dcp$ has been varied in the range $[-180^\circ:180^\circ]$ while keeping all other parameters fixed. Among the NSI parameters, only the effect of $\epsilon_{e\tau}$ has been considered. The black (red) ellipses represent the SM (NSI) case with the best-fit points indicated by black square (red circle). The ellipses and the best-fit points have been determined by fitting the combined data from \nova and T2K. The black circle with error bars represent the experimental data.}}
\label{bievents-et}
\end{figure}

As before, we have emphasized our argument about resolution of the tension between \nova and T2K with the inclusion of NSI due to $\epsilon_{e\tau}$ through bi-event plots in Fig.~\ref{bievents-et}. At the best-fit point of the combined analysis, the value of $|\epsilon_{e\tau}|$ is $0.73$. However, there is a nearly degenerate best-fit point at $|\epsilon_{e\tau}|=0.19$ with $\dchsq=0.13$. Because of the stronger constraint against $\epsilon_{e\tau}$ to be large from IceCube data \cite{Ehrhardt:2019}, we consider the combined best-fit point at $|\epsilon_{e\tau}|=0.19$. It is clear that for NH, inclusion of NSI due to $\epsilon_{e\tau}$ brings the expected $\nu_e$ and $\bar{\nu}_e$ appearance events for both the experiments at their combined best-fit point closer to the observed event numbers. For IH, the change in expected $\nu_e$ and $\bar{\nu}_e$ appearance events for T2K at the combined best-fit point is negligible. For NO$\nu$A, after the inclusion of NSI due to $\epsilon_{e\tau}$ the expected $\bar{\nu}_e$ event numbers at the combined IH best-fit point comes closer to the observed event number, whereas the change in expected $\nu_e$ appearance event number due to the same is quite small.

\begin{figure}[H]
\includegraphics[width=85 mm,scale=2.0]{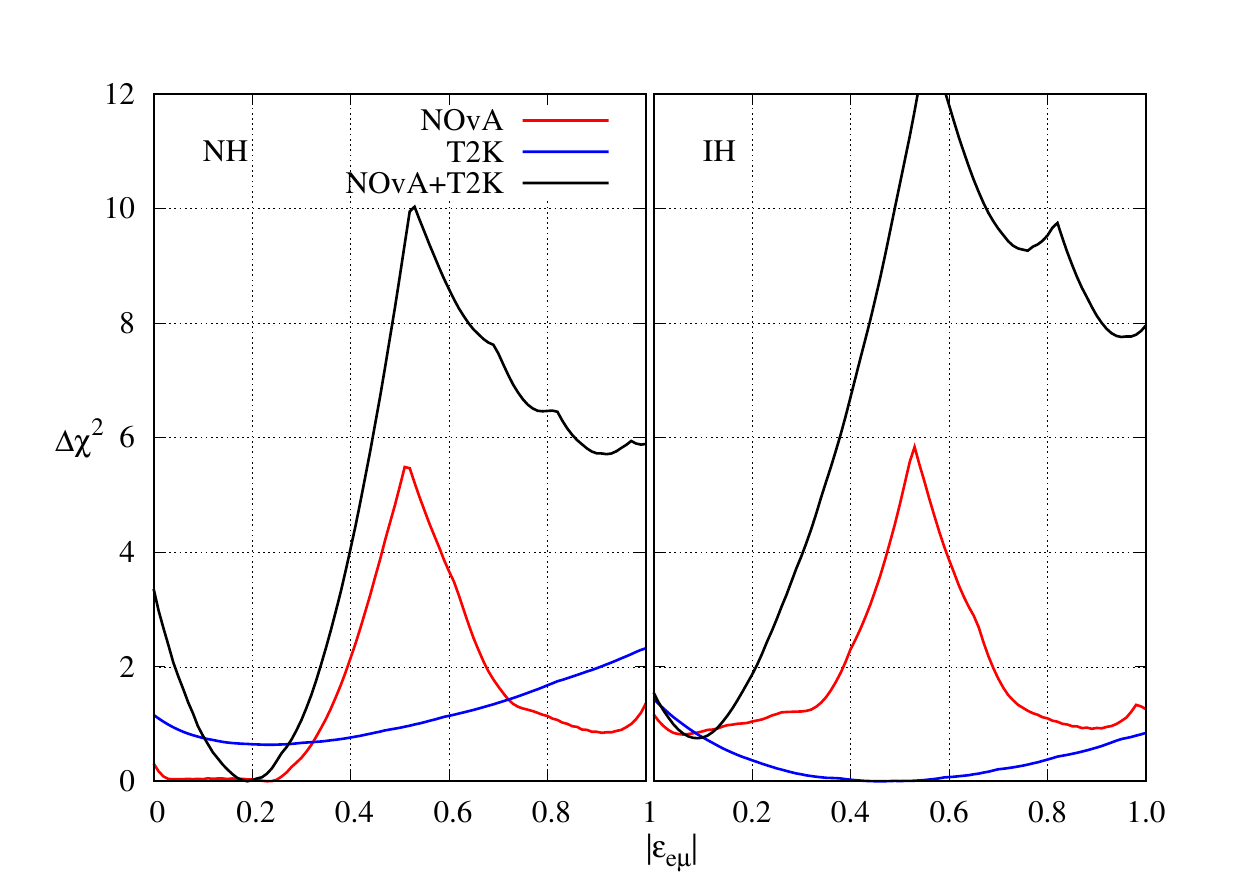}
\includegraphics[width=85 mm,scale=2.0]{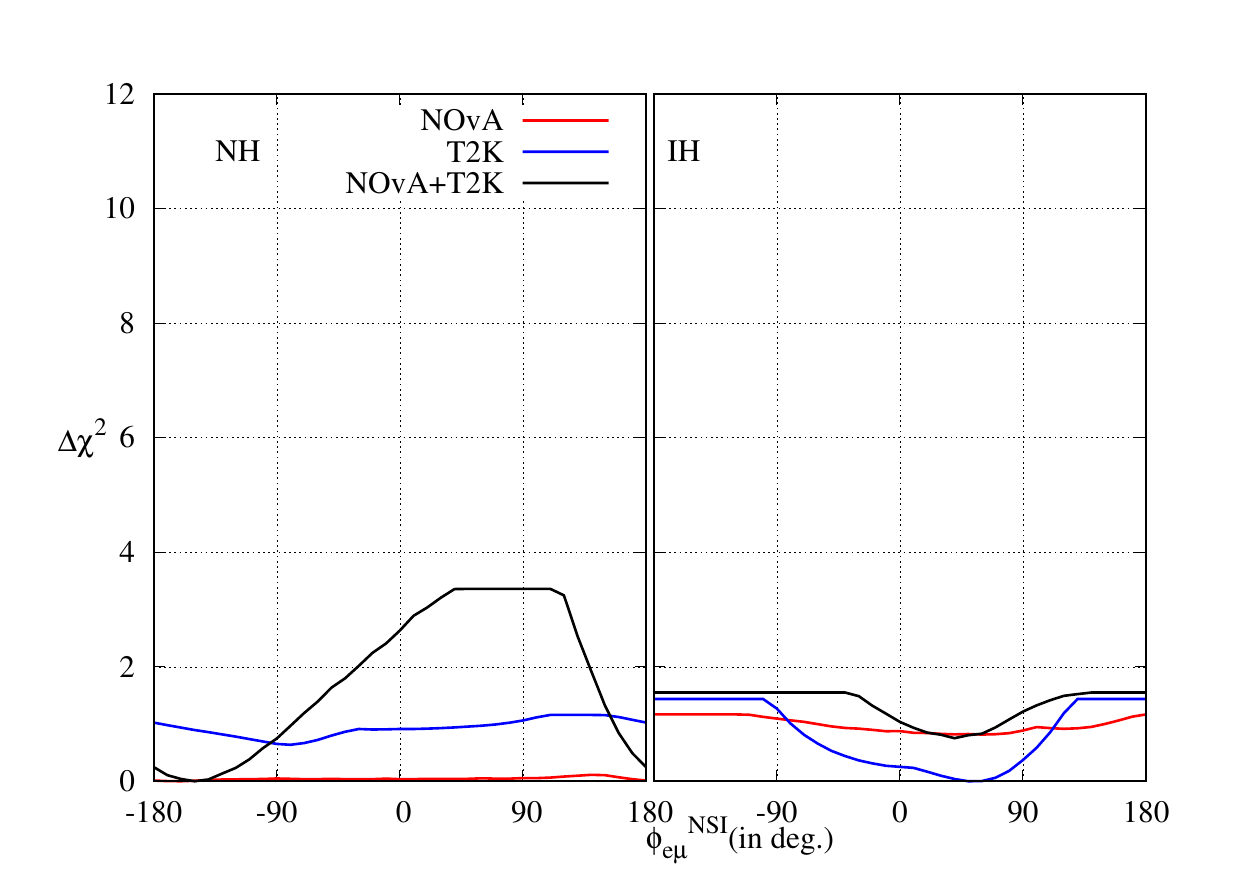}
\vskip -0.5cm
\includegraphics[width=85 mm,scale=2.0]{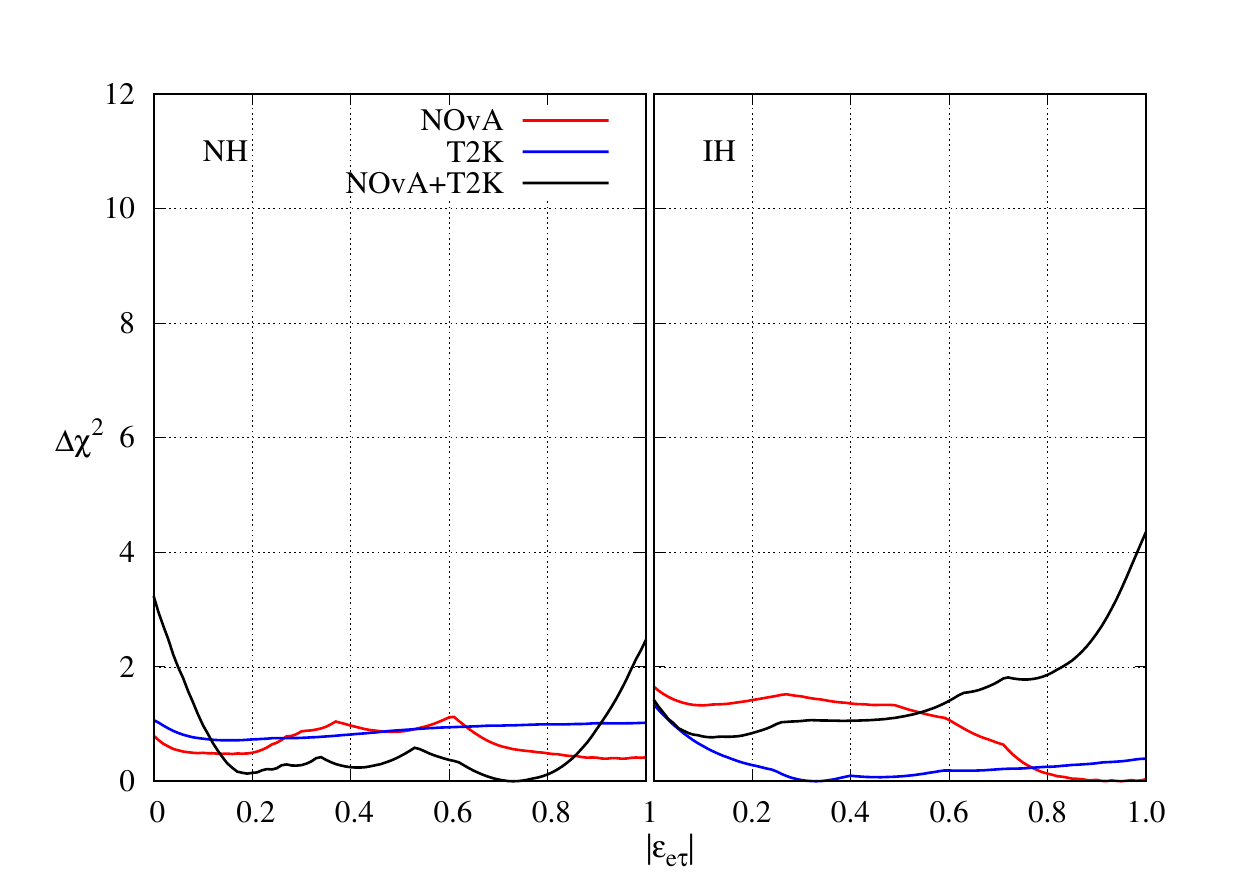}
\includegraphics[width=85 mm,scale=2.0]{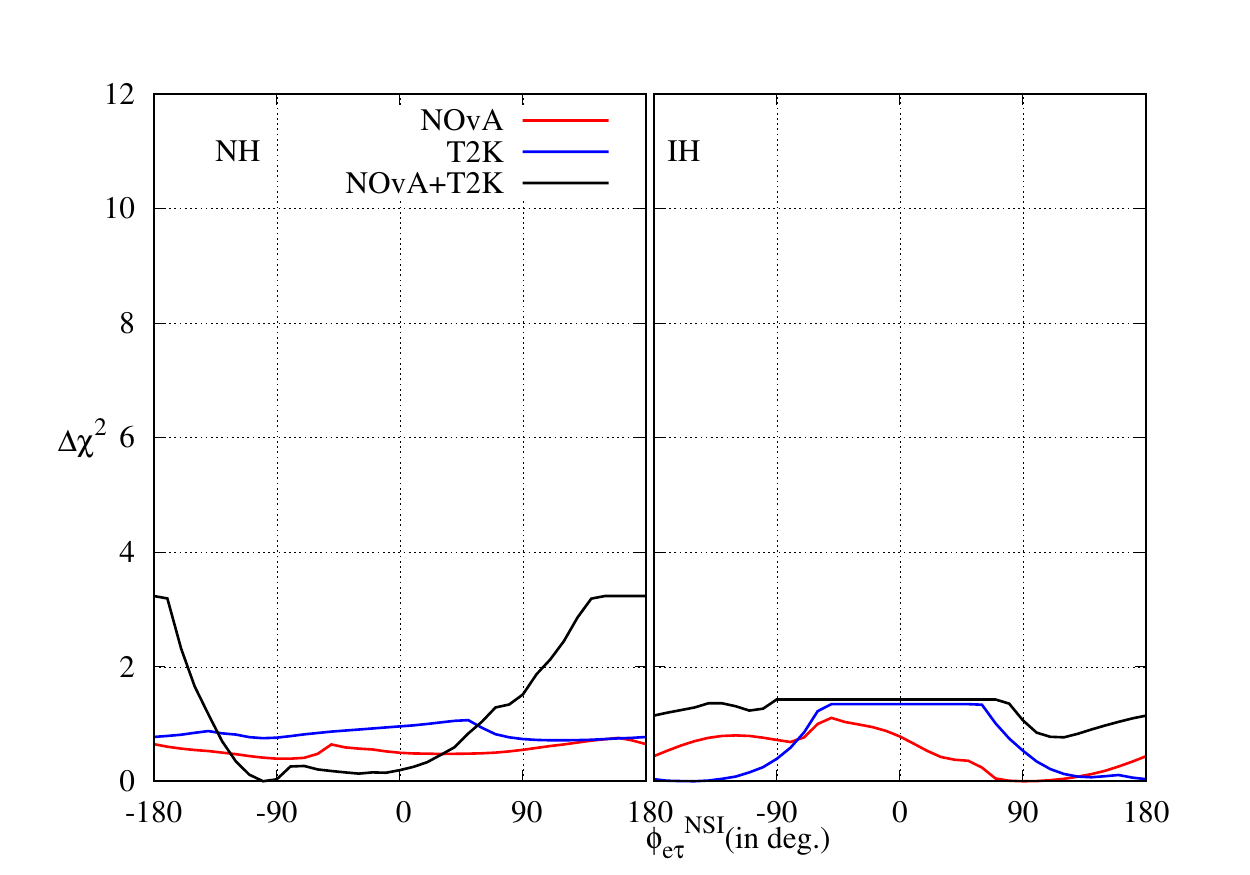}
\caption{\footnotesize{$\dchsq$ as a function of individual NSI parameters.}}
\label{param-NSI}
\end{figure}

In Fig.~\ref{param-NSI}, we have presented the precision plots for the NSI parameters. It can be concluded that when we consider the effect of $\epsilon_{e\mu}$, the present \nova data cannot make any preference between SM and NSI. However, both T2K and the combined data rule out SM at $1\, \sigma$ C.L. When the effect of $\epsilon_{e\tau}$ is considered, all three cases -- NO$\nu$A, T2K, and their combined data -- rule out SM at $1\, \sigma$ C.L.

The best-fit values for various non-standard parameters discussed in this section have been listed in table~\ref{BSM}. The $90\%$ C.L.\ limit for $1$ degree of freedom (d.o.f.) of these parameters have been mentioned as well in the parenthesis. When the $90\%$ limit falls beyond the studied range of a parameter, NA has been mentioned instead of a number.
\begin{table}
\hspace* {-25 mm}
{\footnotesize
  \begin{tabular}{|l|l|l|l|l|l|l|}
    \hline
    {Parameters} &
      \multicolumn{2}{c|}{NO$\nu$A} &
      \multicolumn{2}{c|}{T2K} &
       \multicolumn{2}{c|}{NO$\nu$A+T2K} \\

    & NH& IH  & NH & IH  & NH &IH \\
    \hline
      $\alpha_{00}$  &  $0.84$ (NA) & $0.72$ (NA) & $0.80$ (NA) & $0.80$ (NA)& $0.70$ (NA) & $0.76$ (NA) \\
     
     \hline
     $|\alpha_{10}|$ & $0.080$ (NA) & $0.0.12^{+0.06}_{-0.12}$ (NA) & $0.082^{+0.108}_{-0.082}$ ($<0.190$)  & $0.080^{+0.110}_{-0.060}$ (NA) &$0.125^{+0.025}_{-0.085}$ ($<0.170$)&$0.110^{+0.040}_{-0.070}$ ($<0.155$) \\
     \hline
   $\alpha_{11}$& $0.97^{+0.02}_{-0.03}$ ($>0.92$)&$0.96^{+0.03}_{-0.03}$ ($>0.92$)&$0.98$ ($>0.95$)&$0.98^{+0.02}_{-0.03}$ ($>0.95$)& $0.98^{+0.01}_{-0.02}$ ($>0.95$)& $0.98^{+0.01}_{-0.02}$ ($>0.95$)\\
   
    \hline

    $\phi_{10}/^\circ$&$-(125.68^{+54.32}_{-305.68})$&$76.15^{+103.85}_{-86.4}$&$54.77^{+97.10}_{-60.54}$&$112.69^{+42.38}_{-79.93}$&$120.41^{+59.57}_{-300.33}$ & $4.31^{+162.71}_{-181.51}$ \\
    
    \hline
    \hline
   
    $\frac{|a_{e\mu}|}{10^{-23}{\rm GeV}}$ &$4.81$($<8.19$)&$2.22$ ($<7.78$)&$4.60$ ($<15.25$)&$6.17_{-6.02}^{+4.86}$ ($<14.92$)&$1.86^{+2.57}_{-1.86}$ ($<4.80$)&$1.52$ ($<3.80$)\\
    \hline
    $\frac{|a_{e\tau}|}{10^{-23}{\rm GeV}}$ &$2.52$ ($<3.18$)&$5.33^{+9.20}_{-5.33}$ ($<15.71$) &$11.14$ (NA)&$8.06$ (NA)& $0.57$ ($<6.70$) & $4.16$ ($<9.50$)\\

    \hline
    $\phi_{e\mu}/^\circ$ &$-114.52$ & $141.18$ & $64.29$&$-77.34$ & $-115.72$ & $110.08$\\

    \hline
    $\phi_{e\tau}/^\circ$ & $-145.02$&$25.04$ &$-153.24$&$158.07$ & $84.36$ & $-89.27$\\

    \hline
    \hline
     $|\epsilon_{e\mu}|$ &$0.23$ ($<0.40$)&$0.06^{+0.10}_{-0.04}$ ($<0.42$) &$0.24_{-0.21}^{+0.28}$ (NA) &$0.46$ (NA) &$0.19_{-0.10}^{+0.11}$ ($<0.36$)&$0.09_{-0.05}^{+0.05}$ ($<0.24$)\\
    \hline
    $|\epsilon_{e\tau}|$ &$0.16$ (NA)&$0.95$ (NA) &$0.15$ (NA)&$0.33$ (NA) & $0.19_{-0.09}^{+0.72}$ (NA) & $0.12_{-0.08}^{+0.13}$ ($<0.91$) \\

    \hline
    $\phi_{e\mu}^{\rm NSI}/^\circ$ &$-160$ & $60$ & $-80$&$50$ & $-150$ & $40$\\

    \hline
    $\phi_{e\tau}^{\rm NSI}/^\circ$ & $30$&$90$ &$120$&$-150$ & $-30$ & $120$\\

    \hline
  \end{tabular}
  }
  \caption{Best-fit values of several BSM parameters discussed in section \ref{resolution} along with the $1\,\sigma$ error bar, where available. The $90\%$ C.L.\ limits for 1 d.o.f.\ have been mentioned in the parenthesis. When the $90\%$ limit falls beyond the studied range of a parameter, NA has been mentioned.}
  \label{BSM}
\end{table}

\section{Summary and discussion}
\label{summary}
A tension between the best-fit points of T2K and \nova existed from the very beginning,
which became only stronger with time. This tension arises mostly from the $\nu_e$ appearance data of the two experiments. T2K observes a large excess in the $\nu_e$ appearance events compared to the expected event number at the reference point of vacuum oscillation, $\tz=\pi/4$, and $\dcp=0$ (referred to as $000$). This large excess dictates that $\dcp$ be anchored around $-90^\circ$ and that $\tz$ be in HO, for both the hierarchies ($+++$, and $-++$ with former being the best-fit point). The appearance events observed by \nova show a very 
different pattern. They are moderately larger than the expectation from the reference point in
the $\nu_e$ channel and are consistent with it in the $\bar{\nu}_e$ channel. These two facts, when combined together, lead to two possible degenerate solutions for \nova: A. NH - $\tz$ in HO - $\dcp$ in UHP ($++-$), and B. IH - $\tz$ in HO - $\dcp$ in LHP ($-++$). A fit of the combined T2K + \nova data to standard three flavour oscillation framework, has the 
best-fit point as IH - $\tz$ in HO - $\dcp$ in LHP which is reasonably close to the IH
best-fit points of T2K and NO$\nu$A. If NH is assumed to be the true hierarchy, there is 
almost no allowed region within $1\,\sigma$, even though the best-fit point of each experiment picks NH. This is the essential tension between the two experiments. 

Several studies have been done to resolve this tension with BSM physics. Three different BSM scenarios have been considered in the literarture: 1. non-unitary mixing, 2. Lorentz invariance violation, and 3. non-standard interaction during neutrino propagation. All these three scenarios bring the expected event numbers at the combined best-fit point at NH closer to the observed $\nu_e$, and $\bar{\nu}_e$ event numbers of both the experiments, and thus reduce the tension between them. 

T2K and \nova individually prefer non-unitary mixing over unitary mixing at $1\, \sigma$ C.L. Combined data from both of them prefer non-unitary mixing at $2\,\sigma$ C.L. Both the experiments lose hierarchy and octant sensitivity when analysed with non-unitary mixing. There is a large overlap between the $1\, \sigma$ allowed regions on the $\sin^2\tz-\dcp$ plane of the two experiments. 

In the case of LIV, T2K data prefers LIV over standard 3-flavour oscillation at $1\, \sigma$. \nova data cannot make any preference between the two hypothesis at $1\, \sigma$. The combined analysis rules out standard oscillation at $1\, \sigma$ C.L. Just like, non-unitary mixing, both the experiments lose hierarchy, and octant sensitivity in the case of LIV too. In this case also, there is a large overlap between the $1\, \sigma$ allowed region on the $\sin^2\tz-\dcp$ plane of the two experiments. 

In the case of NSI, we considered the effects of $\epsilon_{e\mu}=|\epsilon_{e\mu}|e^{i\phi_{e\mu}^{\rm NSI}}$, and $\epsilon_{e\tau}=|\epsilon_{e\tau}|e^{i\phi_{e\tau}^{\rm NSI}}$ one at a time. In case of $\epsilon_{e\mu}$, T2K data rule out standard oscillation
at $1\,\sigma$ C.L., whereas \nova data cannot make a preference between the two hypothesis. The combined data from \nova and T2K rule out standard oscillation at more than $1.5\, \sigma$ C.L. In the case of $\epsilon_{e\tau}$, data from both \nova and T2K rule out standard oscillation at $1\, \sigma$, whereas their combined data rule out standard oscillation at $1.5\, \sigma$ C.L. As before, in the case of NSI also, each of the two experiments loses their hierarchy and octant sensitivity. A large overlap between the $1\,\sigma$ allowed regions on the $\sin^2\tz-\dcp$ plane of the two experiments exists.

T2K and \nova continue to take data. The additional data may either sharpen or reduce
the tension. If the tension becomes sharper, then we need to explore which new physics
scenario can best relieve this tension. We also need to test the predictions of the
preferred new physics scenario at future neutrino oscillation experiments, such as
T2HK \cite{Hyper-Kamiokande:2016srs} and DUNE \cite{Abi:2018alz, Abi:2018dnh, Abi:2018rgm}.

\acknowledgments
U.R.\ and S.R.\ were supported by a grant from the Research Council of the University of Johannesburg. 

\bibliographystyle{apsrev}
\bibliography{referenceslist}

\end{document}